\documentclass[prd,aps,showpacs,nofootinbib,tightenlines]{revtex4}
\usepackage{mathrsfs}
\usepackage{amsmath}
\usepackage{amssymb}
\usepackage{epsfig}
\usepackage{graphicx}
\usepackage{booktabs}
\usepackage{multirow}
\usepackage{subfigure}
\usepackage{color}
\begin{document}
\newcommand{\psl}{ p \hspace{-1.8truemm}/ }
\newcommand{\nsl}{ n \hspace{-2.2truemm}/ }
\newcommand{\vsl}{ v \hspace{-2.2truemm}/ }
\newcommand{\epsl}{\epsilon \hspace{-1.8truemm}/\,  }


\title{Four-body decays $B_{(s)} \rightarrow  (K\pi)_{S/P} (K\pi)_{S/P}$ in the perturbative QCD
approach}
\author{Zhou Rui$^1$}\email{jindui1127@126.com}
\author{Ya Li$^2$}\email{liyakelly@163.com}
\author{Hsiang-nan Li$^3$}\email{ hnli@phys.sinica.edu.tw}
\affiliation{$^1$  College of Sciences, North China University of Science and Technology, Tangshan, Hebei 063210,   China}
\affiliation{$^2$ Department of Physics, College of Science, Nanjing Agricultural University, Nanjing, Jiangsu 210095, China}
\affiliation{$^3$ Institute of Physics, Academia Sinica, Taipei, Taiwan 115, Republic of China}
\date{\today}
\begin{abstract}
We investigate the four-body decays $B_{(s)} \rightarrow (K\pi)_{S/P} (K\pi)_{S/P}$ in the $K\pi$-pair
invariant mass region around the $K^*(892)$ resonance in the perturbative QCD (PQCD) approach,
where $(K\pi)_{S/P}$ denotes a $S$- or $P$-wave $K\pi$ configuration.
The contributions from the $P$-wave resonance $K^*(892)$ and from possible $S$-wave components
are included into the parametrization of the $K\pi$ two-meson distribution amplitudes, which are the
nonperturbative inputs in this formalism.
We calculate the branching ratio for each resonance and observe sizable $S$-wave contributions to
the $B_s$ modes, well consistent with recently available LHCb data.
We also deduce the polarization fractions for the $K^*\bar{K}^*$ final states, and compare our
predictions to those in previous PQCD analyses of the corresponding two-body $B\to K^*\bar{K}^*$ decays.
The polarization puzzle associated with the two $U$-spin related channels
$B^0\rightarrow K^{*0}\bar{K}^{*0}$ and $B^0_s\rightarrow K^{*0}\bar{K}^{*0}$  still exists.
In addition to direct $CP$ asymmetries, triple-product asymmetries and $S$-wave induced direct $CP$ asymmetries
originating from the interference
among various helicity amplitudes, are presented.
It is shown that true triple-product asymmetries are rather small,
while direct $CP$ asymmetries and $S$-wave-induced direct $CP$ asymmetries are significant in some decays.
Our results will be subject to stringent tests with precise data
from $B$ factories in the near future.

\end{abstract}

\pacs{13.25.Hw, 12.38.Bx, 14.40.Nd}

\maketitle

\section{Introduction}
The charge conjugation ($C$), parity ($P$), and time reversal ($T$) are three fundamental discrete
transformations in nature. The combination of the $C$ and $P$ transformations is
particularly interesting, because $CP$ violation is a key requirement for generating
the  matter-antimatter asymmetry in the universe~\cite{196732,rmp761}. The presence
of $CP$ violation, attributed to an irreducible phase in the Cabibbo-Kobayashi-Maskawa (CKM)
quark-mixing matrix~\cite{prl10531,ptp49652} in the Standard Model (SM), has been well
established experimentally in the $K$ and $B$ meson systems~\cite{pdg2018}. $CP$ violation has
been observed not only in two-body $B$ meson decays, but also in multi-body channels.
In addition to direct and mixing-induced $CP$ asymmetries,
triple-product asymmetries (TPAs)~\cite{prd393339,npbps13487,ijmpa192505}
can be defined in multi-body $B$ meson decays, which may reveal potential signals of $CP$ violation. The
construction of a  scalar triple product requires a final state with at least four particles,
which endow three independent momenta in the rest frame of a $B$ meson.
That is, four-body $B$ meson decays are rich in $CP$ violation phenomena in the quark sector,
and the extra asymmetry observables provide more opportunities to search for new physics beyond the SM.

The $B\rightarrow V_1(\rightarrow P_1P_1')V_2(\rightarrow P_2P_2')$ decay, where
four pseudoscalar mesons are produced through two intermediate vector resonances,
gives rise to three helicity amplitudes commonly referred to as the $P$-wave amplitudes.
There may exist a background from the decay of a scalar resonance $S\rightarrow PP'$
or from a scalar nonresonant $PP'$ production~\cite{prd79074024,jhep09074}, such that
three more amplitudes contribute:
two single $S$-wave amplitudes (the scalar and vector combination)
and one scalar-scalar double $S$-wave amplitude. The interference
among the above helicity amplitudes generate TPAs, which can be extracted from an
angular analysis for the decay. A scalar triple product takes a generic form
$TP\propto\vec{v}_1\cdot (\vec{v}_2 \times \vec{v}_3)$,
$\vec{v}_i$ being the spin or momentum of a final-state particle.
This triple product is odd under the time reversal transformation, and also under the $CP$
transformation due to the $CPT$ invariance.
A TPA $A_T$ represents an asymmetry between the decay rates $\Gamma$ with positive and negative
values of $TP$,
\begin{eqnarray}
A_T\equiv \frac{\Gamma(TP>0)-\Gamma(TP<0)}{\Gamma(TP>0)+\Gamma(TP<0)}.
\end{eqnarray}
In the presence of the scalar background,
$S$-wave induced direct $CP$ asymmetries,  besides TPAs, are accessible in untagged decays~\cite{prd88016007}.
Phenomenological investigations on these asymmetry observables have been conducted intensively in the
literature~\cite{prd84096013,plb701357,prd86076011,prd88016007,prd92076013,prd87116005}.

$B$ meson decays into various four-body charmless hadronic final states in certain
two-body invariant mass regions have been observed
by the Belle~\cite{prd80051103,prd81071101,prd88072004}, BaBar~\cite{prd78092008,prl100081801,prl101201801},
and LHCb~\cite{plb70950,jhep110922013,jhep050692014,jhep071662015,
jhep100532015,jhep031402018,jhep050262019,jhep070322019} Collaborations.
Branching ratios for many partial waves 
have been measured for the first time or with greatly improved accuracy.
The $B^0_s\rightarrow (K^+\pi^-)(K^-\pi^+)$ mode was first seen by the LHCb~\cite{plb70950},
whose signals mainly come from $K^*\bar{K^*}$ channels with some $S$-wave contributions,
as indicated by $K\pi$ mass distributions. The involved amplitudes were
determined later for the $K^+\pi^-$ and $K^-\pi^+$ invariant masses around the $K^*(892)$ resonance
in the window of $\pm 150 $ MeV~\cite{jhep071662015}. The polarization fractions, TPAs, and $S$-wave
fractions were also extracted through a combined angular and mass analysis. Subsequently,
a flavor-tagged decay-time-dependent amplitude study of this
mode in a $K^{\pm}\pi^{\mp}$ mass window extending from  750 to 1600 MeV was presented~\cite{jhep031402018},
where not only the $K^*$ and $\bar{K}^*$ resonances, but also several scalar, vector and tensor
components in the $K\pi$ system were examined.
More recently, the LHCb performed a combined amplitude analysis for
the $B^0_s\rightarrow (K^+\pi^-)(K^-\pi^+)$ and $B^0\rightarrow (K^+\pi^-)(K^-\pi^+)$ decays,
which are related by the $U$-spin symmetry, and found different polarization patterns:
the latter is dominated by the longitudinal polarization, while the low longitudinal polarization
contribution to the former deviates from theoretical expectations
in the SM~\cite{prd91054033,prd76074018,npb93517,prd80114026,npb77464,prd96073004,epjc77333}.

It is known that multi-body decays of heavy mesons contain more complicated strong dynamics
than two-body ones, as they receive both resonant and nonresonant contributions, and suffer
substantial final-state interactions.
A factorization formalism that describes multi-body $B$ meson decays in full phase space
is not yet available at the moment. It has been proposed~\cite{Chen:2002th} that
a simpler factorization theorem can be constructed for leading-power regions of
a Dalitz plot, where two final-state hadrons are roughly collimated to each other.
The production of the hadron pair can be modelled by nonperturbative two-hadron distribution amplitudes
(DAs)~\cite{G,G1,DM,Diehl:1998dk,Diehl:1998dk1,Diehl:1998dk2,MP},
which collect both resonant and nonresonant contributions~\cite{Chen:2002th}.
This proposal suggests a theoretical framework for studying resonant
contributions from the quasi-two-body-decay mechanism~\cite{Wang:2015uea,npb899247,prd96113003}.
That is, a multi-body final state  is mainly preceded by intermediate resonances,
as justified by ample phenomenological and experimental evidences.
This formalism has been successfully applied to numerous hadronic three-body $B$ meson decays
in the perturbative QCD (PQCD)~\cite{epjc77199,prd99093007,epjc79792,cpc43073103,epjc80394,epjc80517,epjc8191,prd103013005,
prd95056008,epjc7937,prd103016002,epjc80815,prd101111901,jpg46095001}
and QCD factorization (QCDF) approaches~\cite{prd76094006,prd88114014,prd94094015,prd89074025,prd89094007,prd79094005,prd99076010,jhep102017117,jhep112020103}.

Recently, the localized $CP$ violation and branching fraction of the four-body decay
$\bar{B}^0\rightarrow K^-\pi^+\pi^+\pi^-$ were calculated by employing a quasi-two-body QCDF
approach in Refs.~\cite{191211874,200808458}.
Similar to the handling of three-body $B$ meson decays in the PQCD approach,
four-body processes are assumed to proceed dominantly with two intermediate resonances,
which then strongly decay into two light meson pairs.
The PQCD factorization formalism for four-body $B$ meson decays is thus simplified
to the one for two-body decays~\cite{prl744388,plb348597}. A decay amplitude is then written as
\begin{eqnarray}
A \propto \Phi_B \otimes H\otimes \Phi_{P_1P_1'} \otimes \Phi_{P_2P_2'},
\end{eqnarray}
where $\Phi_B$ is the $B$ meson DA, and the two-meson DAs $\Phi_{P_1P_1'}$ and $\Phi_{P_2P_2'}$
absorb the nonperturbative dynamics involved in the meson pairs $P_1P_1'$ and $P_2P_2'$,
respectively. The corresponding hard kernel $H$ is evaluated at
the quark level in perturbation theory. The symbol $\otimes$ stands for the convolution
of all the perturbative and nonperturbative factors in parton momenta.

In the present work we focus on the four-body decays $B_{(s)}\rightarrow (K\pi)_{S/P}(K\pi)_{S/P}$
with $(K\pi)_{S/P}$ denoting a $S$- or $P$-wave $K\pi$ configuration,
and derive their decay amplitudes in the PQCD approach.
Six quasi-two-body channels are computed, which arise from
different combinations of the $K\pi$ pairs in $S$ and $P$ waves.
The $K\pi$ two-meson DAs have been constrained to some extent in previous PQCD studies of
three-body $B$ meson decays~\cite{prd97033006,cpc44073102,prd101016015}. With these universal nonperturbative
inputs, we make quantitative predictions for physical observables in the above modes, including
branching ratios, polarization fractions, $S$-wave fractions, direct $CP$ asymmetries, and
TPAs. Note that a nonzero TPA can be generated by either final-state interactions
or $CP$ violation. To justify a ``true" $CP$ violation signal,
we compare a TPA with that in the corresponding $CP$ conjugate channel.
In accordance with the LHCb measurements~\cite{jhep071662015,jhep070322019},
the invariant masses of the $K\pi$ pairs are restricted to the region around the $K^*(892)$ resonance
in the window of 150 MeV. The formalism presented here, ready to be extended to
other four-body hadronic $B$ meson decays, will have many potential applications.

The paper is organized as below.
In Sec.~\ref{sec:framework} we define the kinematic variables for four-body $B$ meson decays and
the two-meson DAs for the $S$- and $P$-wave $K\pi$ configurations. The
$B_{(s)}\rightarrow (K\pi)_{S/P}(K\pi)_{S/P}$ helicity amplitudes are formulated
in Sec.~\ref{sec:framework1}, in terms of which the polarization fractions and direct $CP$
asymmetries are expressed. The various TPAs and $S$-wave-induced direct $CP$ asymmetries
to be investigated are specified in Sec.~\ref{sec:TPAs}.
Numerical results are presented and discussed in Sec.~\ref{sec:results}, which is followed by
the conclusion in Sec.~\ref{sec:sum}. The explicit PQCD factorization formulas of all the helicity
amplitudes are collected in Appendix~\ref{sec:ampulitude}.

\section{kinematics and two-meson distribution amplitudes}\label{sec:framework}
\begin{figure}[!htbh]
\begin{center}
\vspace{2.0cm} \centerline{\epsfxsize=12 cm \epsffile{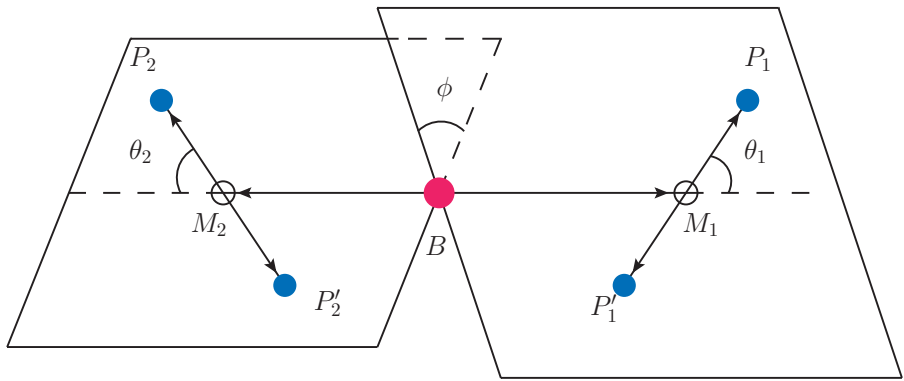}}
\vspace{2.3cm}
\caption{Helicity angles $\theta_1$, $\theta_2$ and $\phi$ for the $B\rightarrow M_1M_2$ decay, with each
intermediate resonance  decaying into two pseudoscalars, $M_1\rightarrow P_1P_1'$ and $M_2\rightarrow P_2P_2'$.}
 \label{fig:plane}
 \end{center}
\end{figure}

Consider the $B$ meson decay into two intermediate states $M_1$ and $M_2$, which
further strongly decay into the pseudoscalar pairs $P_1P_1'$ and $P_2P_2'$, respectively,
\begin{eqnarray}
B(p_B)\rightarrow M_1(p)M_2(q)\rightarrow P_1(p_1)P_1'(p_1')P_2(p_2)P_2'(p_2'),
\end{eqnarray}
with the meson momenta $p_B=p+q$, $p=p_1+p_1'$, and $q=p_2+p_2'$.
We work in the rest frame of the $B$ meson, and choose the momentum
$ p_B=(M/\sqrt{2})(1,1,\textbf{0}_{T})$ in the light-cone coordinates, $M$ being the $B$ meson mass.
The momenta of the  two  meson pairs can be parametrized as
 \begin{eqnarray}\label{eq:pq}
p=\frac{M}{\sqrt{2}}(g^+,g^-,\textbf{0}_{T}),\quad
q=\frac{M}{\sqrt{2}}(f^-,f^+,\textbf{0}_{T}),
\end{eqnarray}
where the factors
 \begin{eqnarray}\label{eq:epsilon}
g^{\pm}&=&\frac{1}{2}\left[1+\eta_1-\eta_2\pm\sqrt{(1+\eta_1-\eta_2)^2-4\eta_1}\right],\nonumber\\
f^{\pm}&=&\frac{1}{2}\left[1-\eta_1+\eta_2\pm\sqrt{(1+\eta_1-\eta_2)^2-4\eta_1}\right],
\end{eqnarray}
with the mass ratios $\eta_{1,2}=\omega_{1,2}^2/M^2$,
are related to the invariant masses of the meson pairs via  $p^2=\omega_1^2$, $q^2=\omega_2^2$, and $p_B=p+q$.
For the $P$-wave pairs, the corresponding longitudinal polarization vectors are defined as
 \begin{eqnarray}\label{eq:pq1}
\epsilon_{p}=\frac{1}{\sqrt{2\eta_1}}(g^+,-g^-,\textbf{0}_{T}),\quad
\epsilon_{q}=\frac{1}{\sqrt{2\eta_2}}(-f^-,f^+,\textbf{0}_{T}),
\end{eqnarray}
which satisfy the normalization $\epsilon_{p}^2=\epsilon_{q}^2=-1$  and the orthogonality
$\epsilon_{p}\cdot p=\epsilon_{q}\cdot q=0$.

We derive the momentum of each final-state meson,
\begin{eqnarray}\label{eq:p1p2}
p_1&=&\left(\frac{M}{\sqrt{2}}(\zeta_1+\frac{r_1-r_1'}{2\eta_1})g^+,\frac{M}{\sqrt{2}}(1-\zeta_1+\frac{r_1-r_1'}{2\eta_1})g^-,\textbf{p}_{T}\right),\nonumber\\
p_1'&=&\left(\frac{M}{\sqrt{2}}(1-\zeta_1-\frac{r_1-r_1'}{2\eta_1})g^+,\frac{M}{\sqrt{2}}(\zeta_1-\frac{r_1-r_1'}{2\eta_1})g^-,-\textbf{p}_{T}\right),\nonumber\\
p_2&=&\left(\frac{M}{\sqrt{2}}(1-\zeta_2+\frac{r_2-r_2'}{2\eta_2})f^-,\frac{M}{\sqrt{2}}(\zeta_2+\frac{r_2-r_2'}{2\eta_2})f^+,\textbf{q}_{T}\right),\nonumber\\
p_2'&=&\left(\frac{M}{\sqrt{2}}(\zeta_2-\frac{r_2-r_2'}{2\eta_2})f^-,\frac{M}{\sqrt{2}}(1-\zeta_2-\frac{r_2-r_2'}{2\eta_2})f^+,-\textbf{q}_{T}\right),
\end{eqnarray}
from the relations $p=p_1+p_1'$ and $q=p_2+p_2'$, and the on-shell conditions $p_i^{(\prime)2}=m_i^{(\prime)2}$
for the mesons $P_i^{(\prime)}$, $i=1,2$, with the mass ratios $r_i^{(\prime)}=m_i^{(\prime)2}/M^2$.
Equations~(\ref{eq:pq}) and~(\ref{eq:p1p2}) give the  meson momentum fractions
\begin{eqnarray}
\frac{p_{1}^+}{p^+}=\zeta_{1}+\frac{r_{1}-r_{1}'}{2\eta_{1}},\quad
\frac{p_{2}^-}{q^-}=\zeta_{2}+\frac{r_{2}-r_{2}'}{2\eta_{2}},
\end{eqnarray}
where the variables $\zeta_i$ bear the meaning of the meson momentum fractions up to corrections
from the meson masses. The transverse momenta $\textbf{p}_{T}$ and $\textbf{q}_{T}$
are written as
\begin{eqnarray}\label{eq:trans}
|\textbf{p}_{T}|^2=\omega_1^2[\zeta_1(1-\zeta_1)+\alpha_1],\quad |\textbf{q}_{T}|^2
=\omega_2^2[\zeta_2(1-\zeta_2)+\alpha_2],
\end{eqnarray}
with the factors
\begin{eqnarray}
\alpha_{i}=\frac{(r_{i}-r_{i}')^2}{4\eta^2_{i}}-\frac{r_{i}+r_{i}'}{2\eta_{i}}.
\end{eqnarray}

Alternatively, one can define the polar angles $\theta_{i}$ of the mesons $P_{i}$ in the $P_iP_i'$ rest frames
and the azimuthal angle $\phi$ between the decay planes of $M_1$  and $M_2$ as illustrated
in Fig.~\ref{fig:plane} to describe the sequential decay. The transformation connecting the $B$ meson rest frame
and the meson pair rest frame leads to the relations between the meson momentum fractions $\zeta_i$ and
the helicity angles $\theta_i$,
\begin{eqnarray}\label{eq:cos}
2\zeta_{i}-1&=& \sqrt{1+4\alpha_{i}}\text{cos} \theta_{i},
\end{eqnarray}
with the bounds
\begin{eqnarray}
 \zeta_{i}\in \left[\frac{1-\sqrt{1+4\alpha_{i}}}{2},\frac{1+\sqrt{1+4\alpha_{i}}}{2}\right].
\end{eqnarray}
Note that Eq.~(\ref{eq:cos}) reduces to the conventional expression in~\cite{npb905373,prd96051901}
when the two mesons in a pair have equal masses, and that it
maintains the decoupling between $S$- and $P$-wave contributions to a total decay rate
in the case with arbitrary final-state meson masses:
\begin{eqnarray}\label{or}
\int_{\zeta_{\min}}^{\zeta_{\max}}(2\zeta-1)d\zeta\propto\int_0^\pi\cos\theta d\theta=0.
\end{eqnarray}
We emphasize that the parametrization with the exact dependence on the final-state meson masses in
Eq.~(\ref{eq:p1p2}) is crucial for establishing Eq.~(\ref{eq:cos}), and then Eq.~(\ref{or}).

\begin{figure}[!htbh]
\begin{center}
\vspace{0cm} \centerline{\epsfxsize=8 cm \epsffile{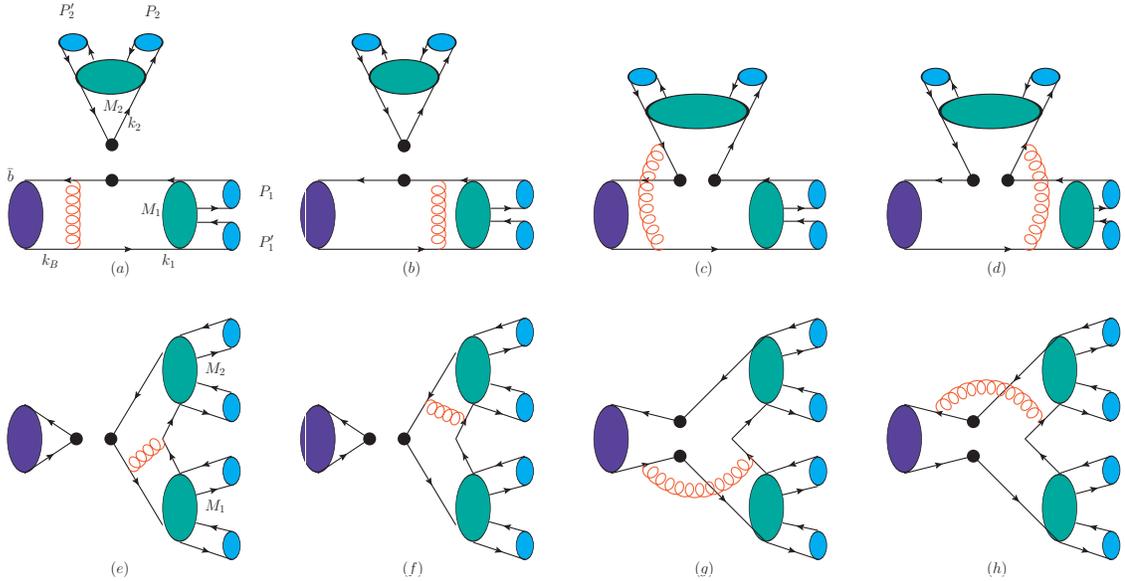}}
\vspace{1.6cm} \caption{Leading-order diagrams for the $B\rightarrow M_1M_2\rightarrow P_1P_1'P_2P_2'$
decays, where the symbol $\bullet$ denotes a weak vertex,
and the cyan oval represents either a scalar or vector resonance.}
 \label{fig:fym}
 \end{center}
\end{figure}

The Feynman diagrams contributing to the hard kernels for the
$B\rightarrow M_1M_2\rightarrow P_1P_1'P_2P_2'$ decays in the PQCD approach
are displayed in Fig.~\ref{fig:fym}, each of which contains a single virtual gluon exchange
at leading order in the strong coupling $\alpha_s$. The diagrams in the first (second) row
correspond to the emission (annihilation) type, which are further classified into the factorizable
ones with hard gluons attaching to the quarks in the same mesons,
and the nonfactorizable ones with hard gluons attaching to the quarks in different mesons.
To evaluate the hard kernels,
we parametrize the three valence quark momenta labelled by $k_B$, $k_1$, and $k_2$ in Fig.~\ref{fig:fym}(a) as
 \begin{eqnarray}\label{eq:kt}
k_B&=&(0,x_Bp_B^-,\textbf{k}_{BT}),\quad k_1=(x_1p^+,0,\textbf{k}_{1T}),\quad k_2=(0,x_2q^-,\textbf{k}_{2T}),
\end{eqnarray}
with the parton momentum fractions $x_i$ and the parton transverse momenta $k_{iT}$, $i=B,1,2$.
Since $k_1$ ($k_2$) is aligned with the meson pair in the plus (minus) direction,
its small minus (plus) component has been neglected. We also drop the plus component $k_B^+$, because
it does not appear in the hard kernels for dominant factorizable contributions.
The next-to-leading-order corrections to the hard kernels, such as those from quark loops~\cite{Li:2005kt},
will be taken into account in the future.

The light-cone hadronic matrix element for a $B$ meson is decomposed as \cite{ppnp5185}
\begin{eqnarray}\label{b}
\Phi_{B}(x,b)=\frac{i}{\sqrt{2N_c}}[(\rlap{/}{p_B}+M)\gamma_5\phi_{B}(x,b)],
\end{eqnarray}
with $b$ being the impact parameter conjugate to the parton transverse momentum $k_{T}$, and $N_c$
being the number of colors. A model-independent determination of the $B$ meson DA from an Euclidean lattice
was attempted very recently~\cite{prd102011502}. Various models of $\phi_{B}$ are available in the
literature~\cite{prd70074030,Li:2012md}, among which
we adopt the conventional one~\cite{ppnp5185,prd65014007},
\begin{eqnarray}
\phi_{B}(x,b)=N_B x^2(1-x)^2\exp\left[-\frac{x^2M^2}{2\omega^2_b}-\frac{\omega^2_bb^2}{2}\right],
\end{eqnarray}
with the shape parameter $\omega_b=0.40$ GeV for $B_{u,d}$ mesons and
$\omega_b=0.48$ GeV for a $B_s$ meson~\cite{201215074}. The normalization constant
$N_B$ is related to the $B$ meson decay constant $f_B$ via the normalization
\begin{eqnarray}
\int_0^1\phi_{B}(x,b=0)d x=\frac{f_{B}}{2\sqrt{2N_c}}.
\end{eqnarray}
Note that the above model $\phi_{B}(x,b=0)$ corresponds to the inverse moment $\lambda_B=241$ ($\lambda_{B_s}=287$)
MeV for the $B$ ($B_s$) meson with the mass $M_B=5.28$ ($M_{B_s}=5.37$) GeV. This $\lambda_B$ value is lower
than $460 \pm 110$ MeV evaluated at 1 GeV scale in QCD sum rules~\cite{prd69034014}, but consistent with
the experimental bound $\lambda_B > 238$ MeV~\cite{prd91112009}.

The two- and three-parton $B$ meson DAs have been decomposed according to definite twist and
conformal spin assignments up to twist 6 in Ref.~\cite{Braun:2017liq}. Here
only the $B$ meson DA associated with the leading Lorentz structure in
Eq.~(\ref{b}) is considered, while other power-suppressed pieces are negligible within the accuracy
of the current work~\cite{prd65014007,prd66054013}. It has been shown explicitly in the PQCD approach that
the next-to-leading-power $B$ meson DA contributes about 17\% of the $B\to\pi$ transition form factors
at large recoil~\cite{Yang:2020xal}, consistent with the naive estimate for the power suppression effect
$\bar \Lambda/m_b\equiv (M_B-m_b)/m_b\sim 10\%$, $m_b$ being the $b$ quark mass.
We will include the power-suppressed $B$ meson DAs, together with other subleading
contributions~\cite{Li:2014xda,Li:2012nk}, for a complete
analysis of four-body $B$ meson decays in the future, when aiming at precision better than 10\%.

As stated before, we focus on the leading-power regions of phase space, where
the four-body decays $B_{(s)}\rightarrow (K\pi)_{S/P}(K\pi)_{S/P}$ proceed mainly via
quasi-two-body processes. Besides the $B$ meson DA, the two-meson DAs, which absorb strong
interaction related to the production of the $K\pi$ meson pairs, are the necessary inputs to PQCD
calculations. In what follows the subscripts $S$ and $P$ label the
corresponding partial waves. The light-cone matrix element for a $S$-wave $K\pi$ pair in
the plus direction is decomposed as~\cite{prd97033006,prd101016015}
\begin{eqnarray}\label{eq:swave}
\Phi_{S}(x,\omega)=\frac{1}{\sqrt{2N_c}}[\rlap{/}{p}\phi_S(x,\omega)+
\omega\phi^s_S(x,\omega)+\omega(\rlap{/}{n}\rlap{/}{v}-1)\phi^t_S(x,\omega)],
\end{eqnarray}
with the dimensionless vectors $n=(1,0,\textbf{0}_{\text{T}})$
and $v=(0,1,\textbf{0}_{\text{T}})$, and the $K\pi$ invariant mass $\omega$.
The explicit forms of the twist-2 DA $\phi_S$ and twist-3 DAs $\phi^s_S$ and $\phi^t_S$
are given by~\cite{prd97033006}:
\begin{eqnarray}\label{eq:phi0st}
\phi_S(x,\omega)&=&\frac{3}{\sqrt{2N_c}}F_s(\omega^2)x(1-x) \left[\frac{1}{\mu_S}+B_13(1-2x)
+B_3 \frac{5}{2}(1-2x)(7(1-2x)^2-3)\right], \nonumber\\
\phi^s_S(x,\omega)&=&\frac{1}{2\sqrt{2N_c}}F_s(\omega^2),\nonumber\\
\phi^t_S(x,\omega)&=&\frac{1}{2\sqrt{2N_c}}F_s(\omega^2)(1-2x),
\end{eqnarray}
with the ratio $\mu_S=\omega/(m_{s}-m_{q})$, where $m_s(1\;{\rm GeV})=119$ MeV~\cite{prd73014017,prd77014034}
is the running strange quark mass and the light quark masses $m_q$, $q=u,d$, are set to zero.
We take the Gegenbauer moments $B_1=-0.57\pm 0.13$ and $B_3=-0.42\pm 0.22$
at the 1 GeV scale from scenario I in the QCD sum rule analysis~\cite{prd73014017,prd77014034},
which have been widely employed in the PQCD studies of the $B\rightarrow K^*_0(1430)$
transitions~\cite{prd79014013,jpg40025002,jhep032020162}. Another possible choice of the Gegenbauer moments
as those for the $\kappa(800)$ meson DA does not work well for accommodating experimental data.
Because available data are not yet sufficiently precise for fixing more Gegenbauer moments,
the asymptotic forms have been assumed for the two twist-3 DAs.

For the scalar form factor $F_s(\omega^2)$, we follow the LASS line shape~\cite{npb296493},
which consists of the $K_0^*(1430)$ resonance as well as an effective-range nonresonant component,
\begin{eqnarray}\label{eq:sform}
F_s(\omega)&=&\frac{\omega}{k(\omega)}\cdot\frac{1}{\cot \delta_B-i}+
e^{2i \delta_B}\frac{m_0^2\Gamma_0/k(m_0)}{m_0^2-\omega^2-im_0^2
\frac{\Gamma_0}{\omega}\frac{k(\omega)}{k(m_0)}}, \nonumber\\
\cot\delta_B&=&\frac{1}{ak(\omega)}+\frac{1}{2}bk(\omega),
\end{eqnarray}
with the $K^*_0(1430)$ mass (width) $m_0=1450\pm80$ MeV ($\Gamma_0=400\pm230$ MeV) \cite{prd92012012}.
The kaon three-momentum $k(\omega)$ is written, in the $K\pi$ center-of-mass frame, as
\begin{eqnarray}\label{eq:kq}
k(\omega)=\frac{\sqrt{[\omega^2-(m_K+m_{\pi})^2][\omega^2-(m_K-m_{\pi})^2]}}{2\omega},
\end{eqnarray}
$m_{K(\pi)}$ being the kaon (pion) mass.
For the scattering length $a$ and effective-range parameters $b$, we take the values
$a=3.2\pm1.8$ GeV$^{-1}$ and $b=0.9\pm1.1$ GeV$^{-1}$~\cite{prd92012012}.

The light-cone matrix elements for longitudinal and transverse $P$-wave $K\pi$ pairs are organized,
in analogy with the dipion case~\cite{prd98113003}, into
\begin{eqnarray}\label{eq:pwavekpi}
\Phi_{P}^{L}(x,\zeta,\omega)&=&\frac{1}{\sqrt{2N_c}}
\left[\omega\rlap{/}{\epsilon_p}\phi_P(x,\omega)+\omega \phi^s_P(x,\omega)+
\frac{\rlap{/}{p}_1\rlap{/}{p}'_1-\rlap{/}{p}'_1\rlap{/}{p}_1}{\omega(2\zeta-1)}
\phi^t_P(x,\omega)\right](2\zeta-1),\nonumber\\
\Phi_{P}^{T}(x,\zeta,\omega)&=&\frac{1}{\sqrt{2N_c}}
\left[\gamma_5\rlap{/}{\epsilon}_T\rlap{/}{p} \phi^T_P(x,\omega)
+\omega \gamma_5\rlap{/}{\epsilon}_T \phi^a_P(x,\omega)
+i\omega\frac{\epsilon^{\mu\nu\rho\sigma}\gamma_{\mu}
\epsilon_{T\nu}p_{\rho}n_{-\sigma}}{p\cdot n_-} \phi^v_P(x,\omega)\right]\sqrt{\zeta(1-\zeta)+\alpha},
\end{eqnarray}
respectively. Note that the factor $2\zeta-1$ in $\Phi_{P}^{L}$ corresponds exactly to
the structure for a time-like vector form factor,
\begin{eqnarray}
(p_1-p'_1)_{\mu}-\frac{m_1^2-m_{1}^{'2}}{p^2}p_{\mu}=(2\zeta_1-1)\omega_1\epsilon_{p\mu},
\end{eqnarray}
with the kinematic variables in Eqs.~(\ref{eq:pq}), (\ref{eq:pq1}) and (\ref{eq:p1p2}) being inserted.
This is why the first Lorentz structure for the longitudinal $K\pi$ pair has been set to the
polarization vector $\epsilon_p$, different from the parametrization in~\cite{cpc44073102}.
The factor $\sqrt{\zeta(1-\zeta)+\alpha}$ for the transverse $K\pi$ pair comes from Eq.~(\ref{eq:trans}).

The twist-2 DAs $\phi_P$ and $\phi^T_P$, and the other twist-3 DAs appearing in Eq.~(\ref{eq:pwavekpi})
are expanded in terms of the Gegenbauer polynomials~\cite{LRL},
\begin{eqnarray}
\phi_P(x,\omega)&=&\frac{3F_{K\pi}^{\parallel}(\omega^2)}{\sqrt{2N_c}} x(1-x)
\bigg\{1+a_{1K^*}^{\parallel}3(1-2x)+a_{2K^*}^{\parallel}\frac{3}{2}(5(1-2x)^2-1)
\bigg\}\;,\label{eqphi0}\\
\phi^s_P(x,\omega)&=&\frac{3F_{K\pi}^{\perp}(\omega^2)}{2\sqrt{2N_c}}(1-2x) \;,\label{eqphis}\\
\phi^t_P(x,\omega)&=&\frac{3F_{K\pi}^{\perp}(\omega^2)}{2\sqrt{2N_c}}(1-2x)^2 \;,\label{eqphit}\\
\phi^T_P(x,\omega)&=&\frac{3F_{K\pi}^{\perp}(\omega^2)}{\sqrt{2N_c}}x(1-x)
\bigg\{1+a_{1K^*}^{\perp}3(1-2x)+a_{2K^*}^{\perp}\frac{3}{2}(5(1-2x)^2-1)
\bigg\}\;,\label{eqphi}\\
\phi^a_P(x,\omega)&=&\frac{3F_{K\pi}^{\parallel}(\omega^2)}{4\sqrt{2N_c}}(1-2x)\;,\label{eqphia}\\
\phi^v_P(x,\omega)&=&\frac{3F_{K\pi}^{\parallel}(\omega^2)}{8\sqrt{2N_c}}\left[1+(1-2x)^2\right],\label{eqphiv}
\end{eqnarray}
with the Gegenbauer moments~\cite{LRL}
\begin{eqnarray}\label{eq:a1ka2k}
a_{1K^*}^{\parallel}=a_{1K^*}^{\perp}=0.31\pm 0.16,
\quad a_{2K^*}^{\parallel}=a_{2K^*}^{\perp}=1.188\pm0.098.
\end{eqnarray}
That is, we do not distinguish the Gegenbauer moments for the longitudinal and transverse polarizations.
The time-like vector form factor $F_{K\pi}^{\parallel}$
is parametrized as the relativistic  Breit-Wigner (RBW) function~\cite{cpc44073102,prd101016015}
\begin{eqnarray}
F_{K\pi}^{\parallel}(\omega^2)=\frac{m_{K^*}^2}{m_{K^*}^2-\omega^2-im_{K^*}\Gamma(\omega^2)},
\end{eqnarray}
with the mass-dependent width
\begin{eqnarray}
\Gamma(\omega^2)=\Gamma_{K^*}\frac{k^3(\omega)}{k^3(m_{K^*})}\frac{m_{K^*}}{\omega} \frac{1+r^2k^2(m_{K^*})}{1+r^2k^2(\omega)},
\end{eqnarray}
where $m_{K^*}$ ($\Gamma_{K^*}$) is the $K^*(892)$ mass (width), and $r=3.0$
GeV$^{-1}$~\cite{jhep071662015} is the interaction radius. Here only the $K^*(892)$ resonance is
included, and other higher resonances, such as $K^*(1410)$ and $K^*(1680)$
with pole masses much above the considered mass window, are neglected.

We point out that a phase difference causing the interference between the $S$- and $P$-wave
form factors is expected, which may be a complicated function of the $K\pi$ invariant mass.
Concentrating on the region near the $K^*$ resonance,
we perform the Taylor expansion of this complicated function around $m_{K^*}$,
and keep only the leading nontrivial term proportional to $m_{K^*}^2-\omega^2$.
The first term in the expansion, giving rise to an overall constant phase, does not contribute to
the interference effect. The vector form factor is thus replaced by
\footnote{It is equivalent to associate the phase with the scalar form factor in Eq.~(\ref{eq:sform}).}
\begin{eqnarray}\label{eq:phase}
F_{K\pi}^{\parallel}(\omega^2)\rightarrow e^{i\beta(m_{K^*}^2-\omega^2)} F_{K\pi}^{\parallel}(\omega^2),
\end{eqnarray}
where the tunable parameter $\beta$ renders the phase dimensionless. As shown in Sec. V,
the choice $\beta=(7.5\pm 2.5)$ GeV$^{-2}$ improves the agreement between theoretical results and data.
For another form factor $F_{K\pi}^{\perp}$, we assume the approximate relation
$F_{K\pi}^{\perp}/ F_{K\pi}^{\parallel}\sim f^{T}_{K^*}/f_{K^*}$~\cite{Wang:2016rlo}
with the tensor (vector) decay constant of the $K^*(892)$ meson
$f^{T}_{K^*}=0.185$ GeV ($f_{K^*}=0.217$ GeV)~\cite{jhep020342006} derived in the narrow-width limit
in QCD sum rules. Note that the tensor decay constant
is renormalization-scale and renormalization-scheme dependent, whose value is taken at the typical scale 1 GeV,
and in the $\overline{MS}$ scheme.
We assume the naive relation between the $K\pi$ time-like form factors
$F_{K\pi}^{\parallel}$ and $F_{K\pi}^{\perp}$, due to lack of rigorous theoretical and
experimental information. For example, the establishments of the QCD sum rules for the form factors
depend on their parametrization~\cite{jhep122019083}, and the determinations
of the decay constants are altered by finite-width effects from
a $K^*$ meson~\cite{jhep122019083}.

\section{Helicity amplitudes}\label{sec:framework1}

The differential rate for the decays $B_{(s)}\rightarrow (K\pi)_{S/P}(K\pi)_{S/P}$
in the $B_{(s)}$ meson rest frame is written as
\begin{eqnarray}\label{eq:decayrate}
\frac{d^5\Gamma}{d\Omega}=\frac{k(\omega_1)k(\omega_2)k(\omega_1,\omega_2)}{16(2\pi)^6M^2} |A|^2, 
\end{eqnarray}
where $d \Omega$ with $\Omega\equiv \{\theta_1, \theta_2,\phi,\omega_1,\omega_2\}$ stands for the
five-dimensional measure spanned by the three helicity angles and the two invariant masses, and
\begin{eqnarray}
k(\omega_1,\omega_2)&=&\frac{\sqrt{[M^2-(\omega_1+\omega_2)^2][M^2-(\omega_1-\omega_2)^2]}}{2M},
\end{eqnarray}
is the momentum of one of the $K\pi$ pairs in the $B_{(s)}$ meson rest frame.
The four-body phase space has been derived in the analyses of the
$K\rightarrow \pi\pi l\nu$ decay~\cite{pr1681858}, the
semileptonic $\bar{B}\rightarrow D(D^*)\pi l \nu$ decays~\cite{prd483204},
semileptonic baryonic decays~\cite{prd85094019,plb780100}, and four-body baryonic decays~\cite{plb770348}.
One can confirm that Eq.~(\ref{eq:decayrate}) is equivalent to those in Refs.~\cite{prd85094019,plb770348}
by appropriate variable changes.
Replacing the helicity angles $\theta_i$ by the meson momentum fractions $\zeta_i$ via Eq.~(\ref{eq:cos}),
we turn Eq.~(\ref{eq:decayrate}) into
\begin{eqnarray}\label{eq:decayrate1}
\frac{d^5\Gamma}{d\zeta_1d\zeta_2d \omega_1d \omega_2d\phi}=
\frac{k(\omega_1)k(\omega_2)k(\omega_1,\omega_2)}{4(2\pi)^6M^2\sqrt{1+4\alpha_1}\sqrt{1+4\alpha_2}}|A|^2.
\end{eqnarray}

The $B_{(s)}\rightarrow (K\pi)_{S/P}(K\pi)_{S/P}$ decays involve six helicity
amplitudes $A_h$ with $h=VV$ (three), $VS$, $SV$,
and $SS$. The first three, commonly referred to as the $P$-wave amplitudes, correspond to
the final states, where both $K\pi$ pairs are from intermediate vector mesons.
A $P$-wave decay amplitude can be decomposed into three components in the transversity basis:
$A_0$, for which the polarizations of the vector mesons are longitudinal with respect
to their momenta, and $A_{\parallel}$ ($A_\perp$), for which the polarizations
are transverse to the momenta and parallel (perpendicular) to each other.
As the $S$-wave $K\pi$ pair arises from $M_1$ ($M_2$) in Fig.~\ref{fig:fym}(a),
the resultant single $S$-wave amplitude is denoted as $A_{SV}$ ($A_{VS}$).
The double $S$-wave amplitude $A_{SS}$ is associated with the final state,
where both $K\pi$ pairs are produced in the $S$ wave.
Specifically, these helicity amplitudes represent the following $B_s^0\rightarrow (K^+\pi^-)(K^-\pi^+)$
channels
\begin{eqnarray}
A_{VV}&:& B_s^0 \rightarrow \bar{K}^{*0}(\rightarrow K^-\pi^+) K^{*0}(\rightarrow K^+\pi^-), \nonumber\\
A_{VS}&:& B_s^0 \rightarrow \bar{K}^{*0}(\rightarrow K^-\pi^+)(K^+\pi^-)_0  ,\nonumber\\
A_{SV}&:& B_s^0 \rightarrow \overline{(K^-\pi^+)}_0 K^{*0}(\rightarrow K^+\pi^-),\nonumber\\
A_{SS}&:& B_s^0 \rightarrow \overline{(K^-\pi^+)}_0(K^+\pi^-)_0,
\end{eqnarray}
where the notation $(K\pi)_0$ labels the  $S$-wave $K\pi$ configuration to be modelled
by the LASS line shape.

It is more convenient to introduce the linear combinations
\begin{eqnarray}\label{eq:s-s+}
A_{S^{-}}=\frac{A_{SV}-A_{VS}}{\sqrt{2}},\quad A_{S^{+}}=\frac{A_{VS}+A_{SV}}{\sqrt{2}},
\end{eqnarray}
for the discussion of asymmetry observables below. Note that the amplitude $A_{SV}$ ($A_{VS}$) in our
convention corresponds to $A_{VS}$ ($A_{SV}$) in Ref.~\cite{jhep071662015}, so that the above definitions of
$A_{S^{\pm}}$ are the same as theirs. Note that the amplitudes $A_\perp$ and $A_{S^{+}}$ are $CP$-odd,
and $A_0$, $A_{\parallel}$, $A_{S^-}$ and $A_{SS}$ are $CP$-even for the neutral $B_{(s)}^0$ modes.
Including the $\zeta_i$ and azimuth-angle dependencies, we have the full decay
amplitude in Eq.~(\ref{eq:decayrate1}) as a coherent sum of the $P$-, $S$-, and double $S$-wave components,
\begin{eqnarray}\label{eq:allampli}
A&=&\frac{2\zeta_1-1}{\sqrt{1+4\alpha_1}}\frac{2\zeta_2-1}{\sqrt{1+4\alpha_2}}A_0
+2\sqrt{2}\sqrt{\frac{\zeta_1(1-\zeta_1)+\alpha_1}{1+4\alpha_1}}
\sqrt{\frac{\zeta_2(1-\zeta_2)+\alpha_2}{1+4\alpha_2}}\cos(\phi)A_{\parallel}
\nonumber\\&&
+i2\sqrt{2}\sqrt{\frac{\zeta_1(1-\zeta_1)+\alpha_1}{1+4\alpha_1}}
\sqrt{\frac{\zeta_2(1-\zeta_2)+\alpha_2}{1+4\alpha_2}}\sin(\phi)A_{\perp}
+A_{SS}+\frac{2\zeta_2-1}{\sqrt{1+4\alpha_2}}A_{SV}+\frac{2\zeta_1-1}{\sqrt{1+4\alpha_1}}A_{VS}\nonumber\\&=&
\frac{2\zeta_1-1}{\sqrt{1+4\alpha_1}}\frac{2\zeta_2-1}{\sqrt{1+4\alpha_2}}A_0
+2\sqrt{2}\sqrt{\frac{\zeta_1(1-\zeta_1)+\alpha_1}{1+4\alpha_1}}
\sqrt{\frac{\zeta_2(1-\zeta_2)+\alpha_2}{1+4\alpha_2}}\cos(\phi)A_{\parallel}
\nonumber\\&&
+i2\sqrt{2}\sqrt{\frac{\zeta_1(1-\zeta_1)+\alpha_1}{1+4\alpha_1}}
\sqrt{\frac{\zeta_2(1-\zeta_2)+\alpha_2}{1+4\alpha_2}}\sin(\phi)A_{\perp}
+A_{SS}\nonumber\\&&+\frac{1}{\sqrt{2}}\left(\frac{2\zeta_2-1}{\sqrt{1+4\alpha_2}}
-\frac{2\zeta_1-1}{\sqrt{1+4\alpha_1}}\right)A_{S^-}
+\frac{1}{\sqrt{2}}\left(\frac{2\zeta_2-1}{\sqrt{1+4\alpha_2}}
+\frac{2\zeta_1-1}{\sqrt{1+4\alpha_1}}\right)A_{S^+}.
\end{eqnarray}

To calculate the helicity amplitudes, we start from the effective Hamiltonian~\cite{rmp681125}
\begin{eqnarray}
\mathcal{H}_{eff}=\frac{G_F}{\sqrt{2}}\{ V^*_{ub}V_{uq}[C_1(\mu)O_1(\mu)+C_2(\mu)O_2(\mu)]
-V^*_{tb}V_{tq}\sum_{i=3}^{10}C_i(\mu)O_i^{(q)}(\mu)\},
\end{eqnarray}
with the quark $q=d,s$, the Fermi coupling constant $G_F$, the CKM matrix elements $V_{ub},V_{uq},\cdots$,
and the Wilson coefficients $C_i(\mu)$ at the renormalization scale $\mu$. The effective local
four-quark operators $O_i(\mu)$ are given by~\cite{rmp681125}
\begin{eqnarray}\label{eq:operator}
O_1&=&\bar{b}_{\alpha}\gamma_{\mu}(1-\gamma_5)u_{\beta}\otimes\bar{u}_{\beta}\gamma^{\mu}(1-\gamma_5)q_{\alpha}, \nonumber\\
O_2&=&\bar{b}_{\alpha}\gamma_{\mu}(1-\gamma_5)u_{\alpha}\otimes\bar{u}_{\beta}\gamma^{\mu}(1-\gamma_5)q_{\beta}, \nonumber\\
O_3&=&\bar{b}_{\alpha}\gamma_{\mu}(1-\gamma_5)q_{\alpha}\otimes \sum_{q'}\bar{q}'_{\beta}\gamma^{\mu}(1-\gamma_5)q'_{\beta}, \nonumber\\
O_4&=&\bar{b}_{\alpha}\gamma_{\mu}(1-\gamma_5)q_{\beta}\otimes \sum_{q'}\bar{q}'_{\beta}\gamma^{\mu}(1-\gamma_5)q'_{\alpha},\nonumber\\
O_5&=&\bar{b}_{\alpha}\gamma_{\mu}(1-\gamma_5)q_{\alpha}\otimes \sum_{q'}\bar{q}'_{\beta}\gamma^{\mu}(1+\gamma_5)q'_{\beta}, \nonumber\\
O_6&=&\bar{b}_{\alpha}\gamma_{\mu}(1-\gamma_5)q_{\beta}\otimes \sum_{q'}\bar{q}'_{\beta}\gamma^{\mu}(1+\gamma_5)q'_{\alpha},\nonumber\\
O_7&=&\frac{3}{2}\bar{b}_{\alpha}\gamma_{\mu}(1-\gamma_5)q_{\alpha}\otimes \sum_{q'}e_{q'}\bar{q}'_{\beta}\gamma^{\mu}(1+\gamma_5)q'_{\beta}, \nonumber\\
O_8&=&\frac{3}{2}\bar{b}_{\alpha}\gamma_{\mu}(1-\gamma_5)q_{\beta}\otimes \sum_{q'}e_{q'}\bar{q}'_{\beta}\gamma^{\mu}(1+\gamma_5)q'_{\alpha},\nonumber\\
O_9&=&\frac{3}{2}\bar{b}_{\alpha}\gamma_{\mu}(1-\gamma_5)q_{\alpha}\otimes \sum_{q'}e_{q'}\bar{q}'_{\beta}\gamma^{\mu}(1-\gamma_5)q'_{\beta}, \nonumber\\
O_{10}&=&\frac{3}{2}\bar{b}_{\alpha}\gamma_{\mu}(1-\gamma_5)q_{\beta}\otimes \sum_{q'}e_{q'}\bar{q}'_{\beta}\gamma^{\mu}(1-\gamma_5)q'_{\alpha},
\end{eqnarray}
where $\alpha,\beta$ are color indices and  $e_{q'}$ is the electric charges of the quark $q'$
in units of $|e|$. The sum over $q'$ runs over the quark fields active at the $b$ quark mass
scale, i.e., $q'=u,d,s,c,b$.

As stated in the Introduction, a helicity amplitude is expressed
as a convolution of a hard kernel with relevant meson DAs in the PQCD approach.
Parton transverse momenta in the $k_T$ factorization theorem, on which this approach is based,
regularize end-point singularities from the kinematic regions with small momentum fractions $x$.
The resultant double logarithms $\alpha_s\ln^2 k_T$ are organized
into the Sukakov factor $e^{-S}$ through the $k_T$ resummation technique. The double logarithms
$\alpha_s\ln^2 x$ appearing in a hard kernel, being important in the end-point regions, are
summed to all orders into the threshold resummation factor $S_t$.
The typical PQCD factorization formula then reads
\begin{eqnarray}\label{eq:ampli1}
A_h=X_h\int d x_B dx_1dx_2 b_B db_B b_1db_1b_2db_2 {\rm Tr}
[C(t)\Phi_B(x_B,b_B)\Phi_{K\pi}(x_1)\Phi_{K\pi}(x_2)H(x_i,b_i,t)S_t(x_i)e^{-S(t)}],
\end{eqnarray}
where $b_i$ are the impact parameters conjugate to $k_{iT}$ in  Eq.~(\ref{eq:kt}), $t$
represents the largest energy scale in the hard kernel $H$, and
``Tr" denotes the trace over all Dirac structures and color indices.
Both the resummaton factors $e^{-S}$ and $S_t$, whose explicit expressions are provided
in Appendix~\ref{sec:ampulitude}, improve the perturbative evaluation of a helicity amplitude. The prefactors
\begin{eqnarray}\label{eq:xh}
X_h=\left\{
\begin{aligned}
&\sqrt{1+4\alpha_{1}}\sqrt{1+4\alpha_{2}} \quad\quad\quad  &h=0,\parallel,\perp \\
&\sqrt{1+4\alpha_{1,2}} \quad\quad\quad  &h=VS,SV \\
&1 \quad\quad\quad  &h=SS, \\
\end{aligned}\right.
\end{eqnarray}
compensated by the denominators $\sqrt{1+4\alpha_{i}}$ in Eq.~(\ref{eq:allampli}),
normalize the helicity amplitudes $A_h$ properly.

The helicity amplitudes for the pure penguin type channel
$B^0_s\rightarrow K^{*0}(\rightarrow K^-\pi^+)\bar{K}^{*0}(\rightarrow K^+\pi^-)$ are written as
\begin{eqnarray}\label{eq:ampli}
A_h(B^0_s\rightarrow K^{*0}(\rightarrow K^+\pi^-)\bar{K}^{*0}(\rightarrow K^-\pi^+))&=&
-\frac{G_F}{\sqrt{2}}V^*_{tb}V_{ts}X_h\{ [C_4+\frac{C_3}{3}-\frac{1}{2}(C_{10}+\frac{C_9}{3})]\mathcal{F}_{e}^{LL,h}\nonumber\\ &&
+[C_6+\frac{C_5}{3}-\frac{1}{2}(C_{8}+\frac{C_7}{3})]\mathcal{F}_{e}^{SP,h}\nonumber\\ &&
+(C_3-\frac{C_9}{2})\mathcal{M}^{LL,h}_{e}+(C_5-\frac{C_7}{2})\mathcal{M}^{LR,h}_{e}\nonumber\\ &&
+\frac{4}{3}[(C_3+C_4-\frac{C_9}{2}-\frac{C_{10}}{2})]\mathcal{F}_{a}^{LL,h}\nonumber\\ &&
+[C_5+\frac{C_6}{3}-\frac{1}{2}(C_{7}+\frac{C_8}{3})]\mathcal{F}_{a}^{LR,h}\nonumber\\ &&
+[C_6+\frac{C_5}{3}-\frac{1}{2}(C_{8}+\frac{C_7}{3})]\mathcal{F}_{a}^{SP,h}\nonumber\\ &&
+[C_4+C_3-\frac{1}{2}(C_{9}+C_{10})]\mathcal{M}_{a}^{LL,h}\nonumber\\ &&
+(C_5-\frac{C_7}{2})\mathcal{M}_{a}^{LR,h}+(C_6-\frac{C_8}{2})\mathcal{M}_{a}^{SP,h}\nonumber\\ &&
+[C_3+\frac{C_4}{3}-\frac{1}{2}(C_9+\frac{C_{10}}{3})]\mathcal{F'}_{a}^{LL,h}\nonumber\\ &&
+[C_5+\frac{C_6}{3}-\frac{1}{2}(C_7+\frac{C_{8}}{3})]\mathcal{F'}_{a}^{LR,h}\nonumber\\ &&
+(C_4-\frac{C_{10}}{2})\mathcal{M'}^{LL,h}_{a}+(C_6-\frac{C_8}{2})\mathcal{M'}^{SP,h}_{a}\},
\end{eqnarray}
where $\mathcal{F}_{e}(\mathcal{M}_{e})$ comes from the factorizable (nonfactorizable) emission diagrams in
Figs.~\ref{fig:fym}(a),(b)(\ref{fig:fym}(c),(d)), and $\mathcal{F}_{a}(\mathcal{M}_{a})$
comes from the factorizable (nonfactorizable) annihilation diagrams in
Figs.~\ref{fig:fym}(e),(f)(\ref{fig:fym}(g),(h)). The function
$\mathcal{F}'_{a}(\mathcal{M}'_{a})$  also arises from Figs.~\ref{fig:fym}(e),(f)(\ref{fig:fym}(g),(h)),
but with the momenta of the two $K\pi$ pairs being exchanged.
The superscripts $LL$, $LR$, and $SP$ label the contributions from the $(V-A)\bigotimes(V-A)$,
$(V-A)\bigotimes(V+A)$ and $(S-P)\bigotimes(S+P)$ operators, respectively. The
explicit expressions of all the above functions can be found in Appendix~\ref{sec:ampulitude}.
Note that the function $\mathcal{F}_{e}^{SP}$  from the operators $O_{5-8}$
vanishes, when a vector resonance is emitted from the weak vertex, because neither the scalar nor
pseudoscalar operator is responsible for vector resonance production.
Equation~(\ref{eq:ampli}) also holds for the
$B^0\rightarrow K^{*0}(\rightarrow K^-\pi^+)\bar{K}^{*0}(\rightarrow K^+\pi^-)$ decay
with the replacement $V^*_{tb}V_{ts}\rightarrow V^*_{tb}V_{td}$.

Additional tree contributions appear in the
$B^0_{(s)}\rightarrow K^{*+}(\rightarrow K^0\pi^+)K^{*-}(\rightarrow \bar{K}^0\pi^-)$
and $B^+\rightarrow K^{*+}(\rightarrow K^0\pi^+)\bar{K}^{*0}(\rightarrow K^+\pi^-)$ decays,
whose helicity amplitudes are given by
\begin{eqnarray}\label{eq:bspm}
A_h(B_s^0\rightarrow K^{*+}(\rightarrow K^0\pi^+)K^{*-}(\rightarrow \bar{K}^0\pi^-))&=&\frac{G_F}{\sqrt{2}}V^*_{ub}V_{us}X_h
\{(C_2+\frac{C_1}{3})\mathcal{F}_{e}^{LL,h}+C_1\mathcal{M}^{LL,h}_{e}\nonumber\\ &&
(C_1+\frac{C_2}{3})\mathcal{F'}_{a}^{LL,h}+C_2\mathcal{M'}^{LL,h}_{a}\}\nonumber\\ &&
-\frac{G_F}{\sqrt{2}}V^*_{tb}V_{ts}X_h\{ (C_4+\frac{C_3}{3}
+C_{10}+\frac{C_9}{3})\mathcal{F}_{e}^{LL,h}\nonumber\\ &&
+(C_6+\frac{C_5}{3}+C_{8}+\frac{C_7}{3})\mathcal{F}_{e}^{SP,h}\nonumber\\ &&
+(C_3+C_9)\mathcal{M}^{LL,h}_{e}+(C_5+C_7)\mathcal{M}^{LR,h}_{e}\nonumber\\ &&
+\frac{4}{3}[(C_3+C_4-\frac{C_9}{2}-\frac{C_{10}}{2})]\mathcal{F}_{a}^{LL,h}\nonumber\\ &&
+[C_5+\frac{C_6}{3}-\frac{1}{2}(C_{7}+\frac{C_8}{3})]\mathcal{F}_{a}^{LR,h}\nonumber\\ &&
+[C_6+\frac{C_5}{3}-\frac{1}{2}(C_{8}+\frac{C_7}{3})]\mathcal{F}_{a}^{SP,h}\nonumber\\ &&
+[C_4+C_3-\frac{1}{2}(C_{9}+C_{10})]\mathcal{M}_{a}^{LL,h}\nonumber\\ &&
+(C_5-\frac{C_7}{2})\mathcal{M}_{a}^{LR,h}+(C_6-\frac{C_8}{2})\mathcal{M}_{a}^{SP,h}\nonumber\\ &&
+[C_3+\frac{C_4}{3}+C_9+\frac{C_{10}}{3}]\mathcal{F'}_{a}^{LL,h}\nonumber\\ &&
+[C_5+\frac{C_6}{3}+C_7+\frac{C_{8}}{3}]\mathcal{F'}_{a}^{LR,h}\nonumber\\ &&
+(C_4+C_{10})\mathcal{M'}^{LL,h}_{a}+(C_6+C_8)\mathcal{M'}^{SP,h}_{a}\},
\end{eqnarray}
\begin{eqnarray}
A_h(B^0\rightarrow K^{*+}(\rightarrow K^0\pi^+)K^{*-}(\rightarrow \bar{K}^0\pi^-))
&=&\frac{G_F}{\sqrt{2}}V^*_{ub}V_{ud}X_h
\{(C_1+\frac{C_2}{3})\mathcal{F'}_{a}^{LL,h}+C_2\mathcal{M'}^{LL,h}_{a}\}\nonumber\\ &&
-\frac{G_F}{\sqrt{2}}V^*_{tb}V_{td}X_h\{[(C_3+\frac{C_4}{3}-\frac{C_9}{2}
-\frac{C_{10}}{6})]\mathcal{F}_{a}^{LL,h}\nonumber\\ &&
+[C_5+\frac{C_6}{3}-\frac{1}{2}(C_{7}+\frac{C_8}{3})]\mathcal{F}_{a}^{LR,h}\nonumber\\ &&
+[C_4-\frac{1}{2}C_{10}]\mathcal{M}_{a}^{LL,h}+(C_6-\frac{C_8}{2})\mathcal{M}_{a}^{SP,h}\nonumber\\ &&
+[C_3+\frac{C_4}{3}+C_9+\frac{C_{10}}{3}]\mathcal{F'}_{a}^{LL,h}\nonumber\\ &&
+[C_5+\frac{C_6}{3}+C_7+\frac{C_{8}}{3}]\mathcal{F'}_{a}^{LR,h}\nonumber\\ &&
+(C_4+C_{10})\mathcal{M'}^{LL,h}_{a}+(C_6+C_8)\mathcal{M'}^{SP,h}_{a}\},
\end{eqnarray}
\begin{eqnarray}
A_h(B^+\rightarrow K^{*+}(\rightarrow K^0\pi^+)\bar{K}^{*0}(\rightarrow K^-\pi^+))
&=&\frac{G_F}{\sqrt{2}}V^*_{ub}V_{ud}X_h
\{(C_2+\frac{C_1}{3})\mathcal{F}_{a}^{LL,h}+C_1\mathcal{M}^{LL,h}_{a}\}\nonumber\\ &&
-\frac{G_F}{\sqrt{2}}V^*_{tb}V_{td}X_h\{[C_4+\frac{1}{3}C_3-\frac{C_{10}}{2}
-\frac{C_{9}}{6}]\mathcal{F}_{e}^{LL,h}\nonumber\\ &&
+[C_6+\frac{C_5}{3}-\frac{1}{2}(C_{8}+\frac{C_7}{3})]\mathcal{F}_{e}^{SP,h}\nonumber\\ &&
+[C_3-\frac{1}{2}C_{9}]\mathcal{M}_{e}^{LL,h}+[C_5-\frac{C_7}{2}]\mathcal{M}_{e}^{LR,h}\nonumber\\ &&
+[C_4+\frac{C_3}{3}+C_{10}+\frac{C_9}{3}]\mathcal{F}_{a}^{LL,h}\nonumber\\ &&
+[C_6+\frac{C_5}{3}+C_8+\frac{C_7}{3}]\mathcal{F}_{a}^{SP,h}\nonumber\\ &&
+[C_3+C_{9}]\mathcal{M}^{LL,h}_{a}+[C_5+C_7]\mathcal{M}^{LR,h}_{a}\}.
\end{eqnarray}
The isospin ratio $\Gamma(K^{*+}\rightarrow K^+\pi^0)/\Gamma(K^{*+}\rightarrow K^0\pi^+)=1/2$
leads to the relations
\begin{eqnarray}\label{a2}
A_h(B^0_{(s)}\rightarrow K^{*+}(\rightarrow K^+\pi^0)K^{*-}(\rightarrow K^-\pi^0))
&=&\frac{1}{2}A_h(B^0_{(s)}\rightarrow K^{*+}(\rightarrow K^0\pi^+)K^{*-}(\rightarrow \bar{K}^0\pi^-)),\nonumber\\
A_h(B^+\rightarrow K^{*+}(\rightarrow K^+\pi^0)\bar{K}^{*0}(\rightarrow K^-\pi^+))
&=&\frac{1}{\sqrt{2}}A_h(B^+\rightarrow K^{*+}(\rightarrow K^0\pi^+)\bar{K}^{*0}(\rightarrow K^-\pi^+)).
\end{eqnarray}

We obtain the branching ratios from Eq.~(\ref{eq:decayrate1}),
\begin{eqnarray}\label{eq:brsss}
\mathcal{B}_h=\frac{\tau_B}{4(2\pi)^6M^2}\frac{2\pi}{9}Y_h
\int d\omega_1d\omega_2k(\omega_1)k(\omega_2)k(\omega_1,\omega_2)|A_h|^2,
\end{eqnarray}
where the invariant masses $\omega_{1,2}$ are integrated over the selected $K\pi$ mass window,
and the coefficients
\begin{eqnarray}\label{eq:radii}
Y_h=\left\{
\begin{aligned}
&1 \quad\quad\quad  &h=0,\parallel,\perp \\
&3 \quad\quad\quad  &h=S^{\pm}, \\
&9 \quad\quad\quad  &h=SS, \\
\end{aligned}\right.
\end{eqnarray}
are the results of the integrations over $\zeta_1,\zeta_2,\phi$. In terms of Eq.~(\ref{eq:brsss}),
we compute the $CP$-averaged branching ratio and the direct $CP$ asymmetry in each component,
\begin{eqnarray}
\mathcal{B}_h^{avg}=\frac{1}{2}(\mathcal{B}_h+\mathcal{\bar{B}}_h),
\quad \mathcal{A}^{\text{dir}}_h=\frac{\mathcal{\bar{B}}_h-\mathcal{B}_h}{\mathcal{\bar{B}}_h+\mathcal{B}_h},
\end{eqnarray}
respectively, where $\mathcal{\bar{B}}_h$ is the branching ratio of the corresponding $CP$-conjugate channel.
The sum of the six components yields the total branching ratio and the overall direct $CP$ asymmetry,
\begin{eqnarray}
\mathcal{B}_{\text{total}}=\sum_h \mathcal{B}_h,\quad
\mathcal{A}^{\text{dir}}_{CP}=\frac{\sum_h \mathcal{\bar{B}}_h
-\sum_h \mathcal{B}_h}{\sum_h \mathcal{\bar{B}}_h+\sum_h \mathcal{B}_h},
\end{eqnarray}
respectively.

To characterize the relative importance of the $S$-wave contributions, we define the $S$-wave
fractions for the different components as
\begin{eqnarray}\label{eq:fss}
f_{SS}&=&\frac{\mathcal{B}_{SS}}{\mathcal{B}_{\text{total}}},\quad
f_{S^{\pm}}=\frac{\mathcal{B}_{S^{\pm}}}{\mathcal{B}_{\text{total}}},
\end{eqnarray}
and the total $S$-wave fraction as
\begin{eqnarray}\label{eq:fss3}
f_{\text{S-wave}}=f_{SS}+f_{S^+}+f_{S^-}.
\end{eqnarray}
The polarization fractions from the $P$-wave  amplitudes are derived according to
\begin{eqnarray}
f_0&=&\frac{\mathcal{B}_0}{\mathcal{B}_{P}},\quad
f_{\parallel}=\frac{\mathcal{B}_{\parallel}}{\mathcal{B}_{P}},\quad
f_{\perp}=\frac{\mathcal{B}_{\perp}}{\mathcal{B}_{P}},
\end{eqnarray}
with $\mathcal{B}_{P}=\mathcal{B}_0+\mathcal{B}_{\parallel}+\mathcal{B}_{\perp}$ being the total $P$-wave
branching ratio.

\section{Triple product asymmetries
and S-wave-induced direct CP asymmetries}\label{sec:TPAs}

In this work we focus on the TPAs associated with $A_{\perp}$ for the $B_{(s)}\rightarrow (K\pi)_{S/P}(K\pi)_{S/P}$
decays, which are defined in terms of the partially integrated differential decay rates as~\cite{prd84096013,jhep071662015}
\begin{eqnarray} \label{eq:ATs}
A_T^1&=&\frac{\Gamma((2\zeta_1-1)(2\zeta_2-1)\sin\phi>0)-\Gamma((2\zeta_1-1)(2\zeta_2-1)\sin\phi<0)}
{\Gamma((2\zeta_1-1)(2\zeta_2-1)\sin\phi>0)+\Gamma((2\zeta_1-1)(2\zeta_2-1)\sin\phi<0)}\nonumber\\&=&
-\frac{2\sqrt{2}}{\pi\mathcal{D}}\int d\omega_1 d\omega_2k(\omega_1)k(\omega_2)k(\omega_1,\omega_2) Im[A_{\perp}A_0^*],\nonumber\\
A_T^2&=&\frac{\Gamma(\sin(2\phi)>0)-\Gamma(\sin(2\phi)<0)}
{\Gamma(\sin(2\phi)>0)+\Gamma(\sin(2\phi)<0)}\nonumber\\&=&
-\frac{4}{\pi\mathcal{D}}\int d\omega_1 d\omega_2k(\omega_1)k(\omega_2)k(\omega_1,\omega_2) Im[A_{\perp}A_{\parallel}^*],\nonumber\\
A_T^3&=&\frac{\Gamma\left(\left(\frac{2\zeta_1-1}{\sqrt{1+4\alpha_1}}-\frac{2\zeta_2-1}{\sqrt{1+4\alpha_2}}\right)\sin\phi>0\right)
-\Gamma\left(\left(\frac{2\zeta_1-1}{\sqrt{1+4\alpha_1}}-\frac{2\zeta_2-1}{\sqrt{1+4\alpha_2}}\right)\sin\phi<0\right)}
{\Gamma\left(\left(\frac{2\zeta_1-1}{\sqrt{1+4\alpha_1}}-\frac{2\zeta_2-1}{\sqrt{1+4\alpha_2}}\right)\sin\phi>0\right)
+\Gamma\left(\left(\frac{2\zeta_1-1}{\sqrt{1+4\alpha_1}}-\frac{2\zeta_2-1}{\sqrt{1+4\alpha_2}}\right)\sin\phi<0\right)}\nonumber\\&=&
\frac{32}{5\pi\mathcal{D}}\int d\omega_1 d\omega_2 k(\omega_1)k(\omega_2)k(\omega_1,\omega_2)Im[A_{\perp}A^*_{S^{-}}],\nonumber\\
A_T^4&=&\frac{\Gamma(\sin\phi>0)-\Gamma(\sin\phi<0)}
{\Gamma(\sin\phi>0)+\Gamma(\sin\phi<0)}\nonumber\\&=&-
\frac{9\pi}{4\sqrt{2}\mathcal{D}}\int d\omega_1 d\omega_2k(\omega_1)k(\omega_2)k(\omega_1,\omega_2) Im[A_{\perp}A_{SS}^*],
\end{eqnarray}
with the denominator
\begin{eqnarray}
\mathcal{D}=\int d\omega_1 d\omega_2 k(\omega_1)k(\omega_2)k(\omega_1,\omega_2)\sum_h Y_h|A_h|^2.
\end{eqnarray}

The above TPAs contain the integrands $Im(A_{\perp}A_h^*)=|A_{\perp}||A_h^*|\sin(\Delta\phi+\Delta \delta)$,
where $\Delta\phi$ and $\Delta\delta$ represent the weak and strong phase differences between the amplitudes
$A_{\perp}$ and $A_h$, respectively. Note that a strong phase difference yields a TPA even
in the absence of weak phases, so a nonzero TPA does not necessarily signal $CP$ violation.
To identify a true $CP$ violation signal, one has to compare the TPAs in  $B$ and $\bar{B}$ meson decays.
The helicity amplitude $\bar{A}_h$, whose weak phases flip sign relative to those in $A_h$, and the
TPAs $\bar{A}_T^i$ for a $CP$-conjugated process can be derived similarly.
We then have the true and fake asymmetries relative to the $CP$-averaged decay rate~\cite{prd84096013}
\begin{eqnarray}\label{eq:ture}
A_T^i(true)&\equiv& \frac{\Gamma(TP>0)-\Gamma(TP<0)+\bar{\Gamma}(TP>0)-\bar{\Gamma}(TP<0)}
{\Gamma(TP>0)+\Gamma(TP<0)+\bar{\Gamma}(TP>0)+\bar{\Gamma}(TP<0)}
 \propto \sin(\Delta\phi)\cos(\Delta \delta),\nonumber\\
A_T^i(fake)&\equiv& \frac{\Gamma(TP>0)-\Gamma(TP<0)-\bar{\Gamma}(TP>0)+\bar{\Gamma}(TP<0)}
{\Gamma(TP>0)+\Gamma(TP<0)+\bar{\Gamma}(TP>0)+\bar{\Gamma}(TP<0)}\propto \cos(\Delta\phi)\sin(\Delta \delta),
\end{eqnarray}
respectively, with $\bar{\Gamma}$ being the decay rate of the $CP$-conjugate process.
Note that $A_T^i(true)$, nonvanishing only in the presence of the weak phase difference,
provides an alternative measure of $CP$ violation. Furthermore, compared with direct $CP$ asymmetries,
$A_T^i(true)$ does not suffer the suppression from the strong phase difference, and reaches a maximum
when the strong phase difference vanishes.
On the contrary, $A_T^i(fake)$ can be nonzero even as the weak phase difference vanishes.
Such a quantity is sometimes referred to as a fake asymmetry, which
reflects the effect of strong phases~\cite{prd84096013}, instead of $CP$ violation.
Since the helicity amplitudes may have different strong phases,
it is likely that fake TPAs do not diminish for all channels.
In the modes with the neutral intermediate states $K^{*0}\bar{K}^{*0}$,
each helicity amplitude involves the same single weak phase in the SM, such that
true TPAs vanish without the weak phase difference.
Therefore, true TPAs in these modes, if observed, are probably signals of new physics.

The interference of the amplitude $A_{S^+}$ with the others generates  more asymmetries~\cite{jhep071662015}:
\begin{eqnarray}\label{eq:TDAs}
A_D^1&=&\frac{\Gamma\left((2\zeta_1-1)(2\zeta_2-1)\left(\frac{2\zeta_1-1}
{\sqrt{1+4\alpha_1}}+\frac{2\zeta_2-1}{\sqrt{1+4\alpha_2}}\right)>0\right)
-\Gamma\left((2\zeta_1-1)(2\zeta_2-1)\left(\frac{2\zeta_1-1}{\sqrt{1+4\alpha_1}}
+\frac{2\zeta_2-1}{\sqrt{1+4\alpha_2}}\right)<0\right)}
{\Gamma\left((2\zeta_1-1)(2\zeta_2-1)\left(\frac{2\zeta_1-1}
{\sqrt{1+4\alpha_1}}+\frac{2\zeta_2-1}{\sqrt{1+4\alpha_2}}\right)>0\right)
+\Gamma\left((2\zeta_1-1)(2\zeta_2-1)\left(\frac{2\zeta_1-1}
{\sqrt{1+4\alpha_1}}+\frac{2\zeta_2-1}{\sqrt{1+4\alpha_2}}\right)<0\right)}
\nonumber\\&=&\frac{3\sqrt{2}}{5\mathcal{D}}\int d\omega_1 d\omega_2 k(\omega_1)k(\omega_2)k(\omega_1,\omega_2)
[3Re(A_{S^+}A_0^*)+5Re(A_{S^+}A_{SS}^*)],\nonumber\\
A_D^2&=&\frac{\Gamma\left(\cos(\phi)\left(\frac{2\zeta_1-1}
{\sqrt{1+4\alpha_1}}+\frac{2\zeta_2-1}{\sqrt{1+4\alpha_2}}\right)>0\right)
-\Gamma\left(\cos(\phi)\left(\frac{2\zeta_1-1}{\sqrt{1+4\alpha_1}}
+\frac{2\zeta_2-1}{\sqrt{1+4\alpha_2}}\right)<0\right)}
{\Gamma\left(\cos(\phi)\left(\frac{2\zeta_1-1}{\sqrt{1+4\alpha_1}}
+\frac{2\zeta_2-1}{\sqrt{1+4\alpha_2}}\right)>0\right)
+\Gamma\left(\cos(\phi)\left(\frac{2\zeta_1-1}{\sqrt{1+4\alpha_1}}
+\frac{2\zeta_2-1}{\sqrt{1+4\alpha_2}}\right)<0\right)}\nonumber\\&=&
\frac{32}{5\pi\mathcal{D}}\int d\omega_1 d\omega_2 k(\omega_1)
k(\omega_2)k(\omega_1,\omega_2)Re[A_{S^+}A_{\parallel}^*],\nonumber\\
A_D^3&=&\frac{\Gamma\left(\frac{2\zeta_1-1}{\sqrt{1+4\alpha_1}}+\frac{2\zeta_2-1}{\sqrt{1+4\alpha_2}}>0\right)
-\Gamma\left(\frac{2\zeta_1-1}{\sqrt{1+4\alpha_1}}+\frac{2\zeta_2-1}{\sqrt{1+4\alpha_2}}<0\right)}
{\Gamma\left(\frac{2\zeta_1-1}{\sqrt{1+4\alpha_1}}+\frac{2\zeta_2-1}{\sqrt{1+4\alpha_2}}>0\right)
+\Gamma\left(\frac{2\zeta_1-1}{\sqrt{1+4\alpha_1}}+\frac{2\zeta_2-1}{\sqrt{1+4\alpha_2}}<0\right)}\nonumber\\&=&
\frac{6\sqrt{2}}{5\mathcal{D}}\int d\omega_1 d\omega_2k(\omega_1)k(\omega_2)k(\omega_1,\omega_2)
[Re(A_{S^+}A_0^*)+5Re(A_{S^+}A_{SS}^*)],\nonumber\\
A_D^4&=&\frac{\Gamma\left(\left(\frac{2\zeta_1-1}{\sqrt{1+4\alpha_1}}\right)^2
-\left(\frac{2\zeta_2-1}{\sqrt{1+4\alpha_2}}\right)^2>0\right)
-\Gamma\left(\left(\frac{2\zeta_1-1}{\sqrt{1+4\alpha_1}}\right)^2
-\left(\frac{2\zeta_2-1}{\sqrt{1+4\alpha_2}}\right)^2<0\right)}
{\Gamma\left(\left(\frac{2\zeta_1-1}{\sqrt{1+4\alpha_1}}\right)^2
-\left(\frac{2\zeta_2-1}{\sqrt{1+4\alpha_2}}\right)^2>0\right)
+\Gamma\left(\left(\frac{2\zeta_1-1}{\sqrt{1+4\alpha_1}}\right)^2
-\left(\frac{2\zeta_2-1}{\sqrt{1+4\alpha_2}}\right)^2<0\right)}\nonumber\\
&=&-\frac{3}{\mathcal{D}}\int d\omega_1 d\omega_2k(\omega_1)
k(\omega_2)k(\omega_1,\omega_2)Re[A_{S^+}A_{S^-}^{*}].
\end{eqnarray}
Similarly, combining the quantities for the corresponding $CP$-conjugate process,
we define the $S$-wave-induced  direct $CP$ asymmetries~\cite{jhep071662015}
\begin{eqnarray}\label{eq:ture2}
\mathcal{A}^{i}_{S}\equiv A_D^i+\bar{A}_D^i \propto Re(A_{S^+}A_h^*-\bar{A}_{S^+}\bar{A}_h^*).
\end{eqnarray}
Note that the above quantities, as the sum  of the asymmetries in the $B$ and $\bar{B}$ meson decays,
can be measured using untagged samples, in which $CP$-conjugate processes need not be
distinguished~\cite{prd92076013}.

\section{Numerical results}\label{sec:results}

In this section we calculate the physical observables for the
$B_{(s)}\rightarrow (K\pi)_{S/P}(K\pi)_{S/P}$ decays raised in the previous sections,
including the $CP$ averaged branching ratios, $S$-wave fractions, polarization fractions, TPAs, and
direct $CP$ asymmetries. The inputs for the numerical analysis are summarized below.
The meson masses are taken as (in units of GeV)~\cite{pdg2018}
\begin{eqnarray}
M_{B_s}&=&5.37, \quad M_B=5.28,  \quad m_{K}=0.494, \quad m_{\pi}=0.14.
\end{eqnarray}
For the CKM matrix elements, we adopt the Wolfenstein parametrization with the parameters~\cite{pdg2018}
\begin{eqnarray}
\lambda &=& 0.22650, \quad  A=0.790,  \quad \bar{\rho}=0.141, \quad \bar{\eta}=0.357.
\end{eqnarray}
The lifetime (in units of ps) and decay constant (in units of GeV) of the $B_{(s)}$ meson are set to
the values~\cite{pdg2018,prd91054033},
\begin{eqnarray}
f_{B_s} &=& 0.24 , \quad  f_B=0.21,  \quad \tau_{B_s}=1.51, \quad  \tau_{B^0}=1.52, \quad  \tau_{B^+}=1.638.
\end{eqnarray}
Other parameters appearing in the $K\pi$ two-meson DAs have been specified before.

We first examine  the $B \rightarrow K\pi$
transition form factors, to which the factorizable emission contributions from Figs.~\ref{fig:fym}(a),(b)
are related. These form factors serve as the nonperturbative inputs to QCDF evaluations of
the factorizable emission amplitudes~\cite{npb899247}. There are in general seven form factors for the
$P$-wave $B \rightarrow K\pi$ transition and three for the $S$-wave one, whose definitions can be found in
Refs.~\cite{jhep122019083,plb730336,prd85034014}.
Here we focus only on the form factor $F_\perp(\omega_1^2,q^2)$ defined via the matrix element
$\langle K\pi|s\gamma^\mu b|B\rangle$ with the momentum transfer  $q$ being given in Eq.~(\ref{eq:pq}).
A complete and systematic PQCD investigation of all the $B \rightarrow K\pi$  form factors, including their
$\omega_1^2$ and $q^2$ dependencies, will be postponed to a future work.
The corresponding PQCD factorization formula for  $F_\perp(\omega_1^2,q^2)$ is also presented in Appendix~\ref{sec:ampulitude},
which yields $F_\perp(m_{K^*(892)}^2,0)=88$ (64) for the $B_{(s)} \rightarrow K\pi$ transition at maximal recoil.
Results for the $B \rightarrow K\pi$ transition from light-cone sum rules in the narrow-width limit, ranging
between 62 and 93~\cite{jhep122019083,jhep012019150,jhep082016098,jhep092010089}, agree well with ours.
Our $B_s\rightarrow K\pi$ transition form factor is also consistent with  $72\pm 7$ derived in Ref~\cite{jhep082016098}.



\subsection{$CP$ averaged four-body branching ratios and  $S$-wave fractions}
\begin{table}[!htbh]
\caption{PQCD predictions for the $CP$-averaged branching ratios of various components and their sum
in the $B_{(s)}\rightarrow (K\pi)_{S/P}(K\pi)_{S/P}$ decays within the $K\pi$ invariant
mass window of 150 MeV around the $K^*(892)$ resonance.}
\label{tab:brs}
\begin{tabular}[t]{lccc}
\hline\hline
Components  & $B_s^0\rightarrow (K^+\pi^-)(K^-\pi^+)$ & $B_s^0\rightarrow (K^0\pi^+)(\bar{K}^{0}\pi^-)$ & \\
\hline
$\mathcal{B}_0$ & $(2.7^{+0.4+0.3+1.3}_{-0.5-0.4-0.8})\times 10^{-6}$ & $(2.1^{+0.4+0.3+1.0}_{-0.3-0.2-0.4})\times 10^{-6}$ &\\
$\mathcal{B}_{\parallel}$ & $(7.7^{+0.1+1.2+3.4}_{-0.2-1.2-2.1})\times 10^{-7}$  & $(7.4^{+0.1+1.3+3.3}_{-0.1-1.2-1.9})\times 10^{-7}$&\\
$\mathcal{B}_{\perp}$ & $(7.7^{+0.3+1.3+3.5}_{-0.2-1.1-2.0})\times 10^{-7}$  & $(7.3^{+0.1+1.2+3.2}_{-0.2-1.2-1.9})\times 10^{-7}$ &\\
$\mathcal{B}_{SS}$ & $(5.1^{+0.6+3.5+1.6}_{-0.6-3.0-1.1})\times 10^{-7}$  &$(4.7^{+0.5+3.3+1.2}_{-0.4-2.5-0.8})\times 10^{-7}$ &\\
$\mathcal{B}_{S^+}$ & $(6.8^{+1.3+1.0+3.2}_{-1.1-1.1-1.7})\times 10^{-7}$ &$(7.5^{+1.4+1.0+2.7}_{-1.5-1.1-1.7})\times 10^{-7}$ &\\
$\mathcal{B}_{S^-}$ & $(2.7^{+0.6+0.5+1.1}_{-0.4-0.3-0.6})\times 10^{-6}$ &$(2.7^{+0.5+0.4+0.9}_{-0.2-0.3-0.5})\times 10^{-6}$&\\
$\mathcal{B}_{\text{total}}$ & $(8.1^{+1.3+1.1+3.6}_{-1.1-1.2-2.2})\times 10^{-6}$ &$(7.5^{+1.2+1.0+2.7}_{-0.9-0.6-1.5})\times 10^{-6}$&\\
\hline
Components &$B^0\rightarrow (K^-\pi^+)(K^+\pi^-)$ &$B^0\rightarrow (K^0\pi^+)(\bar{K}^{0}\pi^-)$&$B^+\rightarrow (K^0\pi^+)(K^-\pi^+)$ \\
\hline
$\mathcal{B}_0$ &$(1.0^{+0.4+0.3+0.5}_{-0.2-0.1-0.2})\times 10^{-7}$ & $(9.3^{+3.2+3.0+0.4}_{-2.4-0.7-0.6})\times 10^{-8}$& $(2.9^{+1.2+0.3+1.0}_{-0.6-0.2-0.6})\times 10^{-7}$  \\
$\mathcal{B}_{\parallel}$ & $(2.1^{+0.2+0.4+0.8}_{-0.1-0.3-0.5})\times 10^{-8}$ & $(1.4^{+0.2+1.5+0.1}_{-0.4-0.9-0.2})\times 10^{-10}$& $(2.0^{+0.3+0.4+0.9}_{-0.1-0.3-0.5})\times 10^{-8}$  \\
$\mathcal{B}_{\perp}$ &$(2.0^{+0.2+0.4+0.9}_{-0.1-0.3-0.5})\times 10^{-8}$& $(2.1^{+0.0+0.4+0.1}_{-0.2-0.8-0.0})\times 10^{-11}$ & $(2.1^{+0.2+0.3+0.9}_{-0.2-0.4-0.5})\times 10^{-8}$ \\
$\mathcal{B}_{SS}$ &$(2.2^{+0.3+1.6+0.8}_{-0.1-1.2-0.4})\times 10^{-8}$ & $(1.2^{+0.3+1.1+0.0}_{-0.3-0.3-0.0})\times 10^{-8}$ & $(2.4^{+0.5+1.6+0.7}_{-0.3-1.1-0.5})\times 10^{-8}$ \\
$\mathcal{B}_{S^+}$&$(2.3^{+0.8+0.4+0.8}_{-0.6-0.2-0.4})\times 10^{-8}$ & $(1.5^{+0.4+1.1+0.2}_{-0.3-0.0-0.1})\times 10^{-8}$ & $(2.5^{+0.9+1.1+1.0}_{-0.5-0.2-0.5})\times 10^{-8}$ \\
$\mathcal{B}_{S^-}$&$(6.1^{+2.5+1.2+2.9}_{-1.5-0.9-1.6})\times 10^{-8}$ & $(4.5^{+1.3+2.2+0.1}_{-0.9-1.0-0.0})\times 10^{-8}$ & $(9.0^{+3.3+1.4+3.8}_{-2.0-1.7-2.0})\times 10^{-8}$ \\
$\mathcal{B}_{\text{total}}$ &$(2.5^{+0.8+0.3+1.1}_{-0.6-0.4-0.7})\times 10^{-7}$ & $(1.7^{+0.5+0.4+0.0}_{-0.3-0.3-0.0})\times 10^{-7}$ &$(4.7^{+1.6+0.7+1.5}_{-1.0-0.5-1.0})\times 10^{-7}$ \\
\hline\hline
\end{tabular}
\end{table}

The integral in Eq.~(\ref{eq:brsss}) over $\omega_{1,2}$ in the selected mass range gives
the  $CP$-averaged branching ratios of various components and their sum for five penguin-dominated channels, which
are given in Table~\ref{tab:brs}. The two $B_s$ modes, induced by the $b\rightarrow s$
transition, have  relatively large  branching ratios of $\mathcal{O}(10^{-6})$. The remaining three
modes, mediated by the $b\rightarrow d$ transition, are generally an order of magnitude smaller.
In particular, the $B^0\rightarrow (K^0\pi^+)(\bar{K}^{0}\pi^-)$ decay, proceeding only through the
power-suppressed weak annihilation, has an even lower rate.
We have considered three sources of theoretical uncertainties:
the shape parameters in the $B_{(s)}$ meson DAs with 10\% variation,
the Gegenbauer moments in the $S$- and $P$-wave $K\pi$ two-meson DAs,
and the hard scales $t$ defined in Appendix~\ref{sec:ampulitude}, that vary from $0.75t$ to $1.25t$.
Their effects are listed in Table~\ref{tab:brs} in order, among which the third one is more dominant.
Higher-order contributions to four-body $B$ meson decays can be included to reduce the
sensitivity to the variation of the hard scales. The Gegenbauer moments also
need to be constrained for further improving the precision of theoretical predictions.

\begin{table}[!htbh]
\caption{$S$-wave fractions in  the $B_{(s)}\rightarrow (K\pi)_{S/P}(K\pi)_{S/P}$  decays within the $K\pi$ invariant
mass window of 150 MeV around the $K^*(892)$ resonance.
The data corresponding to the same invariant mass window are taken from~\cite{jhep071662015,jhep070322019},
where the first uncertainty is statistical and the second is systematic.}
\label{tab:brs3}
\begin{tabular}[t]{lcccc}
\hline\hline
Modes & $f_{SS}(\%)$ & $f_{S^+}(\%)$  & $f_{S^-}(\%)$ & $f_{\text{S-wave}} (\%)$ \\
\hline
$B_s^0\rightarrow (K^+\pi^-)(K^-\pi^+)$& $6.3^{+0.1+3.5+0.2}_{-0.2-3.4-0.4}$& $8.4^{+0.2+1.2+0.4}_{-0.1-0.9-0.0}$
&$33.3^{+1.6+2.6+0.1}_{-1.2-2.4-0.2}$& $48.0^{+1.9+7.3+0.7}_{-1.5-6.7-0.6}$\\
LHCb~\cite{jhep071662015} &$6.6\pm2.2\pm0.7$ &$11.4\pm3.7\pm2.3$&$48.5\pm5.1\pm1.9$&$\cdots$\\
LHCb~\cite{jhep070322019} &$8.7\pm1.1\pm1.1$ &$4.8\pm1.4\pm1.1$&$55.8\pm2.1\pm1.4$&$69.4\pm1.6\pm1.0$\\ \hline
$B^0\rightarrow (K^-\pi^+)(K^+\pi^-)$  & $8.8^{+1.5+5.5+0.2}_{-1.1-4.9-0.4}$& $9.2^{+0.4+1.0+0.2}_{-0.9-0.7-0.2}$
& $24.4^{+1.5+1.7+0.5}_{-1.1-2.0-0.5}$& $42.4^{+3.4+8.2+0.9}_{-3.1-7.6-1.1}$\\
LHCb~\cite{jhep070322019} &$2.3\pm1.4\pm0.4$ &$0.8\pm1.3\pm0.7$&$37.7\pm5.2\pm2.4$&$40.8\pm5.0\pm1.7$\\ \hline
$B_s^0\rightarrow (K^0\pi^+)(\bar{K}^{0}\pi^-)$ &$6.4^{+0.3+4.1+0.1}_{-0.2-3.4-0.3}$&$9.9^{+0.1+0.3+0.0}_{-0.8-0.9-0.5}$&$35.1^{+0.7+1.7+0.0}_{-0.7-3.0-0.6}$&$51.4^{+1.1+6.1+0.1}_{-1.7-7.3-1.4}$\\
$B^0\rightarrow (K^0\pi^+)(\bar{K}^{0}\pi^-)$& $7.1^{+0.0+5.1+0.0}_{-0.4-1.9-0.1}$& $8.8^{+0.2+5.3+0.4}_{-0.1-0.0-0.4}$&$26.5^{+0.0+6.8+0.0}_{-1.0-3.3-0.4}$& $42.4^{+0.2+17.2+0.4}_{-1.5-5.2-0.9}$\\
$B^+\rightarrow (K^0\pi^+)(K^-\pi^+)$& $5.1^{+0.3+3.0+0.0}_{-0.5-2.6-0.2}$& $5.3^{+0.2+1.9+0.5}_{-0.3-0.4-0.3}$& $19.1^{+0.0+1.3+0.9}_{-0.2-3.4-0.0}$& $29.5^{+0.5+6.2+1.4}_{-1.0-6.4-0.5}$\\
\hline\hline
\end{tabular}
\end{table}

\begin{figure}[tbp]
\begin{center}
\setlength{\abovecaptionskip}{0pt}
\centerline{
\hspace{4cm}\subfigure{\epsfxsize=13 cm \epsffile{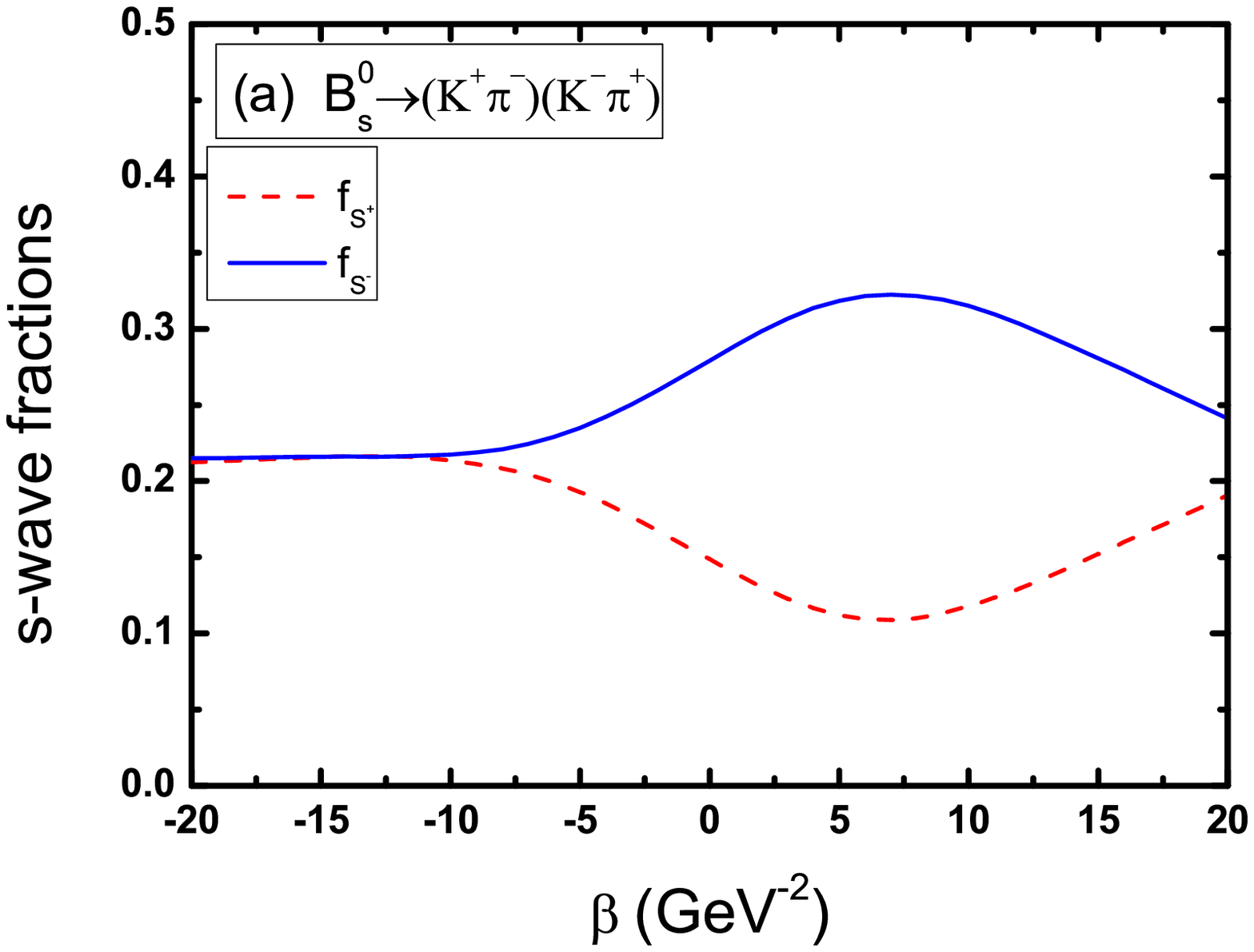} }
\hspace{-6cm}\subfigure{ \epsfxsize=13 cm \epsffile{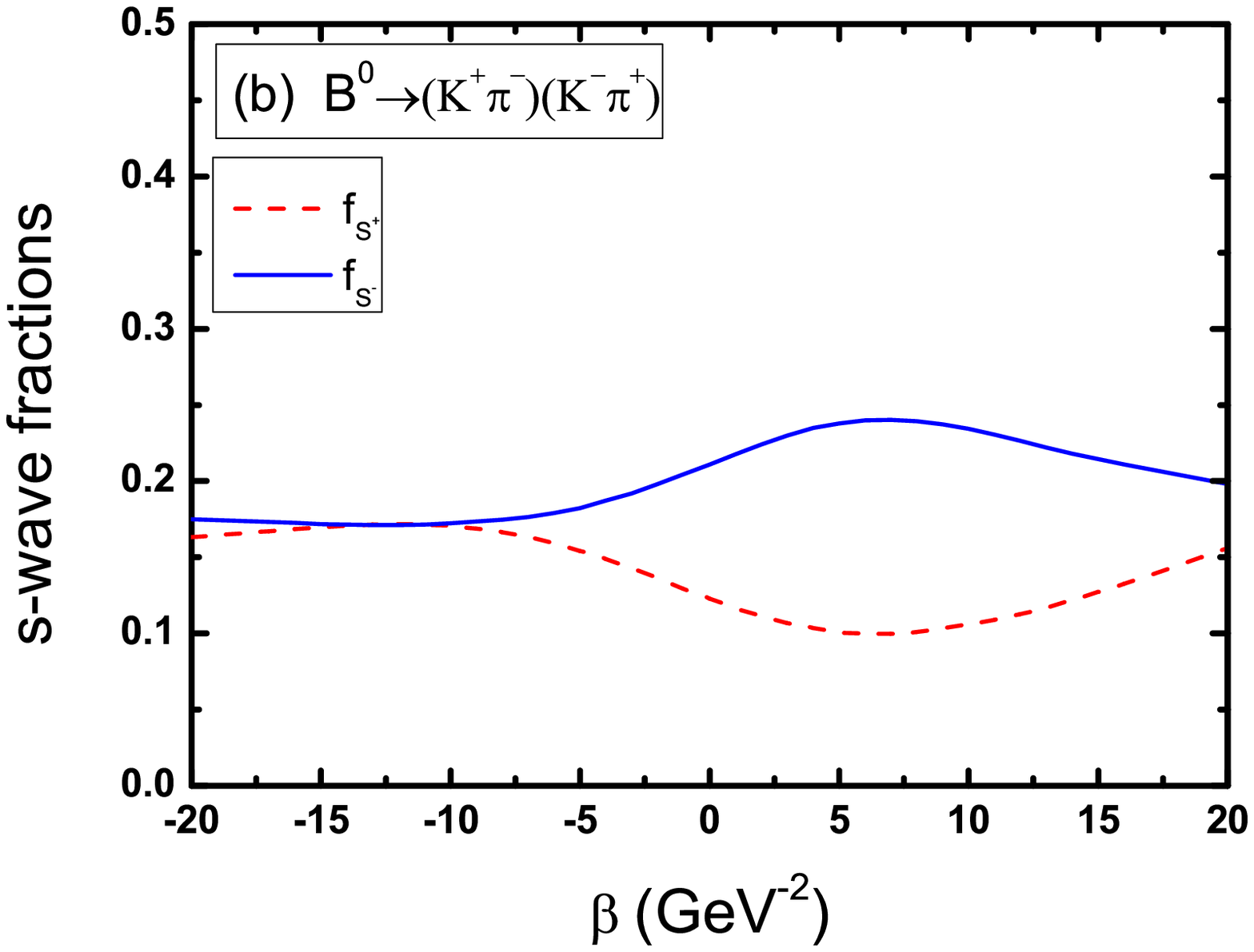}}}
\vspace{-3cm}\caption{Dependencies of the $S$-wave fractions $f_{S^+}$ (dashed red curve) and
$f_{S^-}$ (solid blue curve) on the parameter $\beta$ in
the modes (a) $B_s^0\rightarrow (K^+\pi^-)(K^-\pi^+)$ and (b) $B^0\rightarrow (K^+\pi^-)(K^-\pi^+)$.}
 \label{fig:sfrac}
\end{center}
\end{figure}

It is straightforward to derive from the branching ratios in Table~\ref{tab:brs} and Eq.~(\ref{eq:fss})
the $S$-wave fractions in each mode as presented in Table~\ref{tab:brs3}.
According to Eq.~(\ref{eq:s-s+}), the amplitudes $A_{S^-}$ and $A_{S^+}$, as the linear superposition
of $A_{SV}$ and $A_{VS}$, exhibit the interference between the $S$- and $P$-wave configurations.
Therefore, the phenomenological parameter $\beta$ introduced in Eq.~(\ref{eq:phase}) affects the
relative strength of $f_{S^+}$ and $f_{S^-}$. Figure~\ref{fig:sfrac} illustrates the influence of
$\beta$ on the two single $S$-wave fractions, where the dashed red and solid blue curves describe
$f_{S^+}$ and $f_{S^-}$, respectively. It is found that the $S$-wave fractions
 are more sensitive to $\beta$ in the  region of [-5,20] GeV$^{-2}$.
The value of $\beta$  can be bounded in the range  $\beta=(7.5\pm 2.5)$ GeV$^{-2}$, in which
$f_{S^+}$ and  $f_{S^-}$ have reasonable relative magnitudes to match the data better.
The component $f_{SS}$, which is proportional to the modulus squared of
its corresponding amplitude, is independent of $\beta$.
The total $S$-wave contribution to $f_{\text{S-wave}}$ is also independent of the phase parameter,
since the interference terms  cancel in the sum $f_{S^+}+f_{S^-}$ in Eq.~(\ref{eq:fss3}).

The obtained three $S$-wave fractions for the $B_s^0\rightarrow (K^+\pi^-)(K^-\pi^+)$ decay
are compatible with the two  measurements in Refs.~\cite{jhep071662015,jhep070322019} within errors.
The central values of $f_{SS}$ and  $f_{S^+}$ for the $B^0\rightarrow (K^+\pi^-)(K^-\pi^+)$ decay
are obviously larger, while $f_{S^-}$  is  a bit smaller than the data~\cite{jhep070322019}.
Nevertheless, the total $S$-wave contribution, accounting for about $42.4\%$ of the total
decay rate, is consistent with the data~\cite{jhep070322019}. More precise measurements
and more complete theoretical analyses at the subleading level will help clarify the discrepancy.
We point out  that the $S$-wave contributions could be modified by adding higher Gegenbauer terms
to the twist-3 DAs for the  $S$-wave $K\pi$ pair, and better consistency with the data is expected.
However, such fine tuning will not be attempted in the present work.
The $S$-wave fractions of the other channels in Table~\ref{tab:brs3}
exhibit similar tendency. The predicted total $S$-wave fractions, ranging from
$29.5\%$ to $51.4\%$, can be tested at future experiments.
It is obvious that the $S$-wave contributions to the $B_{(s)}\rightarrow (K\pi)_{S/P}(K\pi)_{S/P}$
decays in the considered mass region are significant.

\subsection{ Two-body branching ratios and  polarization fractions}

The $P$-wave branching ratios in Table~\ref{tab:brs} can be  converted into those of the two-body decays
in the narrow-width limit via
\begin{eqnarray}\label{2body}
\mathcal{B}(B \rightarrow K^*(\rightarrow K\pi)\bar{K}^*(\rightarrow K\pi))\approx
\mathcal{B}(B \rightarrow K^*\bar{K}^*)\times \mathcal{B}(K^*\rightarrow K\pi)\times \mathcal{B}(\bar{K}^*\rightarrow K\pi),
\end{eqnarray}
where $K^*$ denotes $K^{*0}$ or $K^{*+}$, and the $K^*\rightarrow K\pi$
branching ratio $\mathcal{B}(K^*\rightarrow K\pi)=1$~\cite{pdg2018} should be
scaled by the corresponding squared Clebsch-Gordan coefficient, i.e., 2/3 for
$K^{*0}(\bar{K}^{*0})\rightarrow K^+\pi^-(K^-\pi^+)$ and  $K^{*+}\rightarrow K^0\pi^+$.
With the scaled secondary branching fraction $\mathcal{B}(K^{*0}(K^{*+})\rightarrow K^+\pi^-(K^0\pi^+))=2/3$,
we extract the two-body branching ratios  $\mathcal{B}(B \rightarrow K^*\bar{K}^{*})$ from Eq.~(\ref{2body}).
The results are summarized in Table.~\ref{tab:brpwave},
where those from the PQCD approach~\cite{prd91054033,prd76074018,npb93517,prd72054015},
the QCDF approach~\cite{prd80114026,npb77464,prd78094001,prd80114008}, the
SCET~\cite{prd96073004}, and the factorization-assisted topological-amplitude (FAT) approach~\cite{epjc77333}
are also shown for comparison. The world average experimental data are taken from~\cite{pdg2018} whenever available.

\begin{table}[!htbh]
\caption{$CP$-averaged branching ratios and polarization fractions for the two-body $B\rightarrow K^*\bar{K}^*$ decays.
For comparison, we also list the results from
PQCD~\cite{prd91054033,prd76074018,npb93517,prd72054015},
QCDF~\cite{prd80114026,npb77464,prd78094001,prd80114008},
SCET~\cite{prd96073004}, and FAT~\cite{epjc77333}.
The world averages of experimental data are taken from~\cite{pdg2018}.}
\label{tab:brpwave}
\begin{tabular}[t]{lcccc}
\hline\hline
Modes & $\mathcal{B}(10^{-6})$ & $f_0(\%)$ &$f_{\parallel}(\%)$ &$f_{\perp}(\%)$\\ \hline
$B_s^0\rightarrow K^{*0}\bar{K}^{*0}$ & $9.5^{+1.1+1.1+4.6}_{-1.1-1.1-2.6}$ &$63.6^{+2.7+3.3+1.0}_{-4.2-3.9-1.0}$
&$18.2^{+2.1+1.9+0.4}_{-1.5-1.7-0.5}$&$18.2^{+2.0+1.8+0.5}_{-1.3-1.6-0.6}$\\
PQCD-I~\cite{prd91054033} &$5.4^{+3.0}_{-2.4}$ &$38.3^{+12.1}_{-10.5}$&$\cdots$ &$30.0^{+5.3}_{-6.1}$\\
PQCD-II~\cite{prd76074018}& $7.8^{+4.2}_{-2.7}$&$49.7^{+5.7}_{-6.1}$ &$26.8^{+3.3}_{-3.0}$ &$23.5^{+2.8}_{-2.7}$ \\
PQCD-NLO~\cite{npb93517}&$6.7^{+2.9}_{-2.2}$&$43.4^{+12.7}_{-12.9}$&$\cdots$ &$23.5^{+5.8}_{-5.9}$\\
QCDF-I~\cite{prd80114026}& $6.6\pm2.2$ &$56^{+22}_{-27}$ &$\cdots$ &$\cdots$ \\
QCDF-II~\cite{npb77464}& $9.1^{+11.3}_{-6.8}$ &$63^{+42}_{-29}$ &$\cdots$ &$\cdots$ \\
SCET~\cite{prd96073004}&$8.6\pm3.1$&$44.9\pm18.3$&$\cdots$ &$24.9\pm11.1$\\
FAT~\cite{epjc77333}&$14.9\pm3.6$&$34.3\pm12.6$&$\cdots$&$33.2\pm6.9$\\
Data~\cite{pdg2018}& $11.1\pm2.7$ &$24\pm4$ &$30\pm5$ &$38\pm12$ \\ \hline
$B_s^0\rightarrow K^{*+}K^{*-}$ & $8.0^{+0.8+1.2+3.2}_{-0.6-0.8-1.6}$ &$58.8^{+2.9+3.0+1.0}_{-3.4-1.3-0.0}$
&$20.7^{+1.8+0.7+0.0}_{-1.6-0.7-0.5}$&$20.5^{+1.6+0.5+0.0}_{-1.3-1.4-0.5}$\\
PQCD-I~\cite{prd91054033} &$5.4^{+3.3}_{-2.3}$ &$42.0^{+14.2}_{-11.2}$&$\cdots$ &$27.7^{+5.2}_{-7.0}$\\
PQCD-II~\cite{prd76074018}&$6.7^{+3.7}_{-1.9}$&$43.8^{+6.6}_{-4.9}$&$30.1^{+2.4}_{-3.4}$&$26.1^{+2.4}_{-3.2}$\\
PQCD-NLO~\cite{npb93517}&$6.5^{+2.8}_{-2.1}$&$48.1^{+9.7}_{-8.9}$&$\cdots$ &$23.9^{+4.4}_{-5.2}$\\
QCDF-I~\cite{prd80114026}&$7.6^{+2.5}_{-2.1}$  &$52^{+20}_{-21}$ &$\cdots$&$\cdots$\\
QCDF-II~\cite{npb77464}&$9.1^{+10.5}_{-6.3}$&$67^{+31}_{-26}$ &$\cdots$&$\cdots$\\
SCET~\cite{prd96073004}&$11.0\pm3.3$&$55\pm14$&$\cdots$&$20.3\pm8.6$\\
FAT~\cite{epjc77333}&$15.9\pm3.5$&$30.9\pm10.4$&$\cdots$&$34.9\pm5.8$\\ \hline
$B^0\rightarrow K^{*0}\bar{K}^{*0}$ & $0.32^{+0.09+0.09+0.16}_{-0.04-0.04-0.08}$ &$71.5^{+3.8+2.7+1.5}_{-2.6-1.6-0.8}$
&$14.6^{+1.3+0.8+0.4}_{-1.9-1.4-0.7}$&$13.9^{+1.3+0.8+0.4}_{-1.9-1.4-0.8}$\\
PQCD-I~\cite{prd91054033} &$0.34^{+0.16}_{-0.15}$&$58\pm8$&$\cdots$&$19.7^{+4.0}_{-3.6}$\\
PQCD-II~\cite{prd72054015} &$0.35$&$78$&$12$&$10$\\
QCDF-I~\cite{prd78094001,prd80114008}&$0.6^{+0.2}_{-0.3}$  &$52\pm48$ &$\cdots$&$24\pm24$\\
QCDF-II~\cite{npb77464}&$0.6^{+0.5}_{-0.3}$&$69^{+34}_{-27}$ &$\cdots$&$\cdots$\\
SCET~\cite{prd96073004}&$0.48\pm0.16$&$50\pm16$&$\cdots$&$22.9\pm10.0$\\
FAT~\cite{epjc77333}&$0.61\pm0.17$&$58.3\pm11.1$&$\cdots$&$20.8\pm6.0$\\
Data~\cite{pdg2018}& $0.83\pm0.24$ &$74\pm5$ &$\cdots$ &$\cdots$\\
\hline $B^0\rightarrow K^{*+}K^{*-}$ & $0.21^{+0.06+0.06+0.01}_{-0.05-0.03-0.01}$ &$\sim100$&$\sim0.0$&$\sim0.0$\\
PQCD-I~\cite{prd91054033} &$0.21\pm0.10$&$\sim100$&$\sim0.0$&$\sim0.0$\\
PQCD-II~\cite{prd72054015} &$0.064^{+0.005}_{-0.010}$&$99$&$0.5$&$0.4$\\
QCDF-I~\cite{prd78094001,prd80114008}&$0.1\pm0.1$  &$\sim100$ &$\sim0.0$&$\sim0.0$\\
FAT~\cite{epjc77333}&$1.43\pm0.96$ &$\cdots$ &$\cdots$&$\cdots$\\
Data~\cite{pdg2018}& $<2.0$ &$\cdots$ &$\cdots$ &$\cdots$ \\
\hline $B^+\rightarrow K^{*+}\bar{K}^{*0}$ & $0.74^{+0.24+0.06+0.22}_{-0.14-0.08-0.16}$ &$87.6^{+2.3+0.8+0.1}_{-2.3-1.3-0.9}$
&$6.0^{+1.2+0.7+0.5}_{-1.0-0.3-0.0}$&$6.4^{+1.0+0.4+0.3}_{-1.4-0.6-0.2}$\\
PQCD-I~\cite{prd91054033} &$0.56^{+0.26}_{-0.22}$&$74^{+4}_{-5}$&$\cdots$&$12.9^{+1.7}_{-2.4}$\\
PQCD-II~\cite{prd72054015} &$0.48^{+0.12}_{-0.08}$&$81.5$&$9.0$&$9.5$\\
QCDF-I~\cite{prd78094001,prd80114008}&$0.6\pm0.3$  &$45^{+55}_{-38}$ &$\cdots$&$27^{+19}_{-27}$\\
QCDF-II~\cite{npb77464}&$0.5^{+0.4}_{-0.3}$&$62^{+42}_{-33}$ &$\cdots$&$\cdots$\\
SCET~\cite{prd96073004}&$0.52\pm0.18$&$50\pm16$&$\cdots$&$22.9\pm10.0$\\
FAT~\cite{epjc77333}&$0.66\pm0.18$ &$58.3\pm11.1$ &$\cdots$&$20.8\pm6.0$\\
Data~\cite{pdg2018}& $0.91\pm0.29$ &$82^{+15}_{-21}$ &$\cdots$ &$\cdots$ \\
\hline\hline
\end{tabular}
\end{table}

Our predicted branching ratio for the pure penguin mode $B^0_{s}\rightarrow K^{*0} \bar{K}^{*0}$,
being higher than those from the PQCD and QCDF approaches,
agrees with the data. As to the branching ratio of another pure penguin mode
$B^0\rightarrow K^{*0} \bar{K}^{*0}$, our prediction is slightly lower than the PQCD and QCDF ones,
which are all systematically below the data.  It has been found~\cite{prd72054015} that
the factorizable  emission amplitude $\mathcal{F}_e$ gives the dominant contribution to the
$B^0 \rightarrow K^{*0} \bar{K}^{*0}$ decay, and the  annihilation amplitudes
cancel strongly between $\mathcal{F}_a/\mathcal{M}_a$ and $\mathcal{F}'_a/\mathcal{M}'_a$
in Eq.~(\ref{eq:ampli}). On the contrary, such cancellation does not occur in the charged  mode
$B^+ \rightarrow K^{*+} \bar{K}^{*0}$ owing to vanishing $\mathcal{F}'_a/\mathcal{M}'_a$.
Moreover, the charged mode receives a color-allowed tree contribution
from the annihilation diagrams, which is induced by the $(V-A)(V-A)$ operators with the large Wilson
coefficient $C_2+C_1/3$, but subject to the helicity suppression compared with that from the emission diagrams.
The interference between the annihilation and emission amplitudes turns out to increase the branching ratio of
the charged  mode, such that our branching ratio $\mathcal{B}(B^+ \rightarrow K^{*+} \bar{K}^{*0})$
is about twice of the neutral one $\mathcal{B}(B^0 \rightarrow K^{*0} \bar{K}^{*0})$. It is thus
greater than those from the other approaches and more consistent with the data.
The pattern $\mathcal{B}(B^+ \rightarrow K^{*+} \bar{K}^{*0})>\mathcal{B}(B^0 \rightarrow K^{*0} \bar{K}^{*0})$
has been also observed in the previous PQCD investigations~\cite{prd72054015,prd91054033}.

The derived $B^0_s \rightarrow K^{*+} K^{*-}$ branching ratio is close to the
$B^0_{s}\rightarrow K^{*0} \bar{K}^{*0}$ one, because the additional tree contribution
to the former is minor. This feature is common among the predictions from all the approaches.
The predicted branching ratio for the pure annihilation decay $B^0 \rightarrow K^{*+} K^{*-}$
spans a wide range, $(0.064-1.43)\times 10^{-6}$, with large theoretical uncertainties,
as indicated in Table~\ref{tab:brpwave}. Our result is basically the same as the previous PQCD
calculation~\cite{prd91054033}, and about twice the QCDF one~\cite{prd78094001,prd80114008}.
The value from the FAT approach~\cite{epjc77333} is far too large,
but still lower than the upper experimental bound~\cite{pdg2018}.
More precise measurements of this channel are crucial for testing the theoretical expectations.

The heavy quark limit implies that charmless $B\rightarrow VV$  decays mainly produce longitudinally polarized
final states due to the vector-axial structure of weak currents and the quark helicity conservation in
high-energy QCD interactions~\cite{npb77464,plb601151}.
However, data for penguin-dominated $B$ meson decays, contradicting to this speculation in general,
hint that the dynamics of penguin transitions are more complicated.
This so-called polarization puzzle in  penguin-dominated $B\rightarrow VV$ modes has posed a
challenge to the development of theoretical frameworks for $B$ meson decays.

As shown in Table~\ref{tab:brpwave}, the longitudinal polarization fraction $f_0$ for  the
$B^{0(+)}\rightarrow K^{*0(+)}\bar{K}^{*0}$ decays from the PQCD approach (including the present work) are
higher than the transverse one $f_T=f_{\parallel}+f_{\perp}$.
Both the QCDF~\cite{prd78094001,prd80114008} and SCET~\cite{prd96073004}
yield the similar pattern $f_0\sim f_T$, albeit with large uncertainties.
The world average data in Table~\ref{tab:brpwave}~\cite{prd91071101,prd79051102,jhep070322019,prl100081801}
\begin{eqnarray}
f_0(B^0\rightarrow K^{*0}\bar{K}^{*0})&=&\left\{
\begin{aligned}
(&0.724 \pm 0.051(\text{stat})\pm 0.016(\text{syst})) \quad\quad\quad  &\text{LHCb (2019)}, \nonumber\\ 
(&0.80^{+0.10}_{-0.12}(\text{stat})\pm 0.06(\text{syst})) \quad\quad\quad  & \text{BABAR}~(2008).  \nonumber\\ 
\end{aligned}\right. \nonumber\\
f_0(B^+\rightarrow K^{*+}\bar{K}^{*0})&=&\left\{
\begin{aligned}
(&1.06 \pm 0.30(\text{stat})\pm 0.14(\text{syst})) \quad\quad\quad  &\text{Belle (2015)}, \nonumber\\ 
(&0.75^{+0.16}_{-0.26}(\text{stat})\pm 0.03(\text{syst})) \quad\quad\quad  & \text{BABAR}~(2009),   
\end{aligned}\right.
\end{eqnarray}
match the predicted large longitudinal polarization fractions in these two modes.
For the pure annihilation process $B^0\rightarrow K^{*+}K^{*-}$, all the theoretical approaches suggest
that it is extremely dominated by the longitudinal polarization with $f_0 \sim 1$.

The observation  is different for the $B_s$ meson decays: a recent LHCb measurement~\cite{jhep070322019}
revealed an unexpectedly  low longitudinal polarization fraction $f_0\sim 0.2$
for the $B_s^0\rightarrow K^{*0}\bar{K}^{*0}$ mode, which is difficult to explain.
One can see from Table~\ref{tab:brpwave} that various predictions for $f_0(B_s^0\rightarrow K^{*0}\bar{K}^{*0})$
lie in the range 0.343 to 0.636, considerably larger than the data.
The QCDF evaluation gives a low longitudinal polarization fraction
$f_0(B_s^0\rightarrow K^{*0}\bar{K}^{*0})=(27.7^{+8.2+9.5}_{-6.7-18.9})\%$~\cite{epjc77415}
by including weak annihilation corrections with the best-fit endpoint parameters. This abnormal longitudinal
polarization fraction seems to be reconciled, but an obvious tension between the data and the prediction
for the branching ratio $\mathcal{B}({B}_s\rightarrow \phi K^*)$ is invoked accordingly. The PQCD approach yields
a sizable transverse polarization, which is enhanced by
the weak annihilation from the operator $O_6$ and by nonfactorizable contributions~\cite{prd71054025}.
However, these effects are not able to fully account for the above polarization anomaly.
The small $f_0=0.38$ in Ref.~\cite{prd91054033} is ascribed to the inclusion of the higher-power terms
proportional to $r^2$, with $r$ being the mass ratio between the vector and $B$ mesons.
Our predictions for the longitudinal polarization fractions agree with the QCDF ones~\cite{prd80114026,npb77464},
and a bit higher than the previous PQCD results.
It is worth mentioning that we have employed the Gegenbauer moments for
the transversely polarized $K\pi$ DAs the same as for the longitudinal polarized ones (see Eq.~(\ref{eq:a1ka2k}))
in this analysis. It should not be difficult to accommodate both the branching ratio and polarization data for
the $B_s$ modes  simultaneously by tuning the Gegenbauer moments for the former.

It is noted that the longitudinal polarization contributions to the $B_s\rightarrow K^{*0}\bar{K}^{*0}$
and $B^0\rightarrow K^{*0}\bar{K}^{*0}$ decays can be related to each other in the $U$-spin symmetry.
Nevertheless, the data reveal quite different longitudinal polarization contributions in these two modes.
This discrepancy indicates that the $U$-spin relations are not well respected generally.
The following hierarchy pattern for the polarization fractions
\begin{eqnarray}
f_0(B^0\rightarrow K^{*+}K^{*-})>f_0(B^+\rightarrow K^{*+}\bar{K}^{*0})>
f_0(B^0\rightarrow K^{*0}\bar{K}^{*0})>f_0(B_s\rightarrow K^{*+}K^{*-})\sim f_0(B_s\rightarrow K^{*0}\bar{K}^{*0}),
\end{eqnarray}
seems compatible with the PQCD predictions and the current data,
although measurements for the $B_{(s)} \rightarrow K^{*+}K^{*-}$ decays are not yet available.

\subsection{  Triple  product asymmetries }
\begin{table}[!htbh]
\footnotesize
\caption{PQCD predictions for the TPAs ($\%$)
of the considered decays in the $K\pi$ invariant mass window of 150 MeV  around the $K^*(892)$ resonance.
The sources of theoretical errors are the same as in previous tables but added in quadrature.}
\label{tab:TPAs1}
\begin{tabular}[t]{lccccc}
\hline\hline
Asymmetries & $B_s^0\rightarrow (K^+\pi^-)(K^-\pi^+)$ & $B_s^0\rightarrow (K^0\pi^+)(\bar{K}^{0}\pi^-)$
&$B^0\rightarrow (K^-\pi^+)(K^+\pi^-)$ &$B^0\rightarrow (K^0\pi^+)(\bar{K}^{0}\pi^-)$&$B^+\rightarrow (K^0\pi^+)(K^+\pi^-)$ \\ \hline
$A_T^1$                & $11.8^{+0.8}_{-1.1}$  & $9.7^{+0.5}_{-0.6}$  &$10.6^{+1.3}_{-1.7}$  & $\sim 0$  &$8.5^{+0.9}_{-0.3}$\\
$\bar{A}_T^1$          & $-11.8^{+0.8}_{-1.1}$ & $-9.9^{+0.5}_{-0.6}$ &$-10.6^{+1.3}_{-1.7}$ & $\sim 0$   &$-9.2^{+3.5}_{-0.2}$\\
$A_T^1(\text{true})$   & $0$    & $-0.1^{+0.0}_{-0.1}$ &$0$    & $\sim 0$  &$-0.2^{+0.2}_{-0.1}$\\
$A_T^1(\text{fake})$   & $11.8^{+0.8}_{-1.1}$  & $9.8^{+0.3}_{-0.5}$  &$10.6^{+1.3}_{-1.7}$  & $\sim 0$   &$8.7^{+1.5}_{-0.1}$\\ \hline
$A_T^2$                & $0.2^{+0.1}_{-0.1}$  & $0.2^{+0.1}_{-0.1}$  &$0.3^{+0.1}_{-0.1}$  & $\sim 0$ &$0.2^{+0.0}_{-0.1}$\\
$\bar{A}_T^2$          & $-0.2^{+0.1}_{-0.1}$  & $-0.2^{+0.0}_{-0.1}$ &$-0.3^{+0.1}_{-0.1}$ & $\sim 0$ &$0.3^{+0.2}_{-0.2}$\\
$A_T^2(\text{true})$   & $0$    & $0$ &$0$    & $\sim 0$ &$0.25^{+0.10}_{-0.10}$\\
$A_T^2(\text{fake})$   & $0.2^{+0.1}_{-0.1}$  & $0.2^{+0.1}_{-0.1}$ &$0.3^{+0.1}_{-0.1}$  & $\sim 0$ &$-0.05^{+0.00}_{-0.00}$\\ \hline
$A_T^3$                & $2.1^{+1.0}_{-0.6}$ & $3.3^{+0.4}_{-0.7}$ &$2.0_{-1.0}^{+0.9}$  & $\sim 0$    &$3.6^{+0.3}_{-0.7}$\\
$\bar{A}_T^3$          & $-2.1^{+1.0}_{-0.6}$& $-1.0^{+0.0}_{-0.1}$ &$-2.0_{-1.0}^{+0.9}$ & $\sim 0$   &$-0.8^{+0.1}_{-0.2}$\\
$A_T^3(\text{true})$   & $0$    & $1.2^{+0.5}_{-0.2}$  &$0$    & $\sim 0$   &$1.6^{+0.6}_{-0.1}$\\
$A_T^3(\text{fake})$   & $2.1^{+1.0}_{-0.6}$ & $2.2^{+0.7}_{-1.0}$ &$2.0_{-1.0}^{+0.9}$  & $\sim 0$   &$2.3^{+0.1}_{-0.5}$\\\hline
$A_T^4$                & $-4.6^{+2.0}_{-1.5}$  & $-5.1^{+2.0}_{-1.4}$ &$-5.1^{+2.3}_{-1.9}$ & $\sim 0$   &$-2.7^{+0.5}_{-0.5}$\\
$\bar{A}_T^4$          & $4.6^{+2.0}_{-1.5}$ & $4.1^{+1.6}_{-1.4}$&$5.1^{+2.3}_{-1.9}$  & $\sim 0$  &$1.0^{+0.5}_{-0.5}$\\
$A_T^4(\text{true})$   & $0$    & $-0.5^{+0.2}_{-0.2}$&$0$    & $\sim 0$  &$-0.9^{+0.5}_{-0.5}$\\
$A_T^4(\text{fake})$   & $-4.6^{+2.0}_{-1.5}$  & $-4.6^{+1.8}_{-1.5}$ &$-5.1^{+2.3}_{-1.9}$ & $\sim 0$   &$-1.9^{+0.3}_{-0.9}$\\
\hline\hline
\end{tabular}
\end{table}

The TPAs involve the  perpendicular polarization amplitudes $A_{\perp}$,
which are power-suppressed relative to the longitudinal ones in the naive expectation.
As stated in the previous subsection, this suppression is not numerically realized in penguin-dominated
modes, so the corresponding TPAs may be notable.
The predicted TPAs  for the  $B_{(s)}^0\rightarrow (K\pi)(K\pi)$  decays collected in Table~\ref{tab:TPAs1}
imply that the values for most of the TPAs can reach $10^{-2}$ in magnitude.
The smallness of $A_T^2$ is attributed to the suppression from the strong phase
difference  between the perpendicular and parallel polarization amplitudes, which was found to diminish in the PQCD
framework~\cite{prd91054033,prd76074018}.
Hence, observations of $A_T^2$ with large values  would signal new physics beyond
the SM. Because the decay $B^0\rightarrow (K^0\pi^+)(\bar{K}^{0}\pi^-)$ receives a contribution solely from
the weak annihilation, its tiny transverse polarization component (see Table~\ref{tab:brs})
leads to the negligible TPAs as exhibited in Table~\ref{tab:TPAs1}.

The pure penguin decays $B_{(s)}^0\rightarrow (K^+\pi^-)(K^-\pi^+)$ depend on a single weak phase
effectively, so $A_T^i(true)$ are  predicted to be zero under $CP$ conservation.
Namely, $\bar{A}_T^i= -A_T^i$ holds, and $A_T^i$ are fake asymmetries actually.
The searches for the true TPAs in the $B^0_s\rightarrow (K^+\pi^-)(K^-\pi^+)$
decay~\cite{jhep071662015} in the untagged data sample indeed show no manifest deviation from zero.
The diminishing $A_T^2(true)$ for the $B_s^0\rightarrow (K^0\pi^+)(\bar{K}^{0}\pi^-)$ mode can be also
understood through the similarity between the perpendicular and parallel polarization amplitudes.
The predicted nonvanishing fake TPAs are due to the effect of strong interactions~\cite{prd84096013},
which can be tested if flavor tagged measurements are available in the future.

\subsection{Direct $CP$ asymmetries and $S$-wave-induced direct $CP$ asymmetries}
\begin{table}[!htbh]
\caption{PQCD predictions for the direct $CP$ asymmetries ($\%$)
of the considered decays in the $K\pi$ invariant mass window of 150 MeV  around the $K^*(892)$ resonance.}
\label{tab:acp}
\begin{tabular}[t]{lccc}
\hline\hline
$\mathcal{A}_h^{\text{dir}}$ & $B_s^0\rightarrow (K^0\pi^+)(\bar{K}^{0}\pi^-)$
&$B^0\rightarrow (K^0\pi^+)(\bar{K}^{0}\pi^-)$&$B^+\rightarrow (K^0\pi^+)(K^+\pi^-)$ \\ \hline
$h=0$         & $29.9^{+4.3}_{-2.8}$  & $-29.4^{+41.8}_{-12.0}$ &$-10.7^{+11.3}_{-10.1}$  \\
$h=\parallel$ &$-18.1^{+4.6}_{-4.4}$    &$-3.3^{+2.3}_{-3.0}$    & $-10.0^{+1.7}_{-1.4}$ \\
$h=\perp$     & $-18.1^{+4.5}_{-4.0}$   &$6.5^{+8.3}_{-9.2}$   &$-2.2^{+1.9}_{-1.4}$\\
$h=SS$        & $1.6^{+12.6}_{-2.4}$     &$24.4^{+13.4}_{-8.1}$     &$-61.8^{+9.2}_{-15.4}$ \\
$h=SV$        & $-30.1^{+2.8}_{-8.4}$   &$-55.4^{+21.1}_{-29.6}$   & $-4.6^{+15.9}_{-5.4}$ \\
$h=VS$        & $0.3^{+0.1}_{-0.6}$    &$11.0^{+2.2}_{-1.4}$  &$-44.3^{+7.8}_{-13.1}$ \\
\hline\hline
\end{tabular}
\end{table}

As stated before, no direct $CP$ asymmetries are
generated in the pure penguin decays $B^0_{(s)}\rightarrow (K^+\pi^-)(K^-\pi^+)$ in the PQCD approach,
for their decay amplitudes are governed by a single product of the CKM matrix elements. The up and charm
quark penguins may contribute a weak phase difference and a direct $CP$ asymmetry,
but this effect is negligible and not taken into account here~\cite{prd72054015}.
The other three  modes receive the addition tree contributions induced by the operators $O_{1,2}$, so
direct $CP$ asymmetries arise from the interference between the tree and penguin amplitudes.
The numerical results of the direct $CP$ asymmetries $\mathcal{A}_h^{\text{dir}}$
are listed in Tables~\ref{tab:acp}, 
where we have combined the theoretical uncertainties by adding them in quadrature as in Table~\ref{tab:TPAs1}.

The factorizable emission amplitude $\mathcal{F}_e^{LL}$ dominates the $B_s^0\rightarrow (K^0\pi^+)(\bar{K}^{0}\pi^-)$
mode, which exists in both tree and penguin contributions as shown in Eq.~(\ref{eq:bspm}).
Although the  tree amplitude is CKM suppressed relative to the penguin one,
the large Wilson coefficient $C_2+C_1/3$ compensates this  suppression partly, such that
the interference between the tree and penguin contributions may induce sizable direct $CP$ asymmetries.
It is clear from Table~\ref{tab:acp} that the involved components have larger asymmetries
except $A_{SS}$ and $A_{VS}$, whose values are predicted to be about one percent.
For the $B^0$ and $B^+$ meson decays, the tree operators $O_{1,2}$ contribute only via the annihilation
amplitudes, which are power-suppressed compared with the emission ones.
The CKM suppression of the tree contribution is not effective here,
because $|V_{ub}^*V_{ud}|$ and $|V_{tb}^*V_{td}|$ are of similar order of magnitude, ie., $\mathcal{O}(\lambda^3)$.
The $B^0$ meson decay occurs only through the annihilation diagrams,
so the tree contribution, despite of being color-suppressed, is comparable with the  penguin one.
On the contrary, the tree amplitude in the $B^+$ mode is color-favored,
and the factorizable emission diagram also enhances the  penguin contribution.
Consequently, large direct $CP$ asymmetries are expected in some components of these two modes
as seen in Table~\ref{tab:acp}.

Next we compare the direct $CP$ asymmetries extracted for the two-body decays with
the previous studies based on the two-body formalism.
The  numerical outcomes are summarized in Table~\ref{tab:acpp},
which shows many differences among the predictions in various approaches.
Our value for the $B_s^0\rightarrow K^{*+}K^{*-}$ channel is greater than the previous PQCD ones,
but closer to those from the QCDF, SCET and FAT approaches.
It is found that all the predicted direct $CP$ asymmetries have the same sign,
though some of the errors are still large. The two  distinct PQCD results in Refs.~\cite{prd91054033} and
\cite{prd72054015} for the  direct $CP$ asymmetry $\mathcal{A}_{CP}^{\text{dir}}(B^0\rightarrow K^{*+}K^{*-})$
are attributed to the different models of the vector meson DAs.
Our prediction is located in between, and has a minus sign
the same as of the latter. The QCDF analysis~\cite{prd80114008} shows that
the strong phases for the helicity amplitudes with $h=0,\parallel,\perp$ are either 0 or $\pi$,
so the direct $CP$ asymmetry vanishes.
A similar observation was made in  the same framework with
the endpoint parameters being fixed by the best fit~\cite{jpg43105004}.
The time-like penguin annihilation contribution was neglected in the FAT approach~\cite{epjc77333}
due to the lack of enough data for its determination.
Because the $W$-exchange diagrams were neglected, the FAT~\cite{epjc77333} approach yielded
$\mathcal{A}_{CP}^{\text{dir}}(B^0\rightarrow K^{*+}K^{*-})=0$.
The SCET prediction for this quantity is absent, for the annihilation diagrams, counted as a
next-to-leading-power contribution~\cite{prd96073004}, were dropped.
We predict a negative $\mathcal{A}_{CP}^{\text{dir}}(B^+\rightarrow K^{*+}\bar{K}^{*0})$,
which matches those from PQCD~\cite{prd72054015} and FAT~\cite{epjc77333} in sign,
but has a lower magnitude. However, the
PQCD~\cite{prd91054033}, QCDF~\cite{prd80114008}, and SCET~\cite{prd96073004} results
have a positive sign, with the magnitudes ranging from $9.5\%$ to $23.0\%$.
These different predictions can be discriminated  by more precise measurements in the future.
It is worth stressing that the rather distinct $\mathcal{A}_{CP}^{\text{dir}}$  in our and
previous PQCD studies may suggest a stronger sensitivity of direct $CP$ asymmetries to the
widths of resonant states.


\begin{table}[!htbh]
\caption{Direct $CP$ asymmetries ($\%$) in the $B\rightarrow VV$ decays compared with previous analyses.}
\label{tab:acpp}
\begin{tabular}[t]{lccccc}
\hline\hline
Modes & This work & PQCD~\cite{prd91054033,npb93517,prd76074018} & QCDF~\cite{prd80114026,npb77464,prd80114008}
 &SCET~\cite{prd96073004} &FAT~\cite{epjc77333} \\ \hline
$B_s^0\rightarrow K^{*+}K^{*-}$  &$17.0^{+0.0}_{-2.8}$ &$8.8^{+2.6}_{-9.8}$~\cite{prd91054033}
 & $21^{+2.2}_{-4.5}$~\cite{prd80114026} & $20.6\pm23.3$ &$21.1\pm 7.1$\\
&&$9.3^{+3.3}_{-3.6}$~\cite{prd76074018}&$2^{+40}_{-15}$~\cite{npb77464}&&\\
&&$6.8^{+5.4}_{-4.3}$~\cite{npb93517} &&&\\
$B^0\rightarrow K^{*+}K^{*-}$ &$-28.8^{+39.3}_{-10.1}$ &$29.8^{+8.1}_{-12.0}$~\cite{prd91054033}
 &0~\cite{prd80114008} & $\cdots$ &0 \\
 &&-65~\cite{prd72054015}&&&\\
$B^+\rightarrow K^{*+}\bar{K}^{*0}$ &$-10.2^{+9.9}_{-9.1}$ &$23.0^{+4.7}_{-4.9}$~\cite{prd91054033}
&$16^{+17}_{-34}$~\cite{prd80114008}& $9.5\pm 10.6$ &$-24.8\pm 2.6$ \\
&&-15~\cite{prd72054015}&&&\\
\hline\hline
\end{tabular}
\end{table}

The predictions for the asymmetries from the interference with the amplitude $A_{S^+}$ defined
in Eq.~(\ref{eq:TDAs}) and for the $S$-wave-induced direct $CP$ asymmetries are presented in
Table~\ref{tab:acps}. Because the fractions $f_{S^+}$ in Table~\ref{tab:brs3} are less than $10\%$,
most values of $A_D^i$ and $\bar{A}_D^i$  are only a few percent.
Their constructive interference may result in slightly larger $CP$ asymmetries,
such that $A_S^1$ in the $B_s^0\rightarrow (K^0\pi^+)(\bar{K}^{0}\pi^-)$ decay and
$A_S^4$ in the $B^0\rightarrow (K^0\pi^+)(\bar{K}^{0}\pi^-)$ decay are over 15\% in magnitude.
It is encouraged to search for the $S$-wave-induced direct $CP$ asymmetries in these two modes.
The pure penguin decays $B_{(s)}^0\rightarrow (K^+\pi^-)(K^-\pi^+)$ have vanishing
${\cal A}_S^i$ for a reason the same as for their vanishing $A_T^i(\text{true})$ in Table~\ref{tab:TPAs1}.
The difference between $\mathcal{A}_S^1$ and $\mathcal{A}_S^3$ is ascribed to the distinct combinations
of $Re[A_{S^+}A^*_0]$ and $Re[A_{S^+}A^*_{SS}]$ in Eq.~(\ref{eq:TDAs}).
Table~\ref{tab:brs} shows that the parallel component of the pure annihilation mode
$B^0\rightarrow (K^0\pi^+)(\bar{K}^{0}\pi^-)$ is only of order $10^{-10}$,
explaining why $A_D^2$, $\bar{A}_D^2$ and $A_S^2$ are  approximately zero.
Since the $S$-wave-induced direct $CP$  asymmetries in the considered decays
still acquire less theoretical and experimental attention, we will wait for the confrontation with future data.

\begin{table}[!htbh]
\footnotesize
\caption{PQCD predictions for the asymmetries ($\%$) from the interference with the amplitude
$A_{S^+}$ and for the  $S$-wave-induced direct $CP$ asymmetries ($\%$)
of the considered decays in the $K\pi$ invariant mass window of 150 MeV  around the $K^*(892)$ resonance.}
\label{tab:acps}
\begin{tabular}[t]{lccccc}
\hline\hline
Asymmetries &
$B_s^0\rightarrow (K^+\pi^-)(K^-\pi^+)$ & $B_s^0\rightarrow (K^0\pi^+)(\bar{K}^{0}\pi^-)$
&$B^0\rightarrow (K^-\pi^+)(K^+\pi^-)$ &$B^0\rightarrow (K^0\pi^+)(\bar{K}^{0}\pi^-)$&$B^+\rightarrow (K^0\pi^+)(K^+\pi^-)$ \\ \hline
$A_D^1$ & $6.1^{+0.6}_{-1.5}$ & $-1.0^{+1.0}_{-1.6}$   &     $9.6^{+1.6}_{-3.3}$           & $5.4^{+3.0}_{-5.4}$ &$6.5^{+2.6}_{-3.3}$  \\
$\bar{A}_D^1$ & $-6.1^{+0.6}_{-1.5}$& $-15.1^{+1.5}_{-1.3}$  &  $-9.6^{+1.6}_{-3.3}$          & $-7.0^{+10.3}_{-6.7}$ &$-4.2^{+2.7}_{-5.1}$  \\
$\mathcal{A}_S^1$ &$0$ & $-16.1^{+3.0}_{-2.6}$ &  $0$    & $-1.6^{+12.7}_{-12.9}$ &$2.3^{+4.7}_{-6.6}$  \\ \hline
$A_D^2$ & $-1.9^{+1.2}_{-1.5}$ & $-2.5^{+1.0}_{-1.1}$ &      $-0.7^{+1.3}_{-1.0}$              & $\sim 0$ &$-1.3^{+0.3}_{-0.5}$  \\
$\bar{A}_D^2$ &$1.9^{+1.2}_{-1.5}$ & $3.8^{+0.9}_{-0.8}$ &      $0.7^{+1.3}_{-1.0}$          & $\sim 0$ &$-0.8^{+1.2}_{-1.0}$  \\
$\mathcal{A}_S^2$ & $0$& $1.3^{+0.5}_{-0.3}$ &   $0$     & $\sim 0$ &$-2.1^{+0.9}_{-0.8}$  \\ \hline
$A_D^3$ &$3.4^{+0.9}_{-2.0}$ & $0.8^{+0.2}_{-0.2}$ &        $1.6^{+1.7}_{-1.3}$                        & $6.6^{+4.9}_{-3.8}$ &$4.8^{+2.1}_{-1.9}$  \\
$\bar{A}_D^3$ &$-3.4^{+0.9}_{-2.0}$ & $-8.8^{+2.4}_{-2.3}$ &     $-1.6^{+1.7}_{-1.3}$            & $-4.6^{+4.1}_{-9.3}$ &$-6.5^{+4.0}_{-2.5}$  \\
$\mathcal{A}_S^3$ &$0$ & $-8.0^{+2.3}_{-2.6}$ &    $0$          & $2.0^{+7.2}_{-11.9}$ &$-1.7^{+5.9}_{-4.8}$  \\ \hline
$A_D^4$  &  $5.9^{+1.6}_{-1.0}$& $5.8^{+1.4}_{-1.3}$ &       $4.3^{+2.6}_{-2.2}$                     & $-3.1^{+2.7}_{-1.7}$ &$3.2^{+2.2}_{-0.6}$  \\
$\bar{A}_D^4$ & $-5.9^{+1.6}_{-1.0}$& $-9.4^{+1.7}_{-1.5}$ &      $-4.3^{+2.6}_{-2.2}$       & $-12.0^{+6.5}_{-7.2}$ &$1.5^{+2.0}_{-6.6}$  \\
$\mathcal{A}_S^4$ &$0$ & $-3.6^{+1.0}_{-0.8}$ &     $0$        & $-15.1^{+7.1}_{-3.9}$ &$4.7^{+2.5}_{-1.2}$  \\
\hline\hline
\end{tabular}
\end{table}

\section{ conclusion}\label{sec:sum}
In this paper the four-body decays $B_{(s)} \rightarrow  (K\pi)_{S/P} (K\pi)_{S/P}$ have been investigated
under the quasi-two-body approximation in the perturbative QCD framework.
We have focused on the physical observables derived in the $K\pi$ invariant mass window of 150 MeV
around the $K^*(892)$ resonance. The strong dynamics associated with the hadronization of the $K\pi$ pairs
was parametrized into the nonperturbative $K\pi$ two-meson DAs, which include both resonant and
nonresonant contributions. The evaluation of the hard kernels then reduces
to the one for two-body decays. The above simplified formalism, appropriate for the leading-power
regions of phase space, has been applied to three-body $B$ meson decays successfully, and
extended to four-body decays here for the first time.
Because the $K\pi$ pair can be produced in the $S$-wave configuration,
the interference between the $S$- and $P$-wave contributions stimulate rich phenomenology
in these four-body $B$ meson decays. To describe the interference accurately, we have introduced the
meson and parton fractional momenta with the final-state meson mass dependence. The relations between the
final-state meson momentum fractions in the $B$ meson rest frame and the helicity angles in the
$K\pi$ pair rest frame were given, which facilitate the discussion of various asymmetries in
angular distributions.

We have computed six helicity amplitudes allowed by angular momentum conservation, and
the corresponding $CP$-averaged branching ratios. The branching ratios of the $B_s$ modes
are of order $10^{-6}$, and those of the $B$ modes are at least lower by an order of magnitude
because of the smaller CKM matrix elements. The single and double $S$-wave contributions
were found to be substantial in the chosen invariant mass region and consistent with the LHCb data.
We have extracted the two-body $B\rightarrow K^*\bar{K}^*$ branching ratios from the results for the
corresponding four-body decays by applying the narrow-width approximation.
The obtained two-body branching ratios, except ${\cal B}(B^0\rightarrow K^{*0}\bar{K}^{*0})$,
are slightly above the previous derivations in the
two-body formalism within theoretical uncertainties, and agree with
the measured values.
The predicted hierarchy pattern for the longitudinal polarization fractions in the $B_{(s)}$ meson
decays into the $P$-wave $K\pi$ pairs is compatible with the data roughly.
However, the different polarization data between the $B_s\rightarrow K^{*0}\bar{K}^{*0}$
and $B^0\rightarrow K^{*0}\bar{K}^{*0}$ modes, which are related to each other by the $U$-spin symmetry,
remain puzzling, and call for more in-depth explorations.

We have also calculated  the TPAs, direct $CP$ asymmetries,
and  $S$-wave induced direct $CP$ asymmetries  in the $B_{(s)} \rightarrow  (K\pi)_{S/P} (K\pi)_{S/P}$ decays.
It was observed that the true TPAs in most of the considered channels are tiny, of order  $10^{-2}$ or even lower.
The magnitudes of the fake TPAs could be larger, but they do not reflect CP violation in $B$ meson decays.
The direct $CP$ asymmetries in some helicity configurations, such as those of the
$B^0\rightarrow (K^0\pi^+)(\bar{K}^{0}\pi^-)$ and
$B^+\rightarrow (K^0\pi^+)(K^+\pi^-)$  modes, were found to be sizable. The
direct $CP$ asymmetries in the two-body decays were also extracted from the results for the
corresponding four-body ones, some of which differ from the previous PQCD analyses in the two-body framework,
indicating a strong sensitivity of direct $CP$ asymmetries to the widths of resonant states.
At last, we observed significant $S$-wave-induced direct $CP$ asymmetries in the pure annihilation decay
$B^0\rightarrow (K^0\pi^+)(\bar{K}^{0}\pi^-)$.

In our numerical study the representative theoretical uncertainties from
the hadronic parameters in $B$ meson DAs, the
Gegenbauer moments of the $S$- and $P$-wave $K\pi$ two-meson DAs, and the hard scales were taken into account,
among which the hard scales  are identified as the major source. It suggests that higher-order
contributions to four-body $B$ meson decays need to be included to reduce the
sensitivity to the variation of these scales. At the same time,
the Gegenbauer moments should be further constrained to improve the precision of theoretical
predictions, and to sharpen their confrontation with future data.

\begin{acknowledgments}
We thank Profs. H.Y. Cheng, C.D. L\"{u},  W.F. Wang and Y.M. Wang for helpful discussions
on QCDF calculations of direct $CP$ asymmetries, on TPAs, on multi-body kinematics and on
the contribution of the power-suppressed $B$ meson DA, respectively.
This work is supported in part by National Natural Science Foundation
of China under Grant Nos. 12075086, 12005103, 11947013, 11605060 and the Natural Science Foundation of Hebei Province
under Grant No. A2019209449, by the Natural Science Foundation of Jiangsu Province under Grant No.~BK20190508
and the Research Start-up Funding of Nanjing Agricultural University, and
by  MOST of R.O.C. under Grant No. MOST-107-2119-M-001-035-MY3.
\end{acknowledgments}

\begin{appendix}
\section{FACTORIZATION FORMULAS}\label{sec:ampulitude}

Here we present the factorization formulas for the helicity amplitudes $A_h$ in
Eqs.~(\ref{eq:ampli})-(\ref{a2}), starting with the $P$-wave amplitudes.
The factorizable diagrams in Figs.~\ref{fig:fym}(a) and \ref{fig:fym}(b) give the piece in the longitudinal amplitude
\begin{eqnarray}
\mathcal{F}^{LL,0}_e&=& 8 \pi  M^4  C_f
\int_0^1 dx_Bdx_1\int_0^{1/\Lambda} b_Bdb_Bb_1db_1\phi _B(x_B,b_B)\nonumber\\&&
\{(\sqrt{g^- g^+} (f^-+f^+ (2 g^+ x_1-1)) \phi _P^s(x_1,\omega _1)+\sqrt{g^- g^+} (f^+ (2 g^+
   x_1-1)-f^-) \phi _P^t(x_1,\omega _1)\nonumber\\&&-(f^- g^-+f^+ g^+ (g^+ x_1+1)) \phi^0 _P(x_1,\omega _1))
\alpha_s(t_a)e^{-S_{ab}(t_a)}h(\alpha_e,\beta_a,b_B,b_1)S_t(x_1)\nonumber\\&&+
(2 \sqrt{g^- g^+} (f^- (g^--x_B)-f^+ g^+) \phi _P^s(x_1,\omega _1)-g^+ ((f^-+f^+) g^--f^-
   x_B) \phi^0 _P(x_1,\omega _1))\nonumber\\&&
\alpha_s(t_b)e^{-S_{ab}(t_b)}h(\alpha_e,\beta_b,b_1,b_B)S_t(x_B)\},
\end{eqnarray}
with the color factor $C_f=4/3$ and the QCD scale
$\Lambda=0.25\pm 0.05$ GeV~\cite{prd91054033,prd76074018}.
The effect caused by the variation of the QCD scale is minor
than caused by the variation of hard scale $t$, and has been neglected in the analysis of the
theoretical uncertainties. The dependence of our results on the QCD scale is expected to
reduced by including higher-order QCD corrections. The explicit expressions for
the hard function $h$, the threshold resummation factor $S_t$, and the Sudakov exponent $S_{ij}$
will be provided at the end of this appendix.


The nonfactorizable diagrams in Figs.~\ref{fig:fym}(c) and \ref{fig:fym}(d) yield
\begin{eqnarray}
\mathcal{M}^{LL,0}_e&=& -16 \sqrt{\frac{2}{3}} \pi  M^4 C_f
\int_0^1 dx_Bdx_1dx_2\int_0^{1/\Lambda} b_Bdb_Bb_2db_2\phi _B(x_B,b_B)\phi_P^0(x_2,\omega_2)
\nonumber\\&&\{ (-\sqrt{g^- g^+} (f^- x_B+f^+ g^+ x_1+f^+ f^- x_2) \phi _P^s(x_1,\omega
   _1)+\sqrt{g^- g^+} (f^+ g^+ x_1-f^- (x_B+f^+ (x_2-2)))\nonumber\\&& \phi _P^t(x_1,\omega _1)+(f^--f^+) (g^+
   (x_B+f^+ (x_2-1)+g^- x_1)+f^- g^-) \phi _P(x_1,\omega _1))\nonumber\\&&
\alpha_s(t_c)e^{-S_{cd}(t_c)}h(\beta_c,\alpha_e,b_2, b_B)-
(-\sqrt{g^- g^+} (f^- x_B+f^+ g^+ x_1-f^+ f^- x_2) \phi _P^s(x_1,\omega
   _1)\nonumber\\&&-\sqrt{g^- g^+} (f^- (-x_B)+f^+ g^+ x_1+f^+ f^- x_2) \phi _P^t(x_1,\omega _1)+(f^- g^-+f^+ g^+)
   (-x_B+f^+ x_2+g^+ x_1) \nonumber\\&&\phi _P(x_1,\omega _1))
\alpha_s(t_d)e^{-S_{cd}(t_d)}h(\beta_d,\alpha_e,b_2,b_B)\},
\end{eqnarray}
\begin{eqnarray}
\mathcal{M}^{LR,0}_e&=& 16 \sqrt{\frac{2}{3}} \pi  \sqrt{f^- f^+} M^4 C_f
\int_0^1 dx_Bdx_1dx_2\int_0^{1/\Lambda} b_Bdb_Bb_2db_2\phi _B(x_B,b_B)\nonumber\\&&
\{(\sqrt{g^- g^+} \phi _P^s(x_1,\omega _1) ((-x_B+f^--f^+ x_2+f^++g^+ x_1) \phi
   _P^s(x_2,\omega _2)+(x_B+f^-+f^+ (x_2-1)+g^+ x_1) \nonumber\\&&\phi _P^t(x_2,\omega _2))+\sqrt{g^- g^+} \phi
   _P^t(x_1,\omega _1) ((x_B+f^-+f^+ (x_2-1)+g^+ x_1) \phi _P^s(x_2,\omega _2)\nonumber\\&&-(-x_B+f^--f^+ x_2+f^++g^+
   x_1) \phi _P^t(x_2,\omega _2))-\phi_P(x_1,\omega _1) ((g^+ (x_B+f^+ (x_2-1)+g^- x_1)+f^-
   g^-) \nonumber\\&&\phi _P^s(x_2,\omega _2)-(g^+ (-x_B-f^+ x_2+f^++g^- x_1)+f^- g^-) \phi _P^t(x_2,\omega _2)))
\alpha_s(t_c)e^{-S_{cd}(t_c)}h(\beta_c,\alpha_e,b_2,b_B)\nonumber\\&&
-(g^+ \phi_P(x_1,\omega _1) (-(-x_B+f^+ x_2+g^- x_1) \phi _P^t(x_2,\omega
   _2)-(x_B-f^+ x_2+g^- x_1) \phi _P^s(x_2,\omega _2))\nonumber\\&&+\sqrt{g^- g^+} \phi _P^t(x_1,\omega _1) ((x_B-f^+
   x_2+g^+ x_1) \phi _P^s(x_2,\omega _2)-(x_B-f^+ x_2-g^+ x_1) \phi _P^t(x_2,\omega _2))\nonumber\\&&+\sqrt{g^- g^+} \phi
   _P^s(x_1,\omega _1) ((-x_B+f^+ x_2+g^+ x_1) \phi _P^s(x_2,\omega _2)+(x_B-f^+ x_2+g^+ x_1)\nonumber\\&& \phi
   _P^t(x_2,\omega _2)))
   \alpha_s(t_d)e^{-S_{cd}(t_d)}h(\beta_d,\alpha_e,b_2,b_B)\}.
\end{eqnarray}

The annihilation diagrams in Figs.~\ref{fig:fym}(e)-\ref{fig:fym}(h) give
\begin{eqnarray}
\mathcal{F}^{LL(LR),0}_a&=&8 \pi  M^4 f_B C_f
\int_0^1dx_1dx_2\int_0^{1/\Lambda} b_1db_1b_2db_2\nonumber\\&&
\{(2 \sqrt{f^- f^+} \sqrt{g^- g^+} \phi _P^s(x_1,\omega _1) ((f^+ x_2+g^-+g^+) \phi _P^s(x_2,\omega
   _2)+(f^+ x_2+g^--g^+) \phi _P^t(x_2,\omega _2))\nonumber\\&&-g^+ (f^- g^-+f^+ (f^+ x_2+g^-)) \phi_P(x_1,\omega
   _1) \phi_P(x_2,\omega _2))
\alpha_s(t_e)e^{-S_{ef}(t_e)}h(\alpha_a,\beta_e,b_1,b_2)S_t(x_2)\nonumber\\&&
-(-2 \sqrt{f^- f^+} \sqrt{g^- g^+} (f^--f^+-g^--g^+ x_1+g^+) \phi _P^s(x_2,\omega _2) \phi _P^t(x_1,\omega
   _1)\nonumber\\&&+2 \sqrt{f^- f^+} \sqrt{g^- g^+} (f^-+f^++g^--g^+ x_1+g^+) \phi _P^s(x_1,\omega _1) \phi _P^s(x_2,\omega _2)\nonumber\\&&-(f^+
   (g^+)^2 (-(x_1-1))+f^- g^- (f^++g^-)+f^- f^+ g^+) \phi_P(x_1,\omega _1) \phi_P(x_2,\omega
   _2))\nonumber\\&&\alpha_s(t_f)e^{-S_{ef}(t_f)}h(\alpha_a,\beta_f,b_2,b_1)S_t(x_1)\},
\end{eqnarray}
\begin{eqnarray}
\mathcal{F}^{SP,0}_a&=&-8 \pi  M^4 f_B C_f
\int_0^1dx_1dx_2\int_0^{1/\Lambda} b_1db_1b_2db_2\nonumber\\&&
\{(\sqrt{f^- f^+} g^+ \phi_P(x_1,\omega _1) (-(f^+ x_2+2 g^-) \phi _P^t(x_2,\omega _2)+f^+ x_2 \phi
   _P^s(x_2,\omega _2))\nonumber\\&&+2 \sqrt{g^- g^+} (f^- (f^+ x_2+g^-)-f^+ g^+) \phi_P(x_2,\omega _2) \phi
   _P^s(x_1,\omega _1))\alpha_s(t_e)e^{-S_{ef}(t_e)}h(\alpha_a,\beta_e,b_1,b_2)S_t(x_2)\nonumber\\&&
+(-\sqrt{g^- g^+} (f^- g^-+f^+ g^+ (x_1-1)) \phi_P(x_2,\omega _2) \phi _P^s(x_1,\omega _1)\nonumber\\&&+2
   \sqrt{f^- f^+} (f^- g^--g^+ (f^++g^- x_1)) \phi_P(x_1,\omega _1) \phi _P^s(x_2,\omega _2)\nonumber\\&&+\sqrt{g^- g^+} (f^-
   (2 f^++g^-)-f^+ g^+ (x_1-1)) \phi_P(x_2,\omega _2) \phi _P^t(x_1,\omega _1)) \nonumber\\&& \alpha_s(t_f)e^{-S_{ef}(t_f)}h(\alpha_a,\beta_f,b_2,b_1)S_t(x_1)\},
\end{eqnarray}
\begin{eqnarray}
\mathcal{M}^{LL,0}_a&=&16 \sqrt{\frac{2}{3}} \pi  M^4 C_f
 \int_0^1dx_Bdx_1dx_2\int_0^{1/\Lambda} b_1db_1b_Bdb_B\phi _B(x_B,b_B) \nonumber\\&&
\{(\sqrt{f^- f^+} \sqrt{g^- g^+} \phi _P^s(x_1,\omega _1) ((-x_B+f^+ x_2+g^--g^+ x_1+g^+) \phi
   _P^s(x_2,\omega _2)\nonumber\\&&-(-x_B+f^+ x_2+g^-+g^+ (x_1-1)) \phi _P^t(x_2,\omega _2))+\sqrt{f^- f^+} \sqrt{g^- g^+} \phi
   _P^t(x_1,\omega _1) \nonumber\\&&((-x_B+f^+ x_2+g^-+g^+ (x_1-1)) \phi _P^s(x_2,\omega _2)-(-x_B+f^+ x_2+g^--g^+
   x_1+g^+) \phi _P^t(x_2,\omega _2))\nonumber\\&&-(g^--g^+) (f^- (-x_B+f^+ x_2+g^-)+f^+ g^+ (x_1-1)) \phi
   _P(x_1,\omega _1) \phi_P(x_2,\omega _2))\nonumber\\&&\alpha_s(t_g)e^{-S_{gh}(t_g)}h(\beta_g,\alpha_a,b_B,b_1)\nonumber\\&&
-(\sqrt{f^- f^+} \sqrt{g^- g^+} \phi _P^s(x_1,\omega _1) ((x_B+f^+ x_2+g^--g^+ x_1+g^++2) \phi
   _P^s(x_2,\omega _2)\nonumber\\&&+(x_B+f^+ x_2+g^-+g^+ (x_1-1)) \phi _P^t(x_2,\omega _2))+\sqrt{f^- f^+} \sqrt{g^- g^+} \phi
   _P^t(x_1,\omega _1) \nonumber\\&&(-(x_B+f^+ x_2+g^--g^+ x_1+g^+-2) \phi _P^t(x_2,\omega _2)-(x_B+f^+ x_2+g^-+g^+
   (x_1-1)) \phi _P^s(x_2,\omega _2))\nonumber\\&&-(-f^- g^+ (x_B+g^- x_1-1)+f^+ (g^+ x_B-f^- g^+ x_2+g^-+g^+ g^-
   x_1)+(f^+)^2 g^+ x_2) \nonumber\\&&\phi_P(x_1,\omega _1) \phi_P(x_2,\omega _2))
   \alpha_s(t_h)e^{-S_{gh}(t_h)}h(\beta_h,\alpha_a,b_B,b_1)\},
\end{eqnarray}
\begin{eqnarray}
\mathcal{M}^{LR,0}_a&=&-16 \sqrt{\frac{2}{3}} \pi  M^4 C_f
 \int_0^1dx_Bdx_1dx_2\int_0^{1/\Lambda} b_1db_1b_Bdb_B\phi _B(x_B,b_B) \nonumber\\&&
\{(\sqrt{f^- f^+} g^+ \phi_P(x_1,\omega _1) ((x_B-f^+ x_2-g^- x_1) \phi _P^s(x_2,\omega
   _2)+(x_B-f^+ x_2+g^- (x_1-2)) \phi _P^t(x_2,\omega _2))\nonumber\\&&+\sqrt{g^- g^+} (f^- (-x_B+f^+ x_2+g^-)+f^+
   g^+ (x_1-1)) \phi_P(x_2,\omega _2) \phi _P^s(x_1,\omega _1)\nonumber\\&&+\sqrt{g^- g^+} (f^- (-x_B+f^+ x_2+g^-)-f^+ g^+
   (x_1-1)) \phi_P(x_2,\omega _2) \phi _P^t(x_1,\omega _1))\alpha_s(t_g)e^{-S_{gh}(t_g)}h(\beta_g,\alpha_a,b_B,b_1)\nonumber\\&&
+(\sqrt{f^- f^+} \phi_P(x_1,\omega _1) ((g^+ (x_B+f^+ x_2-2)+g^- (g^+
   x_1+2)) \phi _P^s(x_2,\omega _2)\nonumber\\&&-(g^- (g^+ (x_1-2)+2)-g^+ (x_B+f^+ x_2-2)) \phi
   _P^t(x_2,\omega _2))\nonumber\\&&-\sqrt{g^- g^+} (f^- (x_B+f^+ x_2+g^--2)+f^+ (g^+ (x_1-1)+2)) \phi
   _P(x_2,\omega _2) \phi _P^s(x_1,\omega _1)\nonumber\\&&-\sqrt{g^- g^+} (f^- (x_B+f^+ x_2+g^--2)+f^+ (g^+
   (-x_1)+g^+-2)) \phi_P(x_2,\omega _2) \phi _P^t(x_1,\omega _1))\nonumber\\&&\alpha_s(t_h)e^{-S_{gh}(t_h)}h(\beta_h,\alpha_a,b_B,b_1)\},
\end{eqnarray}
\begin{eqnarray}
\mathcal{M}^{SP,0}_a&=&-8 \sqrt{\frac{2}{3}} \pi  M^4 C_f
 \int_0^1dx_Bdx_1dx_2\int_0^{1/\Lambda} b_1db_1b_Bdb_B\phi _B(x_B,b_B) \nonumber\\&&
\{ (\sqrt{f^- f^+} \sqrt{g^- g^+} \phi _P^s(x_1,\omega _1) ((-x_B+f^+ x_2+g^--g^+ x_1+g^+) \phi
   _P^s(x_2,\omega _2)\nonumber\\&&-(x_B-f^+ x_2-g^--g^+ x_1+g^+) \phi _P^t(x_2,\omega _2))+\sqrt{f^- f^+} \sqrt{g^- g^+} \phi
   _P^t(x_1,\omega _1) \nonumber\\&&((x_B-f^+ x_2-g^--g^+ x_1+g^+) \phi _P^s(x_2,\omega _2)-(-x_B+f^+ x_2+g^--g^+ x_1+g^+) \phi
   _P^t(x_2,\omega _2))\nonumber\\&&+(f^--f^+) g^+ (-x_B+f^+ x_2+g^- x_1) \phi_P(x_1,\omega _1) \phi_P(x_2,\omega
   _2))\nonumber\\&&
   \alpha_s(t_g)e^{-S_{gh}(t_g)}h(\beta_g,\alpha_a,b_B,b_1)\nonumber\\&&
-(\sqrt{f^- f^+} \sqrt{g^- g^+} \phi _P^s(x_1,\omega _1) ((x_B+f^+ x_2+g^--g^+ x_1+g^++2) \phi
   _P^s(x_2,\omega _2)\nonumber\\&&-(x_B+f^+ x_2+g^-+g^+ (x_1-1)) \phi _P^t(x_2,\omega _2))+\sqrt{f^- f^+} \sqrt{g^- g^+} \phi
   _P^t(x_1,\omega _1) \nonumber\\&&((x_B+f^+ x_2+g^-+g^+ (x_1-1)) \phi _P^s(x_2,\omega _2)-(x_B+f^+ x_2+g^--g^+
   x_1+g^+-2) \phi _P^t(x_2,\omega _2))\nonumber\\&&-(f^- ((g^--g^+) (x_B+g^-)+g^+)+f^+ (f^- g^- x_2+g^+
   (f^- (-x_2)-g^+ x_1+g^+)+g^-+g^+ g^- (x_1-1)))\nonumber\\&& \phi_P(x_1,\omega _1) \phi_P(x_2,\omega
   _2))\alpha_s(t_h)e^{-S_{gh}(t_h)}h(\beta_h,\alpha_a,b_B,b_1)\}.
\end{eqnarray}
$\mathcal{F'}_a(\mathcal{M'}_a)$ is obtained from $\mathcal{F}_a(\mathcal{M}_a)$ through the exchanges
\begin{eqnarray}
x_1\leftrightarrow x_2,
\quad \omega_1\leftrightarrow \omega_2.
\end{eqnarray}

The factorization formulas for the parallel and perpendicular polarization amplitudes are collected below:
\begin{eqnarray}
\mathcal{F}^{LL,\parallel}_e&=& 8 \pi  M^2 \omega _2 C_f
\int_0^1 dx_Bdx_1\int_0^{1/\Lambda} b_Bdb_Bb_1db_1\phi _B(x_B,b_B)\nonumber\\&&
\{[ (g^+ x_1 \omega _1 \phi _1^a (x_1,\omega _1 )-\omega _1  (g^+ x_1+2 ) \phi _1^v (x_1,\omega _1 )-M
    (-g^-+g^++2 g^+ g^- x_1 ) \phi _1^T (x_1,\omega _1 ) )]\nonumber\\&&
\alpha_s(t_a)e^{-S_{ab}(t_a)}h(\alpha_e,\beta_a,b_B,b_1)S_t(x_1)+
[ ( (-x_B+g^--g^+ ) \phi _1^a (x_1,\omega _1 )\nonumber\\&&- (-x_B+g^-+g^+ ) \phi _1^v (x_1,\omega
   _1 ) )]
\alpha_s(t_b)e^{-S_{ab}(t_b)}h(\alpha_e,\beta_b,b_1,b_B)S_t(x_B)\},
\end{eqnarray}
\begin{eqnarray}
\mathcal{F}^{LL,\perp}_e&=& -8 \pi  M^2 \omega_2 C_f
\int_0^1 dx_Bdx_1\int_0^{1/\Lambda} b_Bdb_Bb_1db_1\phi _B(x_B,b_B)\nonumber\\&&
\{[ (\omega _1  (g^+ x_1+2 ) \phi _1^a (x_1,\omega _1 )-g^+ x_1 \omega _1 \phi _1^v (x_1,\omega _1 )+M
    (g^-+g^+-2 g^+ g^- x_1 ) \phi _1^T (x_1,\omega _1 ) )]\nonumber\\&&
\alpha_s(t_a)e^{-S_{ab}(t_a)}h(\alpha_e,\beta_a,b_B,b_1)S_t(x_1)+
[ ( (-x_B+g^-+g^+ ) \phi _1^a (x_1,\omega _1 )\nonumber\\&&+ (x_B-g^-+g^+ ) \phi _1^v (x_1,\omega
   _1 ) )]
\alpha_s(t_b)e^{-S_{ab}(t_b)}h(\alpha_e,\beta_b,b_1,b_B)S_t(x_B)\},
\end{eqnarray}
\begin{eqnarray}
\mathcal{M}^{LL,\parallel}_e&=& 16 \sqrt{\frac{2}{3}} \pi  M^3 \omega _2 C_f
\int_0^1 dx_Bdx_1dx_2\int_0^{1/\Lambda} b_Bdb_Bb_2db_2\phi _B(x_B,b_B)
\nonumber\\&&\{[\phi _1^T (x_1,\omega _1 )  ( (g^+  (x_B+f^+  (x_2-1 )+g^- x_1 )+f^- g^- ) \phi
   _2^v (x_2,\omega _2 )- \nonumber\\&&(g^+  (-x_B-f^+ x_2+f^++g^- x_1 )+f^- g^- ) \phi _2^a (x_2,\omega _2 ) )]
\alpha_s(t_c)e^{-S_{cd}(t_c)}h(\beta_c,\alpha_e,b_2, b_B)+\nonumber\\&&
[ (2 \omega _1  (-x_B+f^+ x_2+g^+ x_1 ) \phi _1^a (x_1,\omega _1 ) \phi _2^a (x_2,\omega
   _2 )\nonumber\\&&+g^+ M \phi _1^T (x_1,\omega _1 )  ( (x_B-f^+ x_2-g^- x_1 ) \phi _2^a (x_2,\omega _2 )+ (x_B-f^+ x_2+g^- x_1 ) \phi
   _2^v (x_2,\omega _2 ) )\nonumber\\&&+2 \omega _1  (-x_B+f^+ x_2+g^+ x_1 ) \phi _1^v (x_1,\omega _1 ) \phi _2^v (x_2,\omega _2 ) )]
\alpha_s(t_d)e^{-S_{cd}(t_d)}h(\beta_d,\alpha_e,b_2,b_B)\},
\end{eqnarray}
\begin{eqnarray}
\mathcal{M}^{LL,\perp}_e&=& -16 \sqrt{\frac{2}{3}} \pi  M^3 \omega _2 C_f
\int_0^1 dx_Bdx_1dx_2\int_0^{1/\Lambda} b_Bdb_Bb_2db_2\phi _B(x_B,b_B)
\nonumber\\&&\{[\phi _1^T (x_1,\omega _1 )  ( (g^+  (-x_B-f^+ x_2+f^++g^- x_1 )+f^- g^- ) \phi
   _2^v (x_2,\omega _2 )- \nonumber\\&&(g^+  (x_B+f^+  (x_2-1 )+g^- x_1 )+f^- g^- ) \phi _2^a (x_2,\omega _2 ) )]
\alpha_s(t_c)e^{-S_{cd}(t_c)}h(\beta_c,\alpha_e,b_2, b_B)-\nonumber\\&&
[ (2 \omega _1  (-x_B+f^+ x_2+g^+ x_1 ) \phi _2^a (x_2,\omega _2 ) \phi _1^v (x_1,\omega
   _1 )+2 \omega _1  (-x_B+f^+ x_2+g^+ x_1 ) \phi _1^a (x_1,\omega _1 ) \phi _2^v (x_2,\omega _2 )\nonumber\\&&+g^+ M \phi _1^T (x_1,\omega
   _1 )  ( (x_B-f^+ x_2+g^- x_1 ) \phi _2^a (x_2,\omega _2 )+ (x_B-f^+ x_2-g^- x_1 ) \phi _2^v (x_2,\omega
   _2 ) ) )]\nonumber\\&&
\alpha_s(t_d)e^{-S_{cd}(t_d)}h(\beta_d,\alpha_e,b_2,b_B)\},
\end{eqnarray}
\begin{eqnarray}
\mathcal{M}^{LR,\parallel}_e&=& 16 \sqrt{\frac{2}{3}} \pi  M^3 C_f
\int_0^1 dx_Bdx_1dx_2\int_0^{1/\Lambda} b_Bdb_Bb_2db_2\phi _B(x_B,b_B)\nonumber\\&&
\{[ \phi _2^T (x_2,\omega _2 )  (\omega _1  (f^-  (x_B+f^+  (x_2-2 ) )-f^+ g^+ x_1 ) \phi
   _1^a (x_1,\omega _1 )\nonumber\\&&+\omega _1  (f^- x_B+f^+ g^+ x_1+f^+ f^- x_2 ) \phi _1^v (x_1,\omega _1 )+M  (f^- g^+  (-x_B )+f^+ g^-
   g^+ x_1+f^- f^+  (g^--g^+ x_2+g^+ ) ) \nonumber\\&&\phi _1^T (x_1,\omega _1 ) )]
\alpha_s(t_c)e^{-S_{cd}(t_c)}h(\beta_c,\alpha_e,b_2,b_B)+
  [\phi _2^T (x_2,\omega _2 )  (-\omega _1  (f^-  (-x_B )+f^+ g^+ x_1+f^+ f^- x_2 ) \phi
   _1^a (x_1,\omega _1 )\nonumber\\&&+\omega _1  (f^- x_B+f^+ g^+ x_1-f^+ f^- x_2 ) \phi _1^v (x_1,\omega _1 )+g^+ M  (f^-  (-x_B )+f^+ g^-
   x_1+f^+ f^- x_2 ) \phi _1^T (x_1,\omega _1 ) )]\nonumber\\&& \alpha_s(t_d)e^{-S_{cd}(t_d)}h(\beta_d,\alpha_e,b_2,b_B)\},
\end{eqnarray}
\begin{eqnarray}
\mathcal{M}^{LR,\perp}_e&=& 16 \sqrt{\frac{2}{3}} \pi  M^3 C_f
\int_0^1 dx_Bdx_1dx_2\int_0^{1/\Lambda} b_Bdb_Bb_2db_2\phi _B(x_B,b_B)\nonumber\\&&
\{-[ \phi _2^T (x_2,\omega _2 )  (\omega _1  (f^- x_B+f^+ g^+ x_1+f^+ f^- x_2 ) \phi _1^a (x_1,\omega
   _1 )\nonumber\\&&+\omega _1  (f^-  (x_B+f^+  (x_2-2 ) )-f^+ g^+ x_1 ) \phi _1^v (x_1,\omega _1 )-M  (f^- g^+ x_B+f^+ g^- g^+ x_1+f^-
   f^+  (g^-+g^+  (x_2-1 ) ) ) \nonumber\\&&\phi _1^T (x_1,\omega _1 ) )]
\alpha_s(t_c)e^{-S_{cd}(t_c)}h(\beta_c,\alpha_e,b_2,b_B)+
  [\phi _2^T (x_2,\omega _2 )  (-\omega _1  (f^- x_B+f^+ g^+ x_1-f^+ f^- x_2 ) \phi _1^a (x_1,\omega
   _1 )\nonumber\\&&+\omega _1  (f^-  (-x_B )+f^+ g^+ x_1+f^+ f^- x_2 ) \phi _1^v (x_1,\omega _1 )+g^+ M  (f^- x_B+f^+ g^- x_1-f^+ f^-
   x_2 ) \phi _1^T (x_1,\omega _1 ) )]\nonumber\\&& \alpha_s(t_d)e^{-S_{cd}(t_d)}h(\beta_d,\alpha_e,b_2,b_B)\},
\end{eqnarray}
\begin{eqnarray}
\mathcal{F}^{LL(LR),\parallel}_a&=& 8 \pi  M^2 \omega _1 \omega _2 f_B C_f
\int_0^1dx_1dx_2\int_0^{1/\Lambda} b_1db_1b_2db_2\nonumber\\&&
\{[ (-\phi _1^a (x_1,\omega _1 )  ( (f^+ x_2+g^-+g^+ ) \phi _2^a (x_2,\omega _2 )+ (f^+
   x_2+g^--g^+ ) \phi _2^v (x_2,\omega _2 ) )-\nonumber\\&&\phi _1^v (x_1,\omega _1 )  ( (f^+ x_2+g^--g^+ ) \phi _2^a (x_2,\omega
   _2 )+ (f^+ x_2+g^-+g^+ ) \phi _2^v (x_2,\omega _2 ) ) )]\nonumber\\&&
\alpha_s(t_e)e^{-S_{ef}(t_e)}h(\alpha_a,\beta_e,b_1,b_2)S_t(x_2)
\nonumber\\&&+[ (\phi _1^a (x_1,\omega _1 )  ( (f^-+f^++g^--g^+ x_1+g^+ ) \phi _2^a (x_2,\omega
   _2 )+ (-f^-+f^++g^-+g^+  (x_1-1 ) ) \phi _2^v (x_2,\omega _2 ) )\nonumber\\&&+\phi _1^v (x_1,\omega _1 )
    ( (-f^-+f^++g^-+g^+  (x_1-1 ) ) \phi _2^a (x_2,\omega _2 )+ (f^-+f^++g^--g^+ x_1+g^+ ) \phi _2^v (x_2,\omega
   _2 ) ) )]\nonumber\\&&
\alpha_s(t_f)e^{-S_{ef}(t_f)}h(\alpha_a,\beta_f,b_2,b_1)S_t(x_1)\},
\end{eqnarray}
\begin{eqnarray}
-\mathcal{F}^{LR,\perp}_a=\mathcal{F}^{LL,\perp}_a&=&-8 \pi  M^2 \omega _1 \omega _2 f_B C_f
\int_0^1dx_1dx_2\int_0^{1/\Lambda} b_1db_1b_2db_2\nonumber\\&&
\{[ (\phi _1^v (x_1,\omega _1 )  ( (f^+ x_2+g^-+g^+ ) \phi _2^a (x_2,\omega _2 )+ (f^+
   x_2+g^--g^+ ) \phi _2^v (x_2,\omega _2 ) )\nonumber\\&&+\phi _1^a (x_1,\omega _1 )  ( (f^+ x_2+g^--g^+ ) \phi _2^a (x_2,\omega
   _2 )+ (f^+ x_2+g^-+g^+ ) \phi _2^v (x_2,\omega _2 ) ) )]\nonumber\\&&
\alpha_s(t_e)e^{-S_{ef}(t_e)}h(\alpha_a,\beta_e,b_1,b_2)S_t(x_2)
\nonumber\\&&+[ (\phi _2^a (x_2,\omega _2 )  ( (f^--f^+-g^--g^+ x_1+g^+ ) \phi _1^a (x_1,\omega
   _1 )- (f^-+f^++g^--g^+ x_1+g^+ ) \nonumber\\&&\phi _1^v (x_1,\omega _1 ) )-\phi _2^v (x_2,\omega _2 )  ( (f^-+f^++g^--g^+
   x_1+g^+ ) \phi _1^a (x_1,\omega _1 )+ \nonumber\\&&(-f^-+f^++g^-+g^+  (x_1-1 ) ) \phi _1^v (x_1,\omega _1 ) ) )]
\alpha_s(t_f)e^{-S_{ef}(t_f)}h(\alpha_a,\beta_f,b_2,b_1)S_t(x_1)\},
\end{eqnarray}
\begin{eqnarray}
\mathcal{F}^{SP,\parallel}_a&=& -16 \pi  M^3 f_B C_f
\int_0^1dx_1dx_2\int_0^{1/\Lambda} b_1db_1b_2db_2\nonumber\\&&
\{ \omega _1[ \phi _2^T (x_2,\omega _2 )  ( (f^+ g^+-f^-  (f^+ x_2+g^- ) ) \phi _1^v (x_1,\omega _1 )- (f^+
   g^++f^-  (f^+ x_2+g^- ) ) \phi _1^a (x_1,\omega _1 ) )]\nonumber\\&&
\alpha_s(t_e)e^{-S_{ef}(t_e)}h(\alpha_a,\beta_e,b_1,b_2)S_t(x_2)\nonumber\\&&
+\omega _2[\phi _1^T (x_1,\omega _1 )  ( (g^-  (f^--g^+ x_1+2 g^+ )+f^+ g^+ ) \phi _2^a (x_2,\omega
   _2 )+ (-f^- g^-+f^+ g^++g^+ g^- x_1 ) \phi _2^v (x_2,\omega _2 ) )]\nonumber\\&&
   \alpha_s(t_f)e^{-S_{ef}(t_f)}h(\alpha_a,\beta_f,b_2,b_1)S_t(x_1)\},
\end{eqnarray}
\begin{eqnarray}
\mathcal{F}^{SP,\perp}_a&=& -16 \pi  M^3 f_B C_f
\int_0^1dx_1dx_2\int_0^{1/\Lambda} b_1db_1b_2db_2\nonumber\\&&
\{\omega _1[ \phi _2^T (x_2,\omega _2 )  ( (f^- g^--f^+ g^++f^- f^+ x_2 ) \phi _1^a (x_1,\omega _1 )+ (f^- g^-+f^+
   g^++f^- f^+ x_2 ) \phi _1^v (x_1,\omega _1 ) )]\nonumber\\&&
\alpha_s(t_e)e^{-S_{ef}(t_e)}h(\alpha_a,\beta_e,b_1,b_2)S_t(x_2)\nonumber\\&&
-\omega _2[\phi _1^T (x_1,\omega _1 )  ( (f^- g^--g^+  (f^++g^- x_1 ) ) \phi _2^a (x_2,\omega _2 )- (f^-
   g^-+g^+  (f^++g^-  (-x_1 )+2 g^- ) ) \phi _2^v (x_2,\omega _2 ) )]\nonumber\\&&
   \alpha_s(t_f)e^{-S_{ef}(t_f)}h(\alpha_a,\beta_f,b_2,b_1)S_t(x_1)\},
\end{eqnarray}
\begin{eqnarray}
\mathcal{M}^{LL,\parallel}_a&=&16 \sqrt{\frac{2}{3}} \pi  M^2 C_f
 \int_0^1dx_Bdx_1dx_2\int_0^{1/\Lambda} b_1db_1b_Bdb_B\phi _B(x_B,b_B) \nonumber\\&&
\{g^+ M^2[ (f^-  (-x_B+f^+ x_2+g^- )-f^+ g^-  (x_1-1 ) ) \phi _1^T (x_1,\omega _1 ) \phi
   _2^T (x_2,\omega _2 )] \alpha_s(t_g)e^{-S_{gh}(t_g)}h(\beta_g,\alpha_a,b_B,b_1)\nonumber\\&&
   +[ (2 \omega _1 \omega _2 \phi _1^a (x_1,\omega _1 ) \phi _2^a (x_2,\omega _2 )+M^2  (f^+  (-f^- g^+
   x_2+g^-+g^+ g^-  (x_1-1 ) )-f^- g^+  (x_B+g^--1 ) )\nonumber\\&& \phi _1^T (x_1,\omega _1 ) \phi _2^T (x_2,\omega _2 )+2 \omega _1
   \omega _2 \phi _1^v (x_1,\omega _1 ) \phi _2^v (x_2,\omega _2 ) )]
   \alpha_s(t_h)e^{-S_{gh}(t_h)}h(\beta_h,\alpha_a,b_B,b_1)\},
\end{eqnarray}
\begin{eqnarray}
\mathcal{M}^{LL,\perp}_a&=&16 \sqrt{\frac{2}{3}} \pi  M^2 C_f
 \int_0^1dx_Bdx_1dx_2\int_0^{1/\Lambda} b_1db_1b_Bdb_B\phi _B(x_B,b_B) \nonumber\\&&
\{-[ (f^-  (-x_B+f^+ x_2+g^- )+f^+ g^-  (x_1-1 ) ) \phi _1^T (x_1,\omega _1 ) \phi
   _2^T (x_2,\omega _2 )] \alpha_s(t_g)e^{-S_{gh}(t_g)}h(\beta_g,\alpha_a,b_B,b_1)\nonumber\\&&
   +[ (2 \omega _1 \omega _2  (\phi _2^a (x_2,\omega _2 ) \phi _1^v (x_1,\omega _1 )+\phi _1^a (x_1,\omega
   _1 ) \phi _2^v (x_2,\omega _2 ) )+M^2  (f^- g^+  (x_B+g^--1 )+f^+ \nonumber\\&& (f^- g^+ x_2+g^-+g^+ g^-  (x_1-1 ) ) )
   \phi _1^T (x_1,\omega _1 ) \phi _2^T (x_2,\omega _2 ) )]
   \alpha_s(t_h)e^{-S_{gh}(t_h)}h(\beta_h,\alpha_a,b_B,b_1)\},
\end{eqnarray}
\begin{eqnarray}
\mathcal{M}^{LR,\parallel}_a&=&16 \sqrt{\frac{2}{3}} \pi  M^3  C_f
 \int_0^1dx_Bdx_1dx_2\int_0^{1/\Lambda} b_1db_1b_Bdb_B\phi _B(x_B,b_B) \nonumber\\&&
[  (g^+ \omega _2 \phi _1^T (x_1,\omega _1 )  ( (x_B-f^+ x_2+g^-  (x_1-2 ) ) \phi
   _2^a (x_2,\omega _2 )+ (x_B-f^+ x_2-g^- x_1 ) \phi _2^v (x_2,\omega _2 ) )\nonumber\\&&-\omega _1  (f^-  (-x_B+f^+ x_2+g^- )-f^+ g^+
    (x_1-1 ) ) \phi _1^a (x_1,\omega _1 ) \phi _2^T (x_2,\omega _2 )\nonumber\\&&-\omega _1  (f^-  (-x_B+f^+ x_2+g^- )+f^+ g^+
    (x_1-1 ) ) \phi _2^T (x_2,\omega _2 ) \phi _1^v (x_1,\omega _1 ) )]
\alpha_s(t_g)e^{-S_{gh}(t_g)}h(\beta_g,\alpha_a,b_B,b_1)\nonumber\\&&
+[ (\omega _2 \phi _1^T (x_1,\omega _1 )  ( (g^+  (x_B+f^+ x_2-2 )-g^-  (g^+
    (x_1-2 )+2 ) ) \phi _2^a (x_2,\omega _2 )\nonumber\\&&+ (g^+  (x_B+f^+ x_2-2 )+g^-  (g^+ x_1+2 ) ) \phi
   _2^v (x_2,\omega _2 ) )+\omega _1  (f^-  (x_B+f^+ x_2+g^--2 )+f^+  (g^+  (-x_1 )+g^+-2 ) )\nonumber\\&& \phi
   _1^a (x_1,\omega _1 ) \phi _2^T (x_2,\omega _2 )+\omega _1  (f^-  (x_B+f^+ x_2+g^--2 )+f^+  (g^+
    (x_1-1 )+2 ) ) \phi _2^T (x_2,\omega _2 ) \phi _1^v (x_1,\omega _1 ) )]\nonumber\\&&
\alpha_s(t_h)e^{-S_{gh}(t_h)}h(\beta_h,\alpha_a,b_B,b_1),
\end{eqnarray}
\begin{eqnarray}
\mathcal{M}^{LR,\perp}_a&=&16 \sqrt{\frac{2}{3}} \pi  M^3  C_f
 \int_0^1dx_Bdx_1dx_2\int_0^{1/\Lambda} b_1db_1b_Bdb_B\phi _B(x_B,b_B) \nonumber\\&&
[  (g^+ \omega _2 \phi _1^T (x_1,\omega _1 )  ( (x_B-f^+ x_2-g^- x_1 ) \phi _2^a (x_2,\omega
   _2 )+ (x_B-f^+ x_2+g^-  (x_1-2 ) ) \phi _2^v (x_2,\omega _2 ) )\nonumber\\&&+\omega _1  (f^-  (-x_B+f^+ x_2+g^- )+f^+ g^+
    (x_1-1 ) ) \phi _1^a (x_1,\omega _1 ) \phi _2^T (x_2,\omega _2 )\nonumber\\&&+\omega _1  (f^-  (-x_B+f^+ x_2+g^- )-f^+ g^+
    (x_1-1 ) ) \phi _2^T (x_2,\omega _2 ) \phi _1^v (x_1,\omega _1 ) )]
\alpha_s(t_g)e^{-S_{gh}(t_g)}h(\beta_g,\alpha_a,b_B,b_1)\nonumber\\&&
+[ (\omega _2 \phi _1^T (x_1,\omega _1 )  ( (g^+  (x_B+f^+ x_2-2 )+g^-  (g^+ x_1+2 ) )
   \phi _2^a (x_2,\omega _2 )+ (g^+  (x_B+f^+ x_2-2 )\nonumber\\&&-g^-  (g^+  (x_1-2 )+2 ) ) \phi _2^v (x_2,\omega
   _2 ) )-\omega _1  (f^-  (x_B+f^+ x_2+g^--2 )+f^+  (g^+  (x_1-1 )+2 ) )\nonumber\\&& \phi _1^a (x_1,\omega _1 ) \phi
   _2^T (x_2,\omega _2 )-\omega _1  (f^-  (x_B+f^+ x_2+g^--2 )+f^+  (g^+  (-x_1 )+g^+-2 ) ) \phi _2^T (x_2,\omega
   _2 ) \phi _1^v (x_1,\omega _1 ) )]\nonumber\\&&
\alpha_s(t_h)e^{-S_{gh}(t_h)}h(\beta_h,\alpha_a,b_B,b_1),
\end{eqnarray}
\begin{eqnarray}
\mathcal{M}^{SP,\parallel}_a&=&-16 \sqrt{\frac{2}{3}} \pi  M^2  C_f
 \int_0^1dx_Bdx_1dx_2\int_0^{1/\Lambda} b_1db_1b_Bdb_B\phi _B(x_B,b_B) \nonumber\\&&
[g^+ M^2  (f^+ g^-  (x_1-1 )-f^-  (-x_B+f^+ x_2+g^- ) ) \phi _1^T (x_1,\omega _1 ) \phi
   _2^T (x_2,\omega _2 )]\alpha_s(t_g)e^{-S_{gh}(t_g)}h(\beta_g,\alpha_a,b_B,b_1)\nonumber\\&&
   -[ (2 \omega _1 \omega _2 \phi _1^a (x_1,\omega _1 ) \phi _2^a (x_2,\omega _2 )+M^2  (f^+  (-f^- g^+
   x_2+g^-+g^+ g^-  (x_1-1 ) )-f^- g^+  (x_B+g^--1 ) )\nonumber\\&& \phi _1^T (x_1,\omega _1 ) \phi _2^T (x_2,\omega _2 )+2 \omega _1
   \omega _2 \phi _1^v (x_1,\omega _1 ) \phi _2^v (x_2,\omega _2 ) )]
\alpha_s(t_h)e^{-S_{gh}(t_h)}h(\beta_h,\alpha_a,b_B,b_1),
\end{eqnarray}
\begin{eqnarray}
\mathcal{M}^{SP,\perp}_a&=&-16 \sqrt{\frac{2}{3}} \pi  M^2  C_f
 \int_0^1dx_Bdx_1dx_2\int_0^{1/\Lambda} b_1db_1b_Bdb_B\phi _B(x_B,b_B) \nonumber\\&&
[-g^+ M^2  (f^-  (-x_B+f^+ x_2+g^- )+f^+ g^-  (x_1-1 ) ) \phi _1^T (x_1,\omega _1 ) \phi
   _2^T (x_2,\omega _2 )]\alpha_s(t_g)e^{-S_{gh}(t_g)}h(\beta_g,\alpha_a,b_B,b_1)\nonumber\\&&
   +[ (2 \omega _1 \omega _2  (\phi _2^a (x_2,\omega _2 ) \phi _1^v (x_1,\omega _1 )+\phi _1^a (x_1,\omega
   _1 ) \phi _2^v (x_2,\omega _2 ) )\nonumber\\&&+M^2  (f^- g^+  (x_B+g^--1 )+f^+  (f^- g^+ x_2+g^-+g^+ g^-  (x_1-1 ) ) )
   \phi _1^T (x_1,\omega _1 ) \phi _2^T (x_2,\omega _2 ) )]\nonumber\\&&
\alpha_s(t_h)e^{-S_{gh}(t_h)}h(\beta_h,\alpha_a,b_B,b_1).
\end{eqnarray}
$\mathcal{F'}_a(\mathcal{M'}_a)$ is obtained from $\mathcal{F}_a(\mathcal{M}_a)$ through the exchanges
$x_1\leftrightarrow x_2$ and $\omega_1\leftrightarrow \omega_2$.

The single $S$-wave amplitudes receive only the longitudinal polarization contribution from
the $P$-wave $K\pi$ pair. Because of the similar Lorentz structures between the $S$-wave and
$P$-wave two-meson DAs, the factorization formulas for the single $S$-wave amplitudes are similar
to those for the $P$-wave longitudinal amplitudes with some terms flipping signs, which result
from the negative components of the longitudinal polarization vectors in Eq.(\ref{eq:pq1}).
Their explicit expressions are written as
\begin{eqnarray}
\mathcal{F}^{LL,SV}_e&=& 8 \pi  M^4  C_f
\int_0^1 dx_Bdx_1\int_0^{1/\Lambda} b_Bdb_Bb_1db_1\phi _B(x_B,b_B)\nonumber\\&&
\{(\sqrt{g^- g^+} (f^-+f^+ (2 g^+ x_1-1)) \phi _S^s(x_1,\omega _1)+\sqrt{g^- g^+} (f^+ (2 g^+
   x_1-1)-f^-) \phi _S^t(x_1,\omega _1)\nonumber\\&&+(f^- g^--f^+ g^+ (g^+ x_1+1)) \phi^0 _S(x_1,\omega _1))
\alpha_s(t_a)e^{-S_{ab}(t_a)}h(\alpha_e,\beta_a,b_B,b_1)S_t(x_1)\nonumber\\&&+
(2 \sqrt{g^- g^+} (f^- (g^--x_B)-f^+ g^+) \phi _S^s(x_1,\omega _1)+g^+ ((-f^-+f^+) g^-+f^-
   x_B) \phi^0 _S(x_1,\omega _1))\nonumber\\&&
\alpha_s(t_b)e^{-S_{ab}(t_b)}h(\alpha_e,\beta_b,b_1,b_B)S_t(x_B)\},
\end{eqnarray}
\begin{eqnarray}
\mathcal{M}^{LL,SV}_e&=& 16 \sqrt{\frac{2}{3}} \pi  M^4 C_f
\int_0^1 dx_Bdx_1dx_2\int_0^{1/\Lambda} b_Bdb_Bb_2db_2\phi _B(x_B,b_B)\phi_P^0(x_2,\omega_2)
\nonumber\\&&\{ (\sqrt{g^-g^+} (f^- x_B+f^+ g^+ x_1+f^+ f^- x_2) \phi _S^s(x_1,\omega_1)+\sqrt{g^- g^+} (f^- (x_B+f^+ (x_2-2))-f^+ g^+
   x_1) \nonumber\\&&\phi _S^t(x_1,\omega _1)+(f^--f^+) (g^+
   (-x_B-f^+ x_2+f^++g^- x_1)+f^- g^-) \phi _S(x_1,\omega
   _1))\nonumber\\&&
\alpha_s(t_c)e^{-S_{cd}(t_c)}h(\beta_c,\alpha_e,b_2, b_B)-
(\sqrt{g^-
   g^+} (f^- x_B+f^+ g^+ x_1-f^+ f^- x_2) \phi _S^s(x_1,\omega
   _1)\nonumber\\&&+\sqrt{g^- g^+} (f^- (-x_B)+f^+ g^+ x_1+f^+ f^- x_2)
   \phi _S^t(x_1,\omega _1)\nonumber\\&&+(f^- g^--f^+ g^+) (-x_B+f^+
   x_2+g^+ x_1) \phi _S(x_1,\omega _1))
\alpha_s(t_d)e^{-S_{cd}(t_d)}h(\beta_d,\alpha_e,b_2,b_B)\},
\end{eqnarray}
\begin{eqnarray}
\mathcal{M}^{LR,SV}_e&=& 16 \sqrt{\frac{2}{3}} \pi  \sqrt{f^- f^+} M^4 C_f
\int_0^1 dx_Bdx_1dx_2\int_0^{1/\Lambda} b_Bdb_Bb_2db_2\phi _B(x_B,b_B)\nonumber\\&&
\{ (\phi _S(x_1,\omega
   _1) ((g^+ (-x_B-f^+ x_2+f^++g^- x_1)+f^- g^-) \phi
   _P^s(x_2,\omega _2)\nonumber\\&&-(g^+ (x_B+f^+ (x_2-1)+g^-
   x_1)+f^- g^-) \phi _P^t(x_2,\omega _2))\nonumber\\&&+\sqrt{g^- g^+}
   \phi _S^s(x_1,\omega _1) ((-x_B+f^--f^+ x_2+f^++g^+ x_1)
   \phi _P^s(x_2,\omega _2)\nonumber\\&&-(x_B+f^-+f^+ (x_2-1)+g^+
   x_1) \phi _P^t(x_2,\omega _2))\nonumber\\&&+\sqrt{g^- g^+} \phi
   _S^t(x_1,\omega _1) ((x_B+f^-+f^+ (x_2-1)+g^+
   x_1) \phi _P^s(x_2,\omega _2)\nonumber\\&&-(-x_B+f^--f^+ x_2+f^++g^+
   x_1) \phi _P^t(x_2,\omega _2)))
\alpha_s(t_c)e^{-S_{cd}(t_c)}h(\beta_c,\alpha_e,b_2,b_B)\nonumber\\&&
-(g^+ \phi _S(x_1,\omega
   _1) ((-x_B+f^+ x_2+g^- x_1) \phi _P^s(x_2,\omega
   _2)+(x_B-f^+ x_2+g^- x_1) \phi _P^t(x_2,\omega
   _2))\nonumber\\&&+\sqrt{g^- g^+} \phi _S^s(x_1,\omega _1)
   ((-x_B+f^+ x_2+g^+ x_1) \phi _P^s(x_2,\omega
   _2)+(x_B-f^+ x_2+g^+ x_1) \phi _P^t(x_2,\omega
   _2))\nonumber\\&&+\sqrt{g^- g^+} \phi _S^t(x_1,\omega _1)
   ((x_B-f^+ x_2+g^+ x_1) \phi _P^s(x_2,\omega
   _2)-(x_B-f^+ x_2-g^+ x_1) \phi _P^t(x_2,\omega
   _2)))\nonumber\\&&
   \alpha_s(t_d)e^{-S_{cd}(t_d)}h(\beta_d,\alpha_e,b_2,b_B)\}.
\end{eqnarray}
\begin{eqnarray}
\mathcal{F}^{LL(LR),SV}_a&=&8 \pi  M^4 f_B C_f
\int_0^1dx_1dx_2\int_0^{1/\Lambda} b_1db_1b_2db_2\nonumber\\&&
\{(2 \sqrt{f^- f^+} \sqrt{g^- g^+} \phi _S^s(x_1,\omega
   _1) ((f^+ x_2+g^-+g^+) \phi _P^s(x_2,\omega
   _2)+(f^+ x_2+g^--g^+) \phi _P^t(x_2,\omega
   _2))\nonumber\\&&+g^+ (f^- g^--f^+ (f^+ x_2+g^-)) \phi
   _P(x_2,\omega _2) \phi _S(x_1,\omega _1))
\alpha_s(t_e)e^{-S_{ef}(t_e)}h(\alpha_a,\beta_e,b_1,b_2)S_t(x_2)\nonumber\\&&
-(-2 \sqrt{f^- f^+} \sqrt{g^- g^+} (f^--f^+-g^--g^+
   x_1+g^+) \phi _P^s(x_2,\omega _2) \phi _S^t(x_1,\omega
   _1)\nonumber\\&&+2 \sqrt{f^- f^+} \sqrt{g^- g^+} (f^-+f^++g^--g^+ x_1+g^+) \phi
   _P^s(x_2,\omega _2) \phi _S^s(x_1,\omega _1)\nonumber\\&&+(f^+
   (g^+)^2 (x_1-1)+f^- g^- (f^++g^-)-f^- f^+
   g^+) \phi _P(x_2,\omega _2) \phi _S(x_1,\omega
   _1))\nonumber\\&&\alpha_s(t_f)e^{-S_{ef}(t_f)}h(\alpha_a,\beta_f,b_2,b_1)S_t(x_1)\},
\end{eqnarray}
\begin{eqnarray}
\mathcal{F}^{SP,SV}_a&=&16 \pi  M^4 f_B C_f
\int_0^1dx_1dx_2\int_0^{1/\Lambda} b_1db_1b_2db_2\nonumber\\&&
\{ (\sqrt{f^- f^+} g^+ \phi _S(x_1,\omega _1)
   ((f^+ x_2+2 g^-) \phi _P^s(x_2,\omega _2)-f^+ x_2 \phi
   _P^t(x_2,\omega _2))\nonumber\\&&+2 \sqrt{g^- g^+} (f^- (f^+
   x_2+g^-)-f^+ g^+) \phi _P(x_2,\omega _2) \phi
   _S^s(x_1,\omega _1))\alpha_s(t_e)e^{-S_{ef}(t_e)}h(\alpha_a,\beta_e,b_1,b_2)S_t(x_2)\nonumber\\&&
+ (2 \sqrt{f^- f^+} (f^- g^-+g^+ (f^++g^-
   (-x_1)+2 g^-)) \phi _S(x_1,\omega _1) \phi
   _P^s(x_2,\omega _2)\nonumber\\&&+\sqrt{g^- g^+} (f^- g^-+f^+ g^+
   (x_1-1)) \phi _P(x_2,\omega _2) \phi
   _S^s(x_1,\omega _1)\nonumber\\&&-\sqrt{g^- g^+} (f^- (2 f^++g^-)-f^+
   g^+ (x_1-1)) \phi _P(x_2,\omega _2) \phi
   _S^t(x_1,\omega _1))\nonumber\\&& \alpha_s(t_f)e^{-S_{ef}(t_f)}h(\alpha_a,\beta_f,b_2,b_1)S_t(x_1)\},
\end{eqnarray}
\begin{eqnarray}
\mathcal{M}^{LL,SV}_a&=&16 \sqrt{\frac{2}{3}} \pi  M^4 C_f
 \int_0^1dx_Bdx_1dx_2\int_0^{1/\Lambda} b_1db_1b_Bdb_B\phi _B(x_B,b_B) \nonumber\\&&
\{ (\sqrt{f^- f^+} \sqrt{g^- g^+} \phi
   _S^s(x_1,\omega _1) ((-x_B+f^+ x_2+g^--g^+ x_1+g^+) \phi
   _P^s(x_2,\omega _2)\nonumber\\&&-(-x_B+f^+ x_2+g^-+g^+ (x_1-1))
   \phi _P^t(x_2,\omega _2))\nonumber\\&&+\sqrt{f^- f^+} \sqrt{g^- g^+} \phi
   _S^t(x_1,\omega _1) ((-x_B+f^+ x_2+g^-+g^+
   (x_1-1)) \phi _P^s(x_2,\omega _2)\nonumber\\&&-(-x_B+f^+
   x_2+g^--g^+ x_1+g^+) \phi _P^t(x_2,\omega
   _2))+(g^-+g^+) (f^- (-x_B+f^+ x_2+g^-)+f^+
   g^+ (x_1-1)) \nonumber\\&&\phi _P(x_2,\omega _2) \phi
   _S(x_1,\omega _1))\alpha_s(t_g)e^{-S_{gh}(t_g)}h(\beta_g,\alpha_a,b_B,b_1)\nonumber\\&&
-(\sqrt{f^- f^+} \sqrt{g^- g^+} \phi
   _S^s(x_1,\omega _1) ((x_B+f^+ x_2+g^--g^+ x_1+g^++2)
   \phi _P^s(x_2,\omega _2)\nonumber\\&&+(x_B+f^+ x_2+g^-+g^+
   (x_1-1)) \phi _P^t(x_2,\omega _2))\nonumber\\&&+\sqrt{f^- f^+}
   \sqrt{g^- g^+} \phi _S^t(x_1,\omega _1) (-(x_B+f^+ x_2+g^--g^+
   x_1+g^+-2) \phi _P^t(x_2,\omega _2)\nonumber\\&&-(x_B+f^+ x_2+g^-+g^+
   (x_1-1)) \phi _P^s(x_2,\omega _2))\nonumber\\&&+(f^- g^+
   (x_B-g^- (x_1-2)-1)+f^+ (g^+ (f^-
   x_2-x_B)+g^-+g^+ g^- (x_1-2))+(f^+)^2 g^+
   (-x_2)) \nonumber\\&&\phi _P(x_2,\omega _2) \phi _S(x_1,\omega
   _1))
   \alpha_s(t_h)e^{-S_{gh}(t_h)}h(\beta_h,\alpha_a,b_B,b_1)\},
\end{eqnarray}
\begin{eqnarray}
\mathcal{M}^{LR,SV}_a&=&16 \sqrt{\frac{2}{3}} \pi  M^4 C_f
 \int_0^1dx_Bdx_1dx_2\int_0^{1/\Lambda} b_1db_1b_Bdb_B\phi _B(x_B,b_B) \nonumber\\&&
\{ (\sqrt{f^- f^+} g^+ \phi _S(x_1,\omega
   _1) (-(x_B-f^+ x_2-g^- x_1) \phi _P^t(x_2,\omega
   _2)-(x_B-f^+ x_2+g^- (x_1-2)) \phi _P^s(x_2,\omega
   _2))\nonumber\\&&-\sqrt{g^- g^+} (f^- (-x_B+f^+ x_2+g^-)+f^+ g^+
   (x_1-1)) \phi _P(x_2,\omega _2) \phi
   _S^s(x_1,\omega _1)\nonumber\\&&-\sqrt{g^- g^+} (f^- (-x_B+f^+
   x_2+g^-)-f^+ g^+ (x_1-1)) \phi _P(x_2,\omega _2)
   \phi _S^t(x_1,\omega _1))\nonumber\\&&\alpha_s(t_g)e^{-S_{gh}(t_g)}h(\beta_g,\alpha_a,b_B,b_1)\nonumber\\&&
- (\sqrt{f^- f^+} \phi _S(x_1,\omega
   _1) ((g^+ (x_B+f^+ x_2-2)-g^- (g^+
   (x_1-2)+2)) \phi _P^s(x_2,\omega _2)\nonumber\\&&+(g^+
   (x_B+f^+ x_2-2)+g^- (g^+ x_1+2)) \phi
   _P^t(x_2,\omega _2))\nonumber\\&&-\sqrt{g^- g^+} (f^- (x_B+f^+
   x_2+g^--2)+f^+ (g^+ (x_1-1)+2)) \phi
   _P(x_2,\omega _2) \phi _S^s(x_1,\omega _1)\nonumber\\&&-\sqrt{g^- g^+}
   (f^- (x_B+f^+ x_2+g^--2)+f^+ (g^+
   (-x_1)+g^+-2)) \phi _P(x_2,\omega _2) \phi
   _S^t(x_1,\omega _1))\nonumber\\&&\alpha_s(t_h)e^{-S_{gh}(t_h)}h(\beta_h,\alpha_a,b_B,b_1)\},
\end{eqnarray}
\begin{eqnarray}
\mathcal{M}^{SP,SV}_a&=&16 \sqrt{\frac{2}{3}} \pi  M^4 C_f
 \int_0^1dx_Bdx_1dx_2\int_0^{1/\Lambda} b_1db_1b_Bdb_B\phi _B(x_B,b_B) \nonumber\\&&
\{  (\sqrt{f^- f^+} \sqrt{g^- g^+} \phi
   _S^s(x_1,\omega _1) ((-x_B+f^+ x_2+g^--g^+ x_1+g^+) \phi
   _P^s(x_2,\omega _2)\nonumber\\&&-(x_B-f^+ x_2-g^--g^+ x_1+g^+) \phi
   _P^t(x_2,\omega _2))\nonumber\\&&+\sqrt{f^- f^+} \sqrt{g^- g^+} \phi
   _S^t(x_1,\omega _1) ((x_B-f^+ x_2-g^--g^+ x_1+g^+) \phi
   _P^s(x_2,\omega _2)\nonumber\\&&-(-x_B+f^+ x_2+g^--g^+ x_1+g^+) \phi
   _P^t(x_2,\omega _2))\nonumber\\&&+(f^--f^+) g^+ (-x_B+f^+
   x_2-g^- (x_1-2)) \phi _P(x_2,\omega _2) \phi
   _S(x_1,\omega _1))\nonumber\\&&
   \alpha_s(t_g)e^{-S_{gh}(t_g)}h(\beta_g,\alpha_a,b_B,b_1)\nonumber\\&&
-(\sqrt{f^- f^+} \sqrt{g^- g^+} \phi
   _S^s(x_1,\omega _1) ((x_B+f^+ x_2+g^--g^+ x_1+g^++2)
   \phi _P^s(x_2,\omega _2)\nonumber\\&&-(x_B+f^+ x_2+g^-+g^+
   (x_1-1)) \phi _P^t(x_2,\omega _2))\nonumber\\&&+\sqrt{f^- f^+}
   \sqrt{g^- g^+} \phi _S^t(x_1,\omega _1) ((x_B+f^+ x_2+g^-+g^+
   (x_1-1)) \phi _P^s(x_2,\omega _2)\nonumber\\&&-(x_B+f^+
   x_2+g^--g^+ x_1+g^+-2) \phi _P^t(x_2,\omega _2))\nonumber\\&&+(f^-
   ((g^-+g^+) (x_B+g^-)-g^+)+f^+ (f^- g^-
   x_2+f^- g^+ x_2+g^-+(g^+)^2 (x_1-1)+g^- g^+
   (x_1-1)))\nonumber\\&& \phi _P(x_2,\omega _2) \phi
   _S(x_1,\omega _1))\alpha_s(t_h)e^{-S_{gh}(t_h)}h(\beta_h,\alpha_a,b_B,b_1)\}.
\end{eqnarray}
\begin{eqnarray}
\mathcal{F'}^{LL(LR),SV}_a&=&8 \pi  M^4 f_B C_f
\int_0^1dx_1dx_2\int_0^{1/\Lambda} b_1db_1b_2db_2\nonumber\\&&
\{ (2 \sqrt{f^- f^+} \sqrt{g^- g^+} (f^--f^++g^+ x_1) \phi _P^s(x_2,\omega _2) \phi _S^t(x_1,\omega _1)\nonumber\\&&+2
   \sqrt{f^- f^+} \sqrt{g^- g^+} (f^-+f^++g^+ x_1) \phi _P^s(x_2,\omega _2) \phi _S^s(x_1,\omega _1)\nonumber\\&&+f^+ (f^-
   (g^--g^+)-(g^+)^2 x_1) \phi _P(x_2,\omega _2) \phi _S(x_1,\omega _1))
\alpha_s(t'_e)e^{-S_{ef}(t'_e)}h(\alpha'_a,\beta'_e,b_2,b_1)S_t(x_1)\nonumber\\&&
-(2 \sqrt{f^- f^+} \sqrt{g^- g^+} \phi _S^s(x_1,\omega _1) ((f^--f^+ x_2+f^++g^-+g^+) \phi _P^s(x_2,\omega
   _2)\nonumber\\&&+(f^-+f^+ (x_2-1)-g^-+g^+) \phi _P^t(x_2,\omega _2))+((f^-)^2 g^-+f^- g^- g^++f^+ g^+ (f^+
   (x_2-1)-g^-))\nonumber\\&& \phi _P(x_2,\omega _2) \phi _S(x_1,\omega _1))\alpha_s(t'_f)e^{-S_{ef}(t'_f)}h(\alpha'_a,\beta'_f,b_1,b_2)S_t(x_2)\},
\end{eqnarray}
\begin{eqnarray}
\mathcal{M'}^{LL,SV}_a&=& 16 \sqrt{\frac{2}{3}} \pi  M^4 C_f
 \int_0^1dx_Bdx_1dx_2\int_0^{1/\Lambda} b_1db_1b_Bdb_B\phi _B(x_B,b_B) \nonumber\\&&
\{ (\sqrt{f^- f^+} \sqrt{g^- g^+} \phi _S^s(x_1,\omega _1) ((-x_B+f^--f^+ x_2+f^++g^+ x_1) \phi
   _P^s(x_2,\omega _2)\nonumber\\&&-(x_B-f^--f^+ x_2+f^+-g^+ x_1) \phi _P^t(x_2,\omega _2))+\sqrt{f^- f^+} \sqrt{g^- g^+} \phi
   _S^t(x_1,\omega _1) \nonumber\\&&((x_B-f^--f^+ x_2+f^+-g^+ x_1) \phi _P^s(x_2,\omega _2)-(-x_B+f^--f^+ x_2+f^++g^+ x_1) \phi
   _P^t(x_2,\omega _2))\nonumber\\&&+(f^--f^+) (g^- (-x_B)+f^- g^--f^+ g^+ x_2+f^+ g^++g^+ g^- x_1) \nonumber\\&&\phi _P(x_2,\omega
   _2) \phi _S(x_1,\omega _1))\alpha_s(t'_g)e^{-S_{gh}(t'_g)}h(\beta'_g,\alpha'_a,b_B,b_1)\nonumber\\&&
-(-\sqrt{f^- f^+} \sqrt{g^- g^+} \phi _S^s(x_1,\omega _1) ((x_B+f^--f^+ x_2+f^++g^+ x_1+2) \phi
   _P^s(x_2,\omega _2)\nonumber\\&&-(x_B+f^-+f^+ (x_2-1)+g^+ x_1) \phi _P^t(x_2,\omega _2))-\sqrt{f^- f^+} \sqrt{g^- g^+} \phi
   _S^t(x_1,\omega _1) \nonumber\\&&((x_B+f^-+f^+ (x_2-1)+g^+ x_1) \phi _P^s(x_2,\omega _2)-(x_B+f^--f^+ x_2+f^++g^+
   x_1-2) \phi _P^t(x_2,\omega _2))\nonumber\\&&+(f^+ g^+ (x_B+f^- x_2+g^+ x_1)+f^+ g^- (x_B+f^- x_2+g^+ x_1-1)+f^- g^+)\nonumber\\&&
   \phi _P(x_2,\omega _2) \phi _S(x_1,\omega _1))
   \alpha_s(t'_h)e^{-S_{gh}(t'_h)}h(\beta'_h,\alpha'_a,b_B,b_1)\},
\end{eqnarray}
\begin{eqnarray}
\mathcal{M'}^{SP,SV}_a&=&-16\sqrt{\frac{2}{3}} \pi  M^4 C_f
 \int_0^1dx_Bdx_1dx_2\int_0^{1/\Lambda} b_1db_1b_Bdb_B\phi _B(x_B,b_B) \nonumber\\&&
\{  (-\sqrt{f^- f^+} \sqrt{g^- g^+} \phi _S^s(x_1,\omega _1) ((-x_B+f^--f^+ x_2+f^++g^+ x_1) \phi
   _P^s(x_2,\omega _2)\nonumber\\&&-(-x_B+f^-+f^+ (x_2-1)+g^+ x_1) \phi _P^t(x_2,\omega _2))-\sqrt{f^- f^+} \sqrt{g^- g^+} \phi
   _S^t(x_1,\omega _1)\nonumber\\&& ((-x_B+f^-+f^+ (x_2-1)+g^+ x_1) \phi _P^s(x_2,\omega _2)-(-x_B+f^--f^+ x_2+f^++g^+
   x_1) \phi _P^t(x_2,\omega _2))\nonumber\\&&+f^+ (g^-+g^+) (-x_B+f^- x_2+g^+ x_1) \phi _P(x_2,\omega _2) \phi
   _S(x_1,\omega _1))\nonumber\\&&
   \alpha_s(t'_g)e^{-S_{gh}(t'_g)}h(\beta'_g,\alpha'_a,b_B,b_1)\nonumber\\&&
-(\sqrt{f^- f^+} \sqrt{g^- g^+} \phi _S^s(x_1,\omega _1) ((x_B+f^--f^+ x_2+f^++g^+ x_1+2) \phi
   _P^s(x_2,\omega _2)\nonumber\\&&+(x_B+f^-+f^+ (x_2-1)+g^+ x_1) \phi _P^t(x_2,\omega _2))+\sqrt{f^- f^+} \sqrt{g^- g^+} \phi
   _S^t(x_1,\omega _1)\nonumber\\&& (-(x_B+f^--f^+ x_2+f^++g^+ x_1-2) \phi _P^t(x_2,\omega _2)-(x_B+f^-+f^+ (x_2-1)+g^+
   x_1) \phi _P^s(x_2,\omega _2))\nonumber\\&&+(f^+ (f^+ g^+ (x_2-1)-g^- (x_B+g^+ x_1-1))+f^- (g^- x_B-f^+
   (g^-+g^+ (x_2-1))-g^++g^+ g^- x_1)\nonumber\\&&+(f^-)^2 g^-) \phi _P(x_2,\omega _2) \phi _S(x_1,\omega
   _1))\alpha_s(t'_h)e^{-S_{gh}(t'_h)}h(\beta'_h,\alpha'_a,b_B,b_1)\}.
\end{eqnarray}

\begin{eqnarray}
\mathcal{F}^{LL,VS}_e&=& 8 \pi  M^4 C_f
\int_0^1 dx_Bdx_1\int_0^{1/\Lambda} b_Bdb_Bb_1db_1\phi _B(x_B,b_B)\nonumber\\&&
\{ (\sqrt{g^- g^+} (f^+ (2 g^+ x_1-1)-f^-) \phi _P^s(x_1,\omega _1)+\sqrt{g^- g^+} (f^-+f^+ (2 g^+
   x_1-1)) \phi _P^t(x_1,\omega _1)\nonumber\\&&+(f^- g^--f^+ g^+ (g^+ x_1+1)) \phi _P(x_1,\omega _1))
\alpha_s(t_a)e^{-S_{ab}(t_a)}h(\alpha_e,\beta_a,b_B,b_1)S_t(x_1)\nonumber\\&&-
((2 \sqrt{g^- g^+} (f^- (g^--x_B)+f^+ g^+) \phi _P^s(x_1,\omega _1)+g^+ (f^- (x_B-g^-)+f^+
   g^-) \phi _P(x_1,\omega _1))\nonumber\\&&
\alpha_s(t_b)e^{-S_{ab}(t_b)}h(\alpha_e,\beta_b,b_1,b_B)S_t(x_B)\},
\end{eqnarray}
\begin{eqnarray}
\mathcal{M}^{LL,VS}_e&=& 16 \sqrt{\frac{2}{3}} \pi  M^4 C_f
\int_0^1 dx_Bdx_1dx_2\int_0^{1/\Lambda} b_Bdb_Bb_2db_2\phi _B(x_B,b_B)\phi_S(x_2,\omega_2)
\nonumber\\&&\{  (\sqrt{g^- g^+} (f^+ g^+ x_1-f^- (x_B+f^+ (x_2-2))) \phi
   _P^s(x_1,\omega _1)-\sqrt{g^- g^+} (f^- x_B+f^+ g^+ x_1+f^+ f^- x_2)\nonumber\\&& \phi _P^t(x_1,\omega _1)+(f^-+f^+) (g^+
   (x_B+f^+ (x_2-1)+g^- x_1)+f^- g^-) \phi _P(x_1,\omega _1))\nonumber\\&&
\alpha_s(t_c)e^{-S_{cd}(t_c)}h(\beta_c,\alpha_e,b_2, b_B)-
 (\sqrt{g^- g^+} (f^- (-x_B)+f^+ g^+ x_1+f^+ f^- x_2) \phi
   _P^s(x_1,\omega _1)\nonumber\\&&+\sqrt{g^- g^+} (f^- x_B+f^+ g^+ x_1-f^+ f^- x_2) \phi _P^t(x_1,\omega _1)+(f^- g^--f^+ g^+)
   \nonumber\\&&(-x_B+f^+ x_2+g^+ x_1) \phi _P(x_1,\omega _1))
\alpha_s(t_d)e^{-S_{cd}(t_d)}h(\beta_d,\alpha_e,b_2,b_B)\},
\end{eqnarray}
\begin{eqnarray}
\mathcal{M}^{LR,VS}_e&=& 16 \sqrt{\frac{2}{3}} \pi  \sqrt{f^- f^+} M^4 C_f
\int_0^1 dx_Bdx_1dx_2\int_0^{1/\Lambda} b_Bdb_Bb_2db_2\phi _B(x_B,b_B)\nonumber\\&&
\{  (\sqrt{g^- g^+} \phi _P^s(x_1,\omega _1) ((-x_B+f^--f^+ x_2+f^++g^+ x_1) \phi
   _S^s(x_2,\omega _2)\nonumber\\&&-(x_B+f^-+f^+ (x_2-1)+g^+ x_1) \phi _S^t(x_2,\omega _2))+\sqrt{g^- g^+} \phi
   _P^t(x_1,\omega _1) ((x_B+f^-+f^+ (x_2-1)+g^+ x_1)\nonumber\\&& \phi _S^s(x_2,\omega _2)-(-x_B+f^--f^+ x_2+f^++g^+
   x_1) \phi _S^t(x_2,\omega _2))\nonumber\\&&+\phi _P(x_1,\omega _1) ((g^+ (-x_B-f^+ x_2+f^++g^- x_1)+f^- g^-) \phi
   _S^t(x_2,\omega _2)\nonumber\\&&-(g^+ (x_B+f^+ (x_2-1)+g^- x_1)+f^- g^-) \phi _S^s(x_2,\omega _2)))
\alpha_s(t_c)e^{-S_{cd}(t_c)}h(\beta_c,\alpha_e,b_2,b_B)\nonumber\\&&
-(-g^+ \phi _P(x_1,\omega _1) ((x_B-f^+ x_2+g^- x_1) \phi _S^s(x_2,\omega
   _2)+(-x_B+f^+ x_2+g^- x_1) \phi _S^t(x_2,\omega _2))\nonumber\\&&+\sqrt{g^- g^+} \phi _P^s(x_1,\omega _1) ((-x_B+f^+
   x_2+g^+ x_1) \phi _S^s(x_2,\omega _2)+(x_B-f^+ x_2+g^+ x_1) \phi _S^t(x_2,\omega _2))\nonumber\\&&+\sqrt{g^- g^+} \phi
   _P^t(x_1,\omega _1) ((x_B-f^+ x_2+g^+ x_1) \phi _S^s(x_2,\omega _2)+(-x_B+f^+ x_2+g^+ x_1) \phi
   _S^t(x_2,\omega _2)))\nonumber\\&&
   \alpha_s(t_d)e^{-S_{cd}(t_d)}h(\beta_d,\alpha_e,b_2,b_B)\}.
\end{eqnarray}
\begin{eqnarray}
\mathcal{F}^{LL(LR),VS}_a&=&8 \pi  M^4 f_B C_f
\int_0^1dx_1dx_2\int_0^{1/\Lambda} b_1db_1b_2db_2\nonumber\\&&
\{ (2 \sqrt{f^- f^+} \sqrt{g^- g^+} \phi _P^s(x_1,\omega _1) ((f^+ x_2+g^-+g^+) \phi _S^s(x_2,\omega
   _2)+(f^+ x_2+g^--g^+) \phi _S^t(x_2,\omega _2))\nonumber\\&&+g^+ (f^- g^--f^+ (f^+ x_2+g^-)) \phi _P(x_1,\omega
   _1) \phi _S(x_2,\omega _2))
\alpha_s(t_e)e^{-S_{ef}(t_e)}h(\alpha_a,\beta_e,b_1,b_2)S_t(x_2)\nonumber\\&&
-(-2 \sqrt{f^- f^+} \sqrt{g^- g^+} (f^--f^+-g^--g^+ x_1+g^+) \phi _P^t(x_1,\omega _1) \phi _S^s(x_2,\omega
   _2)\nonumber\\&&+2 \sqrt{f^- f^+} \sqrt{g^- g^+} (f^-+f^++g^--g^+ x_1+g^+) \phi _P^s(x_1,\omega _1) \phi _S^s(x_2,\omega _2)\nonumber\\&&+(f^+
   (g^+)^2 (x_1-1)+f^- g^- (f^++g^-)-f^- f^+ g^+) \phi _P(x_1,\omega _1) \phi _S(x_2,\omega _2))\nonumber\\&&\alpha_s(t_f)e^{-S_{ef}(t_f)}h(\alpha_a,\beta_f,b_2,b_1)S_t(x_1)\},
\end{eqnarray}
\begin{eqnarray}
\mathcal{F}^{SP,VS}_a&=&-16 \pi  M^4 f_B C_f
\int_0^1dx_1dx_2\int_0^{1/\Lambda} b_1db_1b_2db_2\nonumber\\&&
\{ (2 \sqrt{g^- g^+} (f^+ g^++f^- (f^+ x_2+g^-)) \phi _S(x_2,\omega _2) \phi _P^s(x_1,\omega
   _1)-\sqrt{f^- f^+} g^+ \phi _P(x_1,\omega _1) \nonumber\\&&(f^+ x_2 \phi _S^s(x_2,\omega _2)-(f^+ x_2+2 g^-) \phi
   _S^t(x_2,\omega _2)))\alpha_s(t_e)e^{-S_{ef}(t_e)}h(\alpha_a,\beta_e,b_1,b_2)S_t(x_2)\nonumber\\&&
+ (\sqrt{g^- g^+} (f^- (2 f^++g^-)-f^+ g^+ (x_1-1)) \phi _S(x_2,\omega _2) \phi _P^s(x_1,\omega
   _1)\nonumber\\&&+2 \sqrt{f^- f^+} (f^- g^--g^+ (f^++g^- x_1)) \phi _P(x_1,\omega _1) \phi _S^s(x_2,\omega _2)\nonumber\\&&-\sqrt{g^- g^+}
   (f^- g^-+f^+ g^+ (x_1-1)) \phi _S(x_2,\omega _2) \phi _P^t(x_1,\omega _1))\nonumber\\&& \alpha_s(t_f)e^{-S_{ef}(t_f)}h(\alpha_a,\beta_f,b_2,b_1)S_t(x_1)\},
\end{eqnarray}
\begin{eqnarray}
\mathcal{M}^{LL,VS}_a&=&16 \sqrt{\frac{2}{3}} \pi  M^4 C_f
 \int_0^1dx_Bdx_1dx_2\int_0^{1/\Lambda} b_1db_1b_Bdb_B\phi _B(x_B,b_B) \nonumber\\&&
\{  (\sqrt{f^- f^+} \sqrt{g^- g^+} \phi _P^s(x_1,\omega _1) ((-x_B+f^+ x_2+g^--g^+ x_1+g^+) \phi
   _S^s(x_2,\omega _2)\nonumber\\&&+(x_B-f^+ x_2-g^--g^+ x_1+g^+) \phi _S^t(x_2,\omega _2))\nonumber\\&&+\sqrt{f^- f^+} \sqrt{g^- g^+} \phi
   _P^t(x_1,\omega _1) ((-x_B+f^+ x_2+g^-+g^+ (x_1-1)) \phi _S^s(x_2,\omega _2)\nonumber\\&&-(-x_B+f^+ x_2+g^--g^+
   x_1+g^+) \phi _S^t(x_2,\omega _2))\nonumber\\&&+(g^--g^+) (f^- (-x_B+f^+ x_2+g^-)-f^+ g^+ (x_1-1)) \phi
   _P(x_1,\omega _1) \phi _S(x_2,\omega _2))\nonumber\\&&\alpha_s(t_g)e^{-S_{gh}(t_g)}h(\beta_g,\alpha_a,b_B,b_1)\nonumber\\&&
-(\sqrt{f^- f^+} \sqrt{g^- g^+} \phi _P^s(x_1,\omega _1) ((x_B+f^+ x_2+g^--g^+ x_1+g^++2) \phi
   _S^s(x_2,\omega _2)\nonumber\\&&+(x_B+f^+ x_2+g^-+g^+ (x_1-1)) \phi _S^t(x_2,\omega _2))\nonumber\\&&-\sqrt{f^- f^+} \sqrt{g^- g^+} \phi
   _P^t(x_1,\omega _1) ((x_B+f^+ x_2+g^-+g^+ (x_1-1)) \phi _S^s(x_2,\omega _2)\nonumber\\&&+(x_B+f^+ x_2+g^--g^+
   x_1+g^+-2) \phi _S^t(x_2,\omega _2))\nonumber\\&&-(f^- g^+ (x_B+g^- x_1-1)+f^+ (g^+ x_B+f^- g^+ x_2+g^-+g^+ g^-
   x_1)+(f^+)^2 g^+ x_2)\nonumber\\&& \phi _P(x_1,\omega _1) \phi _S(x_2,\omega _2))
   \alpha_s(t_h)e^{-S_{gh}(t_h)}h(\beta_h,\alpha_a,b_B,b_1)\},
\end{eqnarray}
\begin{eqnarray}
\mathcal{M}^{LR,VS}_a&=&16 \sqrt{\frac{2}{3}} \pi  M^4 C_f
 \int_0^1dx_Bdx_1dx_2\int_0^{1/\Lambda} b_1db_1b_Bdb_B\phi _B(x_B,b_B) \nonumber\\&&
\{ (\sqrt{f^- f^+} g^+ \phi _P(x_1,\omega _1) ((-x_B+f^+ x_2+g^- x_1) \phi _S^s(x_2,\omega
   _2)-(x_B-f^+ x_2+g^- (x_1-2)) \phi _S^t(x_2,\omega _2))\nonumber\\&&+\sqrt{g^- g^+} (f^- (-x_B+f^+ x_2+g^-)-f^+ g^+
   (x_1-1)) \phi _S(x_2,\omega _2) \phi _P^s(x_1,\omega _1)\nonumber\\&&+\sqrt{g^- g^+} (f^- (-x_B+f^+ x_2+g^-)+f^+ g^+
   (x_1-1)) \phi _S(x_2,\omega _2) \phi _P^t(x_1,\omega _1))\nonumber\\&&\alpha_s(t_g)e^{-S_{gh}(t_g)}h(\beta_g,\alpha_a,b_B,b_1)\nonumber\\&&
- (\sqrt{f^- f^+} \phi _P(x_1,\omega _1) ((g^+ (x_B+f^+ x_2-2)+g^- (g^+
   x_1+2)) \phi _S^s(x_2,\omega _2)\nonumber\\&&+(g^+ (x_B+f^+ x_2-2)-g^- (g^+ (x_1-2)+2)) \phi
   _S^t(x_2,\omega _2))\nonumber\\&&+\sqrt{g^- g^+} (f^- (x_B+f^+ x_2+g^--2)+f^+ (g^+ (-x_1)+g^+-2)) \phi
   _S(x_2,\omega _2) \phi _P^s(x_1,\omega _1)\nonumber\\&&+\sqrt{g^- g^+} (f^- (x_B+f^+ x_2+g^--2)+f^+ (g^+
   (x_1-1)+2)) \phi _S(x_2,\omega _2) \phi _P^t(x_1,\omega _1))\nonumber\\&&\alpha_s(t_h)e^{-S_{gh}(t_h)}h(\beta_h,\alpha_a,b_B,b_1)\},
\end{eqnarray}
\begin{eqnarray}
\mathcal{M}^{SP,VS}_a&=&16 \sqrt{\frac{2}{3}} \pi  M^4 C_f
 \int_0^1dx_Bdx_1dx_2\int_0^{1/\Lambda} b_1db_1b_Bdb_B\phi _B(x_B,b_B) \nonumber\\&&
\{  (\sqrt{f^- f^+} \sqrt{g^- g^+} \phi _P^s(x_1,\omega _1) ((-x_B+f^+ x_2+g^--g^+ x_1+g^+) \phi
   _S^s(x_2,\omega _2)\nonumber\\&&+(-x_B+f^+ x_2+g^-+g^+ (x_1-1)) \phi _S^t(x_2,\omega _2))\nonumber\\&&-\sqrt{f^- f^+} \sqrt{g^- g^+} \phi
   _P^t(x_1,\omega _1) ((-x_B+f^+ x_2+g^-+g^+ (x_1-1)) \phi _S^s(x_2,\omega _2)\nonumber\\&&+(-x_B+f^+ x_2+g^--g^+
   x_1+g^+) \phi _S^t(x_2,\omega _2))-(f^-+f^+) g^+ (-x_B+f^+ x_2+g^- x_1) \nonumber\\&&\phi _P(x_1,\omega _1) \phi
   _S(x_2,\omega _2))
   \alpha_s(t_g)e^{-S_{gh}(t_g)}h(\beta_g,\alpha_a,b_B,b_1)\nonumber\\&&
- (\sqrt{f^- f^+} \sqrt{g^- g^+} \phi _P^s(x_1,\omega _1) ((x_B+f^+ x_2+g^--g^+ x_1+g^++2) \phi
   _S^s(x_2,\omega _2)\nonumber\\&&-(x_B+f^+ x_2+g^-+g^+ (x_1-1)) \phi _S^t(x_2,\omega _2))\nonumber\\&&+\sqrt{f^- f^+} \sqrt{g^- g^+} \phi
   _P^t(x_1,\omega _1) ((x_B+f^+ x_2+g^-+g^+ (x_1-1)) \phi _S^s(x_2,\omega _2)\nonumber\\&&-(x_B+f^+ x_2+g^--g^+
   x_1+g^+-2) \phi _S^t(x_2,\omega _2))+(f^- ((g^--g^+) (x_B+g^-)+g^+)\nonumber\\&&-f^+ (f^- g^-
   (-x_2)+g^+ (f^- x_2-g^+ x_1+g^+)+g^-+g^+ g^- (x_1-1))) \phi _P(x_1,\omega _1) \phi _S(x_2,\omega
   _2))\nonumber\\&&\alpha_s(t_h)e^{-S_{gh}(t_h)}h(\beta_h,\alpha_a,b_B,b_1)\}.
\end{eqnarray}
\begin{eqnarray}
\mathcal{F'}^{LL(LR),VS}_a&=&8 \pi  M^4 f_B C_f
\int_0^1dx_1dx_2\int_0^{1/\Lambda} b_1db_1b_2db_2\nonumber\\&&
\{ (2 \sqrt{f^- f^+} \sqrt{g^- g^+} (f^--f^++g^+ x_1) \phi _P^t(x_1,\omega _1) \phi _S^s(x_2,\omega _2)\nonumber\\&&+2
   \sqrt{f^- f^+} \sqrt{g^- g^+} (f^-+f^++g^+ x_1) \phi _P^s(x_1,\omega _1) \phi _S^s(x_2,\omega _2)\nonumber\\&&+f^+ (f^-
   (g^--g^+)-(g^+)^2 x_1) \phi _P(x_1,\omega _1) \phi _S(x_2,\omega _2))
\alpha_s(t'_e)e^{-S_{ef}(t'_e)}h(\alpha'_a,\beta'_e,b_2,b_1)S_t(x_1)\nonumber\\&&
-(2 \sqrt{f^- f^+} \sqrt{g^- g^+} \phi _P^s(x_1,\omega _1) ((f^--f^+ x_2+f^++g^-+g^+) \phi _S^s(x_2,\omega
   _2)\nonumber\\&&+(f^-+f^+ (x_2-1)-g^-+g^+) \phi _S^t(x_2,\omega _2))+((f^-)^2 g^-+f^- g^- g^++f^+ g^+ (f^+
   (x_2-1)-g^-))\nonumber\\&& \phi _P(x_1,\omega _1) \phi _S(x_2,\omega _2))\alpha_s(t'_f)e^{-S_{ef}(t'_f)}h(\alpha'_a,\beta'_f,b_1,b_2)S_t(x_2)\},
\end{eqnarray}
\begin{eqnarray}
\mathcal{M'}^{LL,VS}_a&=& 16 \sqrt{\frac{2}{3}} \pi  M^4 C_f
 \int_0^1dx_Bdx_1dx_2\int_0^{1/\Lambda} b_1db_1b_Bdb_B\phi _B(x_B,b_B) \nonumber\\&&
\{ (\sqrt{f^- f^+} \sqrt{g^- g^+} \phi _P^s(x_1,\omega _1) ((-x_B+f^--f^+ x_2+f^++g^+ x_1) \phi
   _S^s(x_2,\omega _2)\nonumber\\&&+(-x_B+f^-+f^+ (x_2-1)+g^+ x_1) \phi _S^t(x_2,\omega _2))-\sqrt{f^- f^+} \sqrt{g^- g^+} \phi
   _P^t(x_1,\omega _1) \nonumber\\&&((-x_B+f^-+f^+ (x_2-1)+g^+ x_1) \phi _S^s(x_2,\omega _2)+(-x_B+f^--f^+ x_2+f^++g^+
   x_1) \phi _S^t(x_2,\omega _2))\nonumber\\&&+(f^-+f^+) (g^- (g^+ x_1-x_B)+f^- g^-+f^+ g^+ (x_2-1)) \phi
   _P(x_1,\omega _1) \phi _S(x_2,\omega _2))\nonumber\\&&\alpha_s(t'_g)e^{-S_{gh}(t'_g)}h(\beta'_g,\alpha'_a,b_B,b_1)\nonumber\\&&
- (\sqrt{f^- f^+} \sqrt{g^- g^+} \phi _P^s(x_1,\omega _1) ((x_B+f^-+f^+ (x_2-1)+g^+ x_1)
   \phi _S^t(x_2,\omega _2)\nonumber\\&&-(x_B+f^--f^+ x_2+f^++g^+ x_1+2) \phi _S^s(x_2,\omega _2))+\sqrt{f^- f^+} \sqrt{g^- g^+} \phi
   _P^t(x_1,\omega _1) \nonumber\\&&((x_B+f^--f^+ x_2+f^++g^+ x_1-2) \phi _S^t(x_2,\omega _2)-(x_B+f^-+f^+ (x_2-1)+g^+
   x_1) \phi _S^s(x_2,\omega _2))\nonumber\\&&+(f^+ g^+ (x_B-f^- (x_2-2)+g^+ x_1)-f^+ g^- (x_B-f^- (x_2-2)+g^+
   x_1-1)-f^- g^+) \nonumber\\&&\phi _P(x_1,\omega _1) \phi _S(x_2,\omega _2))
   \alpha_s(t'_h)e^{-S_{gh}(t'_h)}h(\beta'_h,\alpha'_a,b_B,b_1)\},
\end{eqnarray}
\begin{eqnarray}
\mathcal{M'}^{SP,VS}_a&=&16 \sqrt{\frac{2}{3}} \pi  M^4 C_f
 \int_0^1dx_Bdx_1dx_2\int_0^{1/\Lambda} b_1db_1b_Bdb_B\phi _B(x_B,b_B) \nonumber\\&&
\{ (\sqrt{f^- f^+} \sqrt{g^- g^+} \phi _P^s(x_1,\omega _1) ((-x_B+f^--f^+ x_2+f^++g^+ x_1) \phi
   _S^s(x_2,\omega _2)\nonumber\\&&+(x_B-f^--f^+ x_2+f^+-g^+ x_1) \phi _S^t(x_2,\omega _2))+\sqrt{f^- f^+} \sqrt{g^- g^+} \phi
   _P^t(x_1,\omega _1)\nonumber\\&& ((-x_B+f^-+f^+ (x_2-1)+g^+ x_1) \phi _S^s(x_2,\omega _2)-(-x_B+f^--f^+ x_2+f^++g^+
   x_1) \phi _S^t(x_2,\omega _2))\nonumber\\&&+f^+ (g^--g^+) (-x_B-f^- (x_2-2)+g^+ x_1) \phi _P(x_1,\omega _1)
   \phi _S(x_2,\omega _2))
   \alpha_s(t'_g)e^{-S_{gh}(t'_g)}h(\beta'_g,\alpha'_a,b_B,b_1)\nonumber\\&&
-(\sqrt{f^- f^+} \sqrt{g^- g^+} \phi _P^s(x_1,\omega _1) ((x_B+f^--f^+ x_2+f^++g^+ x_1+2) \phi
   _S^s(x_2,\omega _2)\nonumber\\&&+(x_B+f^-+f^+ (x_2-1)+g^+ x_1) \phi _S^t(x_2,\omega _2))-\sqrt{f^- f^+} \sqrt{g^- g^+} \phi
   _P^t(x_1,\omega _1)\nonumber\\&& ((x_B+f^-+f^+ (x_2-1)+g^+ x_1) \phi _S^s(x_2,\omega _2)+(x_B+f^--f^+ x_2+f^++g^+
   x_1-2) \phi _S^t(x_2,\omega _2))\nonumber\\&&+(f^+ (g^- (x_B+g^+ x_1-1)+f^+ g^+ (x_2-1))+f^- (g^- x_B+f^+
   (g^-+g^+ (x_2-1))+g^+\nonumber\\&&+g^+ g^- x_1)+(f^-)^2 g^-)\phi _P(x_1,\omega _1) \phi _S(x_2,\omega
   _2))\alpha_s(t'_h)e^{-S_{gh}(t'_h)}h(\beta'_h,\alpha'_a,b_B,b_1)\}.
\end{eqnarray}

For the double $S$-wave amplitude, we have the factorization formulas
\begin{eqnarray}
\mathcal{F}^{LL,SS}_e&=&  8 \pi  M^4 C_f
\int_0^1 dx_Bdx_1\int_0^{1/\Lambda} b_Bdb_Bb_1db_1\phi _B(x_B,b_B)\nonumber\\&&
\{ (\sqrt{g^- g^+} (f^+ (2 g^+ x_1-1)-f^-) \phi _S^s(x_1,\omega _1)+\sqrt{g^- g^+} (f^-+f^+ (2 g^+
   x_1-1)) \phi _S^t(x_1,\omega _1)\nonumber\\&&-(f^- g^-+f^+ g^+ (g^+ x_1+1)) \phi _S(x_1,\omega _1))
\alpha_s(t_a)e^{-S_{ab}(t_a)}h(\alpha_e,\beta_a,b_B,b_1)S_t(x_1)\nonumber\\&&-
(2 \sqrt{g^- g^+} (f^- (g^--x_B)+f^+ g^+) \phi _S^s(x_1,\omega _1)-g^+ ((f^-+f^+) g^--f^-
   x_B) \phi _S(x_1,\omega _1))\nonumber\\&&
\alpha_s(t_b)e^{-S_{ab}(t_b)}h(\alpha_e,\beta_b,b_1,b_B)S_t(x_B)\},
\end{eqnarray}
\begin{eqnarray}
\mathcal{M}^{LL,SS}_e&=& -16 \sqrt{\frac{2}{3}} \pi  M^4 C_f
\int_0^1 dx_Bdx_1dx_2\int_0^{1/\Lambda} b_Bdb_Bb_2db_2\phi _B(x_B,b_B)\phi_S(x_2,\omega_2)
\nonumber\\&&\{  (\sqrt{g^- g^+} (f^- (x_B+f^+ (x_2-2))-f^+ g^+ x_1) \phi
   _S^s(x_1,\omega _1)+\sqrt{g^- g^+} (f^- x_B+f^+ g^+ x_1+f^+ f^- x_2) \phi _S^t(x_1,\omega _1)\nonumber\\&&+(f^-+f^+) (g^+
   (-x_B-f^+ x_2+f^++g^- x_1)+f^- g^-) \phi _S(x_1,\omega _1))
\alpha_s(t_c)e^{-S_{cd}(t_c)}h(\beta_c,\alpha_e,b_2, b_B)\nonumber\\&&-
 (-\sqrt{g^- g^+} (f^- (-x_B)+f^+ g^+ x_1+f^+ f^- x_2) \phi
   _S^s(x_1,\omega _1)-\sqrt{g^- g^+} (f^- x_B+f^+ g^+ x_1-f^+ f^- x_2) \nonumber\\&&\phi _S^t(x_1,\omega _1)+(f^- g^-+f^+ g^+)
   (-x_B+f^+ x_2+g^+ x_1) \phi _S(x_1,\omega _1))
\alpha_s(t_d)e^{-S_{cd}(t_d)}h(\beta_d,\alpha_e,b_2,b_B)\},
\end{eqnarray}
\begin{eqnarray}
\mathcal{M}^{LR,SS}_e&=&  16 \sqrt{\frac{2}{3}} \pi  \sqrt{f^- f^+} M^4 C_f
\int_0^1 dx_Bdx_1dx_2\int_0^{1/\Lambda} b_Bdb_Bb_2db_2\phi _B(x_B,b_B)\nonumber\\&&
\{ (\sqrt{g^- g^+} \phi _S^s(x_1,\omega _1) ((-x_B+f^--f^+ x_2+f^++g^+ x_1) \phi
   _S^s(x_2,\omega _2)-(x_B+f^-+f^+ (x_2-1)+g^+ x_1) \nonumber\\&&\phi _S^t(x_2,\omega _2))+\sqrt{g^- g^+} \phi
   _S^t(x_1,\omega _1) ((x_B+f^-+f^+ (x_2-1)+g^+ x_1) \phi _S^s(x_2,\omega _2)\nonumber\\&&-(-x_B+f^--f^+ x_2+f^++g^+
   x_1) \phi _S^t(x_2,\omega _2))+\phi _S(x_1,\omega _1) ((g^+ (-x_B-f^+ x_2+f^++g^- x_1)+f^- g^-)\nonumber\\&& \phi
   _S^s(x_2,\omega _2)-(g^+ (x_B+f^+ (x_2-1)+g^- x_1)+f^- g^-) \phi _S^t(x_2,\omega _2)))
\alpha_s(t_c)e^{-S_{cd}(t_c)}h(\beta_c,\alpha_e,b_2,b_B)\nonumber\\&&
-(g^+ \phi _S(x_1,\omega _1) ((-x_B+f^+ x_2+g^- x_1) \phi _S^s(x_2,\omega
   _2)+(x_B-f^+ x_2+g^- x_1) \phi _S^t(x_2,\omega _2))\nonumber\\&&+\sqrt{g^- g^+} \phi _S^s(x_1,\omega _1) ((-x_B+f^+
   x_2+g^+ x_1) \phi _S^s(x_2,\omega _2)+(x_B-f^+ x_2+g^+ x_1) \phi _S^t(x_2,\omega _2))\nonumber\\&&+\sqrt{g^- g^+} \phi
   _S^t(x_1,\omega _1) ((x_B-f^+ x_2+g^+ x_1) \phi _S^s(x_2,\omega _2)+(-x_B+f^+ x_2+g^+ x_1) \phi
   _S^t(x_2,\omega _2)))\nonumber\\&&
   \alpha_s(t_d)e^{-S_{cd}(t_d)}h(\beta_d,\alpha_e,b_2,b_B)\}.
\end{eqnarray}
\begin{eqnarray}
\mathcal{F}^{LL(LR),SS}_a&=& 8 \pi  M^4 f_B C_f
\int_0^1dx_1dx_2\int_0^{1/\Lambda} b_1db_1b_2db_2\nonumber\\&&
\{ (2 \sqrt{f^- f^+} \sqrt{g^- g^+} \phi _S^s(x_1,\omega _1) ((f^+ x_2+g^-+g^+) \phi _S^s(x_2,\omega
   _2)+(f^+ x_2+g^--g^+) \phi _S^t(x_2,\omega _2))\nonumber\\&&-g^+ (f^- g^-+f^+ (f^+ x_2+g^-)) \phi _S(x_1,\omega
   _1) \phi _S(x_2,\omega _2))
\alpha_s(t_e)e^{-S_{ef}(t_e)}h(\alpha_a,\beta_e,b_1,b_2)S_t(x_2)\nonumber\\&&
-(-2 \sqrt{f^- f^+} \sqrt{g^- g^+} (f^--f^+-g^--g^+ x_1+g^+) \phi _S^s(x_2,\omega _2) \phi _S^t(x_1,\omega
   _1)\nonumber\\&&+2 \sqrt{f^- f^+} \sqrt{g^- g^+} (f^-+f^++g^--g^+ x_1+g^+) \phi _S^s(x_1,\omega _1) \phi _S^s(x_2,\omega _2)\nonumber\\&&-(f^+
   (g^+)^2 (-(x_1-1))+f^- g^- (f^++g^-)+f^- f^+ g^+) \phi _S(x_1,\omega _1) \phi _S(x_2,\omega
   _2))\nonumber\\&&\alpha_s(t_f)e^{-S_{ef}(t_f)}h(\alpha_a,\beta_f,b_2,b_1)S_t(x_1)\},
\end{eqnarray}
\begin{eqnarray}
\mathcal{F}^{SP,SS}_a&=&16 \pi  M^4 f_B C_f
\int_0^1dx_1dx_2\int_0^{1/\Lambda} b_1db_1b_2db_2\nonumber\\&&
\{  (\sqrt{f^- f^+} g^+ \phi _S(x_1,\omega _1) ((f^+ x_2+2 g^-) \phi _S^s(x_2,\omega _2)-f^+ x_2 \phi
   _S^t(x_2,\omega _2))\nonumber\\&&-2 \sqrt{g^- g^+} (f^+ g^++f^- (f^+ x_2+g^-)) \phi _S(x_2,\omega _2) \phi
   _S^s(x_1,\omega _1))\alpha_s(t_e)e^{-S_{ef}(t_e)}h(\alpha_a,\beta_e,b_1,b_2)S_t(x_2)\nonumber\\&&
+ (-\sqrt{g^- g^+} (f^- (2 f^++g^-)-f^+ g^+ (x_1-1)) \phi _S(x_2,\omega _2) \phi
   _S^s(x_1,\omega _1)\nonumber\\&&+2 \sqrt{f^- f^+} (f^- g^-+g^+ (f^++g^- (-x_1)+2 g^-)) \phi _S(x_1,\omega _1) \phi
   _S^s(x_2,\omega _2)\nonumber\\&&+\sqrt{g^- g^+} (f^- g^-+f^+ g^+ (x_1-1)) \phi _S(x_2,\omega _2) \phi _S^t(x_1,\omega
   _1))\nonumber\\&& \alpha_s(t_f)e^{-S_{ef}(t_f)}h(\alpha_a,\beta_f,b_2,b_1)S_t(x_1)\},
\end{eqnarray}
\begin{eqnarray}
\mathcal{M}^{LL,SS}_a&=&16 \sqrt{\frac{2}{3}} \pi  M^4 C_f
 \int_0^1dx_Bdx_1dx_2\int_0^{1/\Lambda} b_1db_1b_Bdb_B\phi _B(x_B,b_B) \nonumber\\&&
\{  (\sqrt{f^- f^+} \sqrt{g^- g^+} \phi _S^s(x_1,\omega _1) ((-x_B+f^+ x_2+g^--g^+ x_1+g^+) \phi
   _S^s(x_2,\omega _2)\nonumber\\&&+(x_B-f^+ x_2-g^--g^+ x_1+g^+) \phi _S^t(x_2,\omega _2))+\sqrt{f^- f^+} \sqrt{g^- g^+} \phi
   _S^t(x_1,\omega _1) \nonumber\\&&((-x_B+f^+ x_2+g^-+g^+ (x_1-1)) \phi _S^s(x_2,\omega _2)-(-x_B+f^+ x_2+g^--g^+
   x_1+g^+) \phi _S^t(x_2,\omega _2))\nonumber\\&&-(g^-+g^+) (f^- (-x_B+f^+ x_2+g^-)-f^+ g^+ (x_1-1)) \phi
   _S(x_1,\omega _1) \phi _S(x_2,\omega _2))\nonumber\\&&\alpha_s(t_g)e^{-S_{gh}(t_g)}h(\beta_g,\alpha_a,b_B,b_1)\nonumber\\&&
-(\sqrt{f^- f^+} \sqrt{g^- g^+} \phi _S^s(x_1,\omega _1) ((x_B+f^+ x_2+g^--g^+ x_1+g^++2) \phi
   _S^s(x_2,\omega _2)\nonumber\\&&+(x_B+f^+ x_2+g^-+g^+ (x_1-1)) \phi _S^t(x_2,\omega _2))-\sqrt{f^- f^+} \sqrt{g^- g^+} \phi
   _S^t(x_1,\omega _1) \nonumber\\&&((x_B+f^+ x_2+g^-+g^+ (x_1-1)) \phi _S^s(x_2,\omega _2)+(x_B+f^+ x_2+g^--g^+
   x_1+g^+-2) \phi _S^t(x_2,\omega _2))\nonumber\\&&-(f^- g^+ (x_B-g^- (x_1-2)-1)+f^+ (g^+ (x_B+f^- x_2)-g^-
   (g^+ (x_1-2)+1))+(f^+)^2 g^+ x_2)\nonumber\\&& \phi _S(x_1,\omega _1) \phi _S(x_2,\omega _2))
   \alpha_s(t_h)e^{-S_{gh}(t_h)}h(\beta_h,\alpha_a,b_B,b_1)\},
\end{eqnarray}
\begin{eqnarray}
\mathcal{M}^{LR,SS}_a&=&16 \sqrt{\frac{2}{3}} \pi  M^4 C_f
 \int_0^1dx_Bdx_1dx_2\int_0^{1/\Lambda} b_1db_1b_Bdb_B\phi _B(x_B,b_B) \nonumber\\&&
\{  (\sqrt{f^- f^+} g^+ \phi _S(x_1,\omega _1) ((-x_B+f^+ x_2+g^- x_1) \phi _S^t(x_2,\omega
   _2)-(x_B-f^+ x_2+g^- (x_1-2)) \phi _S^s(x_2,\omega _2))\nonumber\\&&+\sqrt{g^- g^+} (f^- (-x_B+f^+ x_2+g^-)-f^+ g^+
   (x_1-1)) \phi _S(x_2,\omega _2) \phi _S^s(x_1,\omega _1)\nonumber\\&&+\sqrt{g^- g^+} (f^- (-x_B+f^+ x_2+g^-)+f^+ g^+
   (x_1-1)) \phi _S(x_2,\omega _2) \phi _S^t(x_1,\omega _1))\nonumber\\&&\alpha_s(t_g)e^{-S_{gh}(t_g)}h(\beta_g,\alpha_a,b_B,b_1)\nonumber\\&&
- (\sqrt{f^- f^+} \phi _S(x_1,\omega _1) ((g^+ (x_B+f^+ x_2-2)-g^- (g^+
   (x_1-2)+2)) \phi _S^s(x_2,\omega _2)\nonumber\\&&+(g^+ (x_B+f^+ x_2-2)+g^- (g^+ x_1+2)) \phi
   _S^t(x_2,\omega _2))\nonumber\\&&+\sqrt{g^- g^+} (f^- (x_B+f^+ x_2+g^--2)+f^+ (g^+ (-x_1)+g^+-2)) \phi
   _S(x_2,\omega _2) \phi _S^s(x_1,\omega _1)\nonumber\\&&+\sqrt{g^- g^+} (f^- (x_B+f^+ x_2+g^--2)+f^+ (g^+
   (x_1-1)+2)) \phi _S(x_2,\omega _2) \phi _S^t(x_1,\omega _1))\nonumber\\&&\alpha_s(t_h)e^{-S_{gh}(t_h)}h(\beta_h,\alpha_a,b_B,b_1)\},
\end{eqnarray}
\begin{eqnarray}
\mathcal{M}^{SP,SS}_a&=&16\sqrt{\frac{2}{3}} \pi  M^4 C_f
 \int_0^1dx_Bdx_1dx_2\int_0^{1/\Lambda} b_1db_1b_Bdb_B\phi _B(x_B,b_B) \nonumber\\&&
\{   (\sqrt{f^- f^+} \sqrt{g^- g^+} \phi _S^s(x_1,\omega _1) ((-x_B+f^+ x_2+g^--g^+ x_1+g^+) \phi
   _S^s(x_2,\omega _2)\nonumber\\&&+(-x_B+f^+ x_2+g^-+g^+ (x_1-1)) \phi _S^t(x_2,\omega _2))-\sqrt{f^- f^+} \sqrt{g^- g^+} \phi
   _S^t(x_1,\omega _1) \nonumber\\&&((-x_B+f^+ x_2+g^-+g^+ (x_1-1)) \phi _S^s(x_2,\omega _2)+(-x_B+f^+ x_2+g^--g^+
   x_1+g^+) \phi _S^t(x_2,\omega _2))\nonumber\\&&+(f^-+f^+) g^+ (x_B-f^+ x_2+g^- (x_1-2)) \phi _S(x_1,\omega
   _1) \phi _S(x_2,\omega _2))
   \alpha_s(t_g)e^{-S_{gh}(t_g)}h(\beta_g,\alpha_a,b_B,b_1)\nonumber\\&&
- (\sqrt{f^- f^+} \sqrt{g^- g^+} \phi _S^s(x_1,\omega _1) ((x_B+f^+ x_2+g^--g^+ x_1+g^++2) \phi
   _S^s(x_2,\omega _2)\nonumber\\&&-(x_B+f^+ x_2+g^-+g^+ (x_1-1)) \phi _S^t(x_2,\omega _2))+\sqrt{f^- f^+} \sqrt{g^- g^+} \phi
   _S^t(x_1,\omega _1) \nonumber\\&&((x_B+f^+ x_2+g^-+g^+ (x_1-1)) \phi _S^s(x_2,\omega _2)-(x_B+f^+ x_2+g^--g^+
   x_1+g^+-2) \phi _S^t(x_2,\omega _2))\nonumber\\&&-(f^- (g^-+g^+) (x_B+g^-)-f^- g^++f^+ g^+ (f^- x_2-g^+
   x_1+g^+)+f^+ g^- (f^- x_2-g^+ x_1+g^+-1))\nonumber\\&& \phi _S(x_1,\omega _1) \phi _S(x_2,\omega _2))\alpha_s(t_h)e^{-S_{gh}(t_h)}h(\beta_h,\alpha_a,b_B,b_1)\}.
\end{eqnarray}
$\mathcal{F'}_a(\mathcal{M'}_a)$ is obtained from $\mathcal{F}_a(\mathcal{M}_a)$ via the exchanges
$x_1\leftrightarrow x_2$ and $\omega_1\leftrightarrow \omega_2$.



The additional factorizable emission contributions $\mathcal{F}_e^{SP}$ appear in the $VS$ and $SS$ amplitudes:
\begin{eqnarray}
\mathcal{F}^{SP,VS}_e&=& 16 \pi  \sqrt{f^- f^+} M^4 C_f
\int_0^1 dx_Bdx_1\int_0^{1/\Lambda} b_Bdb_Bb_1db_1\phi _B(x_B,b_B)\nonumber\\&&
\{(\sqrt{g^- g^+} (g^+ x_1+2) \phi _P^s(x_1,\omega _1)-g^+ \sqrt{g^- g^+} x_1 \phi _P^t(x_1,\omega
   _1)+(g^++g^- (2 g^+ x_1-1)) \phi _P(x_1,\omega _1))\nonumber\\&&
\alpha_s(t_a)e^{-S_{ab}(t_a)}h(\alpha_e,\beta_a,b_B,b_1)S_t(x_1)\nonumber\\&&+
(g^+ x_B \phi _P(x_1,\omega _1)+2 \sqrt{g^- g^+} (-x_B+g^-+g^+) \phi _P^s(x_1,\omega _1))\nonumber\\&&
\alpha_s(t_b)e^{-S_{ab}(t_b)}h(\alpha_e,\beta_b,b_1,b_B)S_t(x_B)\},
\end{eqnarray}
\begin{eqnarray}
\mathcal{F}^{SP,SS}_e&=& 16 \pi  \sqrt{f^- f^+} M^4 C_f
\int_0^1 dx_Bdx_1\int_0^{1/\Lambda} b_Bdb_Bb_1db_1\phi _B(x_B,b_B)\nonumber\\&&
\{ (\sqrt{g^- g^+} (g^+ x_1+2) \phi _S^s(x_1,\omega _1)-g^+ \sqrt{g^- g^+} x_1 \phi _S^t(x_1,\omega
   _1)+(g^-+g^+-2 g^+ g^- x_1) \phi _S(x_1,\omega _1))\nonumber\\&&
\alpha_s(t_a)e^{-S_{ab}(t_a)}h(\alpha_e,\beta_a,b_B,b_1)S_t(x_1)\nonumber\\&&+
(2 \sqrt{g^- g^+} (-x_B+g^-+g^+) \phi _S^s(x_1,\omega _1)+g^+ (x_B-2 g^-) \phi _S(x_1,\omega
   _1))\nonumber\\&&
\alpha_s(t_b)e^{-S_{ab}(t_b)}h(\alpha_e,\beta_b,b_1,b_B)S_t(x_B)\}.
\end{eqnarray}

The $B\rightarrow K\pi$ transition form factor $F_{\perp}$  
is defined via the hadronic matrix element of a $b\rightarrow s$ vector current~\cite{jhep122019083},
\begin{eqnarray}
i\langle K(p_1)\pi(p_1') |\bar{s}\gamma^{\mu}b|B(p_B) \rangle &=&F_{\perp}(\omega_1^2,q^2)k^{\mu}_{\perp},
\end{eqnarray}
with  the vector $k^{\mu}_{\perp}$ being specified in~\cite{jhep122019083}.
A straightforward calculation based on  Figs.~\ref{fig:fym}(a), (b) leads to
\begin{eqnarray}
F_{\perp}(\omega_1^2,q^2) &=&- 8 \pi  M^2  C_f
\int_0^1 dx_Bdx_1\int_0^{1/\Lambda} b_Bdb_Bb_1db_1\phi _B(x_B,b_B)\nonumber\\&&
\{\omega_1 (x_1 g^++2) \phi _P^a(x_1,\omega _1)+M(g^++g^--2g^+g^-x_1) \phi _P^T(x_1,\omega _1)
-\omega_1x_1g^+ \phi^v_P(x_1,\omega _1))\nonumber\\&&
\alpha_s(t_a)e^{-S_{ab}(t_a)}h(\alpha_e,\beta_a,b_B,b_1)S_t(x_1)\nonumber\\&&+
\omega_1[(g^++g^--x_B) \phi _P^a(x_1,\omega _1)+(g^+-g^-+x_B) \phi^v _P(x_1,\omega _1)]\nonumber\\&&
\alpha_s(t_b)e^{-S_{ab}(t_b)}h(\alpha_e,\beta_b,b_1,b_B)S_t(x_B)\}.
\end{eqnarray}

The hard functions $h$, resulting from the Fourier transformation  of virtual quark and gluon
propagators, read
\begin{eqnarray}\label{h}
h(\alpha,\beta,b_1,b_2)&=&h_1(\alpha,b_1)\times h_2(\beta,b_1,b_2),\nonumber\\
h_1(\alpha,b_1)&=&\left\{\begin{array}{ll}
K_0(\sqrt{\alpha}b_1), & \quad  \quad \alpha >0,\\
K_0(i\sqrt{-\alpha}b_1),& \quad  \quad \alpha<0,
\end{array} \right.\nonumber\\
h_2(\beta,b_1,b_2)&=&\left\{\begin{array}{ll}
\theta(b_1-b_2)I_0(\sqrt{\beta}b_2)K_0(\sqrt{\beta}b_1)+(b_1\leftrightarrow b_2), & \quad   \beta >0,\\
\theta(b_1-b_2)J_0(\sqrt{-\beta}b_2)K_0(i\sqrt{-\beta}b_1)+(b_1\leftrightarrow b_2),& \quad   \beta<0,
\end{array} \right.
\end{eqnarray}
where $J_0$ is the Bessel function, and $K_0$ and $I_0$ are the modified Bessel functions.
The arguments $\alpha$ and $\beta$ of the hard functions, related to the virtuality of
the internal gluon and quark, respectively, have the expressions
\begin{eqnarray}
\alpha_{e}&=&x_Bx_1g^+M^2,\nonumber\\
\alpha_{a}&=&-g^+(1-x_1)(f^+x_2+g^-)M^2,\nonumber\\
\beta_{a}&=&x_1g^+M^2,\nonumber\\
\beta_{b}&=&(x_B-g^-)g^+M^2,\nonumber\\
\beta_{c}&=&[f^-+g^+x_1][f^+(x_2-1)+x_B]M^2,\nonumber\\
\beta_{d}&=&-x_1g^+[f^+x_2-x_B]M^2,\nonumber\\
\beta_{e}&=&-g^+(f^+x_2+g^-)M^2,\nonumber\\
\beta_{f}&=&-(1-g^+x_1)M^2,\nonumber\\
\beta_{g}&=&g^+(x_1-1)(f^+x_2+g^--x_B)M^2,\nonumber\\
\beta_{h}&=&\{(g^+(x_1-1)+1)(f^+x_2+g^-+x_B-1)+1\}M^2,\nonumber\\
\alpha'_{a}&=&f^+(x_2-1)(f^-+g^+x_1)M^2,\nonumber\\
\beta'_{e}&=&-f^+(g^+x_1+f^-)M^2,\nonumber\\
\beta'_{f}&=&-(1-f^+x_2)M^2,\nonumber\\
\beta'_{g}&=&f^+(x_2-1)(g^+x_1+f^--x_B)M^2,\nonumber\\
\beta'_{h}&=&\{(f^+(x_2-1)+1)(g^+x_1+f^-+x_B-1)+1\}M^2.\nonumber\\
\end{eqnarray}
The hard scales $t$ are chosen as the maximal virtuality of the internal particles,
including $1/b_i$, $i=B,1,2$:
\begin{eqnarray}\label{eq:scale}
t_{a,b}&=&\max(\sqrt{\alpha_e},\sqrt{\beta_{a,b}},1/b_B,1/b_1), \nonumber\\
t_{c,d}&=&\max(\sqrt{\alpha_e},\sqrt{\beta_{c,d}},1/b_B,1/b_2),\nonumber\\
t^{(')}_{e,f}&=&\max(\sqrt{\alpha^{(')}_a},\sqrt{\beta^{(')}_{e,f}},1/b_1,1/b_2),\nonumber\\
t^{(')}_{g,h}&=&\max(\sqrt{\alpha^{(')}_a},\sqrt{\beta^{(')}_{g,h}},1/b_B,1/b_1).
\end{eqnarray}

The  Sudakov exponents are given by
\begin{eqnarray}\label{ss}
S_{ab}(t)&=&s(\frac{M}{\sqrt{2}}x_B,b_B)+s(\frac{M}{\sqrt{2}}g^+x_1,b_1)+s(\frac{M}{\sqrt{2}}g^+(1-x_1),b_1)
+\frac{5}{3}\int_{1/b_B}^t\frac{d\mu}{\mu}\gamma_q(\mu)+2\int_{1/b_1}^t\frac{d\mu}{\mu}\gamma_q(\mu),\nonumber\\
S_{cd}(t)&=&s(\frac{M}{\sqrt{2}}x_B,b_B)+s(\frac{M}{\sqrt{2}}g^+x_1,b_B)+s(\frac{M}{\sqrt{2}}g^+(1-x_1),b_B)
+s(\frac{M}{\sqrt{2}}f^+x_2,b_2)+s(\frac{M}{\sqrt{2}}f^+(1-x_2),b_2)\nonumber\\&&
+\frac{11}{3}\int_{1/b_B}^t\frac{d\mu}{\mu}\gamma_q(\mu)+2\int_{1/b_2}^t\frac{d\mu}{\mu}\gamma_q(\mu),\nonumber\\
S_{ef}(t)&=&s(\frac{M}{\sqrt{2}}g^+x_1,b_1)+s(\frac{M}{\sqrt{2}}g^+(1-x_1),b_1)+s(\frac{M}{\sqrt{2}}f^+x_2,b_2)+s(\frac{M}{\sqrt{2}}f^+(1-x_2),b_2)\nonumber\\&&
+2\int_{1/b_1}^t\frac{d\mu}{\mu}\gamma_q(\mu)+2\int_{1/b_2}^t\frac{d\mu}{\mu}\gamma_q(\mu),\nonumber\\
S_{gh}(t)&=&s(\frac{M}{\sqrt{2}}x_B,b_B)+s(\frac{M}{\sqrt{2}}g^+x_1,b_1)+s(\frac{M}{\sqrt{2}}g^+(1-x_1),b_1)
+s(\frac{M}{\sqrt{2}}f^+x_2,b_1)+s(\frac{M}{\sqrt{2}}f^+(1-x_2),b_1)\nonumber\\&&
+\frac{5}{3}\int_{1/b_B}^t\frac{d\mu}{\mu}\gamma_q(\mu)+4\int_{1/b_1}^t\frac{d\mu}{\mu}\gamma_q(\mu),
\end{eqnarray}
where $\gamma_q=-\alpha_s/\pi$ is the quark anomalous dimension, and
the explicit expression of the function $s(Q,b)$ can be found in   \cite{epjc11695}.
The threshold resummation factor $S_t(x)$ is adopted from Ref.~\cite{prd65014007},
\begin{eqnarray}
S_t(x)=\frac{2^{1+2c}\Gamma (3/2+c)}{\sqrt{\pi}\Gamma(1+c)}[x(1-x)]^c,
\end{eqnarray}
with $c=0.3$.

\end{appendix}

\end{document}